\documentclass[aps,prd,nofootinbib,twocolumn,superscriptaddress,preprintnumbers,balancelastpage,longbibliography]{revtex4-1}

\usepackage{pifont}

\usepackage{url}
\usepackage{appendix}
\usepackage{amsmath,amssymb,mathtools,bm}
\usepackage{graphicx, color, hepunits}
\usepackage[dvipsnames]{xcolor}
\usepackage{float}
\usepackage{multirow}
\usepackage{filecontents}
\usepackage[colorlinks=true 
,urlcolor=purple 
,anchorcolor=blue
,citecolor=blue 
,filecolor=magenta  
,linkcolor=blue 
,menucolor=blue
,linktocpage=true
,pdfproducer=medialab
,pdfa=true
]{hyperref}

\usepackage{tabularx}
\usepackage{array}
\newcolumntype{P}[1]{>{\centering\arraybackslash}p{#1}}
\newcommand{\ra}[1]{\renewcommand{\arraystretch}{#1}}
\newcommand{\cmark}{\ding{51}}%
\newcommand{\xmark}{\ding{55}}%

\makeatletter
\def\hlinewd#1{%
\noalign{\ifnum0=`}\fi\hrule \@height #1 \futurelet
\reserved@a\@xhline}
\makeatother
\newcolumntype{L}[1]{>{\raggedright\let\newline\\\arraybackslash\hspace{0pt}}m{#1}}
\newcolumntype{C}[1]{>{\centering\let\newline\\\arraybackslash\hspace{0pt}}m{#1}}
\newcolumntype{R}[1]{>{\raggedleft\let\newline\\\arraybackslash\hspace{0pt}}m{#1}}
\renewcommand{\arraystretch}{1.2}

\newcommand{\es}[2] {\begin{equation} \label{#1} \begin{split} #2 \end{split} \end{equation}}

\begin{document}

\title{Was There a 3.5 keV Line?}

\author{Christopher Dessert}
\email{cd2458@nyu.edu}
\affiliation{Center for Cosmology and Particle Physics, Department of Physics,
New York University, New York, NY 10003, USA}
\affiliation{Center for Computational Astrophysics, Flatiron Institute, New York, NY 10010, USA}

\author{Joshua W. Foster}
\email{jwfoster@mit.edu}
\affiliation{Center for Theoretical Physics, Massachusetts Institute of Technology, Cambridge, Massachusetts 02139, U.S.A}

\author{Yujin Park}
\email{yjpark99@berkeley.edu}
\affiliation{Berkeley Center for Theoretical Physics, University of California, Berkeley, CA 94720, U.S.A.}
\affiliation{Theoretical Physics Group, Lawrence Berkeley National Laboratory, Berkeley, CA 94720, U.S.A.}

\author{Benjamin R. Safdi}
\email{brsafdi@berkeley.edu}
\affiliation{Berkeley Center for Theoretical Physics, University of California, Berkeley, CA 94720, U.S.A.}
\affiliation{Theoretical Physics Group, Lawrence Berkeley National Laboratory, Berkeley, CA 94720, U.S.A.}

\date{\today}
\preprint{MIT-CTP/5600}

\begin{abstract}
The 3.5 keV line is a purported emission line observed in galaxies, galaxy clusters, and the Milky Way whose origin is inconsistent with known atomic transitions and has previously been suggested to arise from dark matter decay. We systematically re-examine the bulk of the evidence for the 3.5 keV line, attempting to reproduce six previous analyses that found evidence for the line.
Surprisingly, we only reproduce one of the analyses; 
in the other five we find 
no significant evidence for a 3.5 keV line when following the described analysis procedures on the original data sets.   
For example, previous results claimed 4$\sigma$ evidence for a 3.5 keV line from the Perseus cluster; we dispute this claim, finding no evidence for a 3.5 keV line. 
We find evidence for background mismodeling in multiple analyses.
We show that analyzing these data in narrower energy windows diminishes the effects of mismodeling but 
returns no evidence for a 3.5 keV line.
We conclude that there is little robust evidence for the existence of the 3.5 keV line. 
 Some of the discrepancy of our results from those of the original works may be due to the earlier reliance on local optimizers, 
which we demonstrate can lead to incorrect results.  For ease of reproducibility, all code and data are publicly available. 
\end{abstract}\maketitle

\section{Introduction}

\begin{figure}[!ht]  
\hspace{0pt}
\vspace{-0.2in}
\begin{center}
\includegraphics[width=0.49\textwidth]{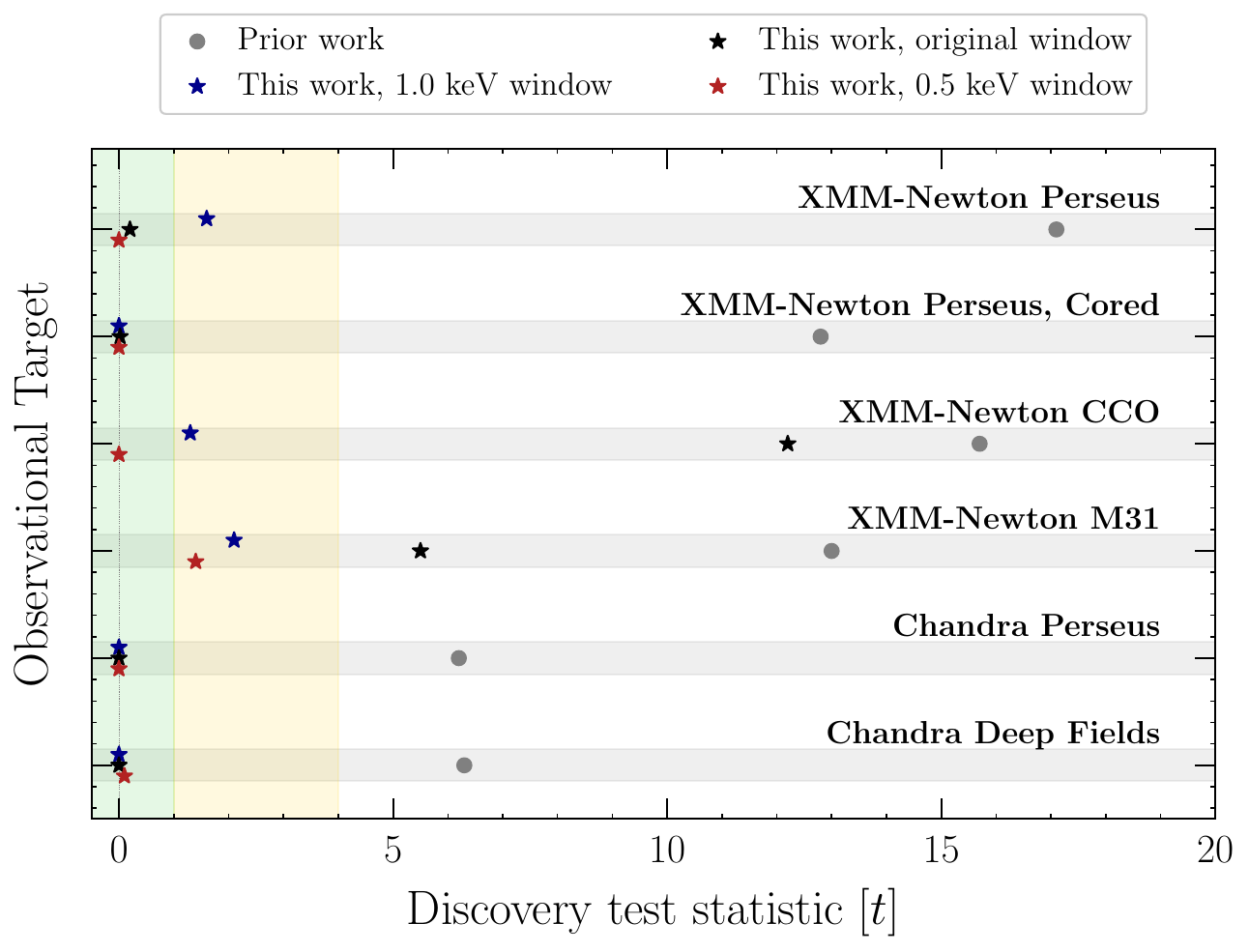}
\caption{The one-sided discovery test statistic ($t$) in favor of the signal model with a 3.5 keV line relative to the null hypothesis. Note that $\sqrt{t}$ is roughly interpreted as the number of ``$\sigma$" in favor of the 3.5 keV line model.  The 1$\sigma$  (2$\sigma$) expectation under the null hypothesis is shaded in green (gold). 
We show the results from the original works as `Prior work' for the six targets considered in this article. We do not, however, reproduce the results of these studies, finding no evidence for a 3.5 keV line in most of our reanalyses.
In all of the analyses there is no evidence for the 3.5 keV line when performing narrower-energy-window analyses, as indicated; these are less subject to background mismodeling. See Tab.~\ref{tab:Results} for a full summary.} 
\label{fig:SigPlot}
\end{center}
\end{figure}

Decaying dark matter (DM) models such as sterile neutrino DM may lead to narrow spectral features in the $X$-ray band from galaxies, galaxy clusters, and otherwise empty regions of the Milky Way. 
For this reason, a significant interest was generated in 2014 by the claimed discovery of an unassociated $X$-ray line (UXL) at an energy near 3.5 keV by the XMM-Newton and Chandra telescopes~\cite{Bulbul:2014sua,Boyarsky:2014jta,Cappelluti:2017ywp} that appeared consistent with arising from decaying DM. In this work, we revisit and perform reanalyses of the following foundational studies on the 3.5 keV UXL: (i) XMM-Newton Perseus cluster, with and without the central core of the cluster masked (Ref.~\cite{Bulbul:2014sua}); (ii) XMM-Newton stacked clusters (Ref.~\cite{Bulbul:2014sua}); (iii) XMM-Newton M31 (Ref.~\cite{Boyarsky:2014jta}); (iv) Chandra Perseus cluster (Ref.~\cite{Bulbul:2014sua}); and (v) Chandra Deep Field (Ref.~\cite{Cappelluti:2017ywp}). We find that most of these results are not reproducible, giving instead no evidence for a 3.5 keV UXL when following the claimed analysis procedures.  
A summary of our key results is provided in Fig.~\ref{fig:SigPlot}, which shows that our reanalyses result in mostly insignificant evidence in favor of the 3.5 keV UXL.

All of the analyses we consider share the common feature that they are searches for a narrow spectral feature at 3.5 keV over an otherwise mostly smooth background, and all use either XMM-Newton or Chandra $X$-ray telescope data. These UXL searches are made difficult by the fact that the continuum background is difficult to model and thus subject to systematic uncertainties. Additionally, instrumental lines and the numerous astrophysical lines 
complicate the background fitting. As a result, many of the previous analyses have dozens of model parameters. These complications motivate two major concerns with regards to previous studies of the 3.5 keV line using parametric frequentist statistics: (i) systematic bias in the recovered 3.5 keV UXL signal strength and claimed statistical evidence due to mismodeling; and (ii) not achieving the correct global likelihood maximum given the large number of model parameters. We find evidence that both of these concerns likely played a role in providing fictitious evidence for the 3.5 keV line in the previous analyses that we revisit. For example, we suspect that the previous studies we revisit 
made the error of using local optimizers that converged away from the global likelihood maximum, given the parameter ranges and model components outlined in those works.

The goals of this work are to (i) reproduce the original evidence for the 3.5 keV line, and then (ii) examine the robustness of the evidence. To achieve (i) we follow as closely as possible the original analysis frameworks using the original data sets, though as we discuss further in Sec.~\ref{sec:discussion}, these analysis strategies are not ideal for looking for decaying DM.  Our inability to reproduce many of the original results likely points to errors in those works, though we are not able to decisively identify the sources of error. However, our use of global instead of local optimization appears important.  To achieve goal (ii), we consider the robustness of our results to shrinking the energy-windows of the analyses, which should help mitigate background mismodeling; as seen in Fig.~\ref{fig:SigPlot}, shrinking the analysis window leads to no evidence for a 3.5 keV line in any of the data sets. To ensure the reproducibility of our results, we provide \href{https://github.com/joshwfoster/3p5}{supplementary code} and data for each step in our data reduction and analysis pipelines~\cite{suppdata}. 

Our work is significant both because of the interest generated by the 3.5 keV line specifically but also, looking towards the future, because it informs how analyses searching for decaying DM and UXLs in the $X$-ray band should be performed. To date, significant effort has been devoted to searching for UXLs with space-based $X$-ray telescopes that may have a DM origin (see~\cite{Drewes:2016upu,Abazajian:2017tcc,Boyarsky:2018tvu} for reviews). Sterile neutrino DM provides an especially motivated target for these searches since these models may explain the active neutrino masses, while also having natural mechanisms to explain the observed DM abundance~\cite{Dodelson:1993je,Shi:1998km,Kusenko:2006rh}. Sterile neutrinos may decay to active neutrinos and mono-energetic $X$-rays at a rate potentially within the reach of the sensitivity of current instruments~\cite{Pal:1981rm}, and are an important new-physics target for upcoming instruments such as eROSITA~\cite{2012arXiv1209.3114M,Dekker:2021bos}, XRISM~\cite{XRISMScienceTeam:2020rvx,Dessert:2023vyl}, {\it Athena}~\cite{2015JPhCS.610a2008B,Neronov:2015kca,Piro:2021oaa}, and LEM~\cite{2022arXiv221109827K,Krnjaic:2023odw}. We show explicitly that the planned analysis techniques for these searches can lead to mismodeling and spurious signals and point to improved methodologies for the future. 

Even prior to this work, however, the decaying DM interpretation of the 3.5 keV UXL appeared strongly disfavored.  A number of extragalactic targets, including Milky Way dwarf galaxies, 
failed to show evidence for a 3.5 keV line at the expected level given the DM abundances in these systems~\cite{Horiuchi:2013noa,Malyshev:2014xqa,Anderson:2014tza,Tamura:2014mta,Jeltema:2015mee,Hitomi:2016mun}.  Furthermore, analyses of archival XMM-Newton data failed to detect evidence for a 3.5 keV UXL in the ambient DM halo of the Milky Way, ruling out the DM interpretation of the UXL by over an order of magnitude in the lifetime~\cite{Dessert:2018qih,Dessert:2020hro,Foster:2021ngm}. Similar analyses of archival Chandra~\cite{Sicilian:2020glg}, {\it NuSTAR}~\cite{Roach:2022lgo}, {\it Swift}~\cite{Sicilian:2022wvm}, and {\it Hitomi}~\cite{Dessert:2023vyl} observations of the Milky Way halo also found no evidence for a DM decay origin of the UXL.  Additionally, a $\sim$7 keV sterile neutrino is now known to dampen small-scale structure at a level inconsistent with Milky Way satellite galaxy counts~\cite{DES:2020fxi}, though in principle the 3.5 keV line could arise from the decay of a completely non-thermal DM candidate such as an axion-like particle.

A number of Standard Model and Beyond the Standard Model explanations of the 3.5 keV UXL have been proposed, beyond the simplest decaying DM paradigm.
For example, DM may decay into an axion-like-particle, which then converts to an $X$-ray in astrophysical magnetic fields~\cite{Conlon:2014xsa,Cicoli:2014bfa}, or an excited DM state may decay into a ground state and an $X$-ray~\cite{Finkbeiner:2014sja}.  Another possibility is that the UXL is not related to DM but rather to poorly-understood astrophysical processes within the galaxies and galaxy clusters that have 3.5 keV excesses. For example, the excess emission may be partially due to contamination from nearby K and Ar lines~\cite{Jeltema:2014qfa}. On the other hand, while recent laboratory measurements support the existence of lines not included in standard $X$-ray databases, they do not appear to have sufficient emissivity to account for the UXL~\cite{Bulbul:2018tbf,Weller_2019,Gall:2019vib}. Charge exchange processes, on the other hand, remain a feasible explanation~\cite{Gu:2015gqm,Shah:2016efh}. 
Our work claims, in contrast, that the 
3.5 keV UXL does not exist as a physical emission line to be understood.

The remainder of this article proceeds as follows. 
In Sec.~\ref{sec:methods} we discuss the methods we use in reanalyzing the XMM-Newton and Chandra data sets. In particular, we discuss in the context of toy examples how narrowing the analysis energy range may help mitigate the effects of mismodeling, and we provide a simple example that emphasizes the importance of global versus local optimization for parametric frequentist inference.  In Sec.~\ref{sec:xmm-cluster} we then revisit the analyses of galaxy clusters, including the Perseus cluster, originally performed in~\cite{Bulbul:2014sua}. Sec.~\ref{sec:xmm-m31} reanalyzes the M31 XMM-Newton data studied first in~\cite{Boyarsky:2014jta}.  Secs.~\ref{sec:chandra-perseus} and~\ref{sec:survey} revisit Chandra data analyses from~\cite{Bulbul:2014sua} towards the Perseus cluster and from~\cite{Cappelluti:2017ywp} towards Milky Way blank sky regions, respectively. We conclude in Sec.~\ref{sec:discussion} with a discussion on the implications of our results both for the 3.5 keV line and for future searches with $X$-ray telescopes for UXLs.

\section{Methods and toy examples}
\label{sec:methods}

Before we consider the real \textit{X}-ray data, we begin with a discussion of toy examples that illustrate the challenges associated with analysis strategies typical of UXL searches.  The UXL search strategies commonly used in the literature, in particular in the context of the 3.5 keV UXL, typically consist of joint frequentist modeling of the UXL of interest in conjunction with a number of continuum and line-like background components. As we discuss further below, this approach brings in two major concerns: (i) mismodeling, and (ii) likelihood optimization. By mismodeling we mean that the null-hypothesis model may not be a true reflection of the data, which could bias our reconstruction of the signal model parameters. We show that narrowing the analysis energy range can help mitigate mismodeling, though at the expense of reduced sensitivity to the signal model parameters. By likelihood optimization we mean that in models with many model parameters it can be difficult to find the global maximum of the likelihood, which is required by the frequentist framework. Global optimizers are necessary in many circumstances, and we show the reliance of local optimizers instead, as has been typical in 3.5 keV studies, can lead to incorrect results.  

Our philosophy in this work is to follow as closely as possible the statistical setups in the various original references. We include the same parametric model components, with the same parameter ranges when applicable, and we use the same data sets, as much as possible.  Our goal is to check the self consistency of the original results by verifying if we can find the claimed evidence for the 3.5 keV UXL and then if so to check whether it is robust to small changes in the analysis framework, such as shrinking the analysis energy range. On the other hand, we emphasize that even though we follow the analysis frameworks in these references, we do not advocate that they are optimal approaches for searches for the UXLs in $X$-ray data sets, particularly those that arise from DM decay. We revisit this point in the Discussion in Sec.~\ref{sec:discussion}. 

\subsection{Mock data for a UXL search with mismodeling}
\label{sec:toy}

\begin{figure*}[htb]  
\hspace{0pt}
\vspace{-0.2in}
\begin{center}
\includegraphics[width=0.9\textwidth]{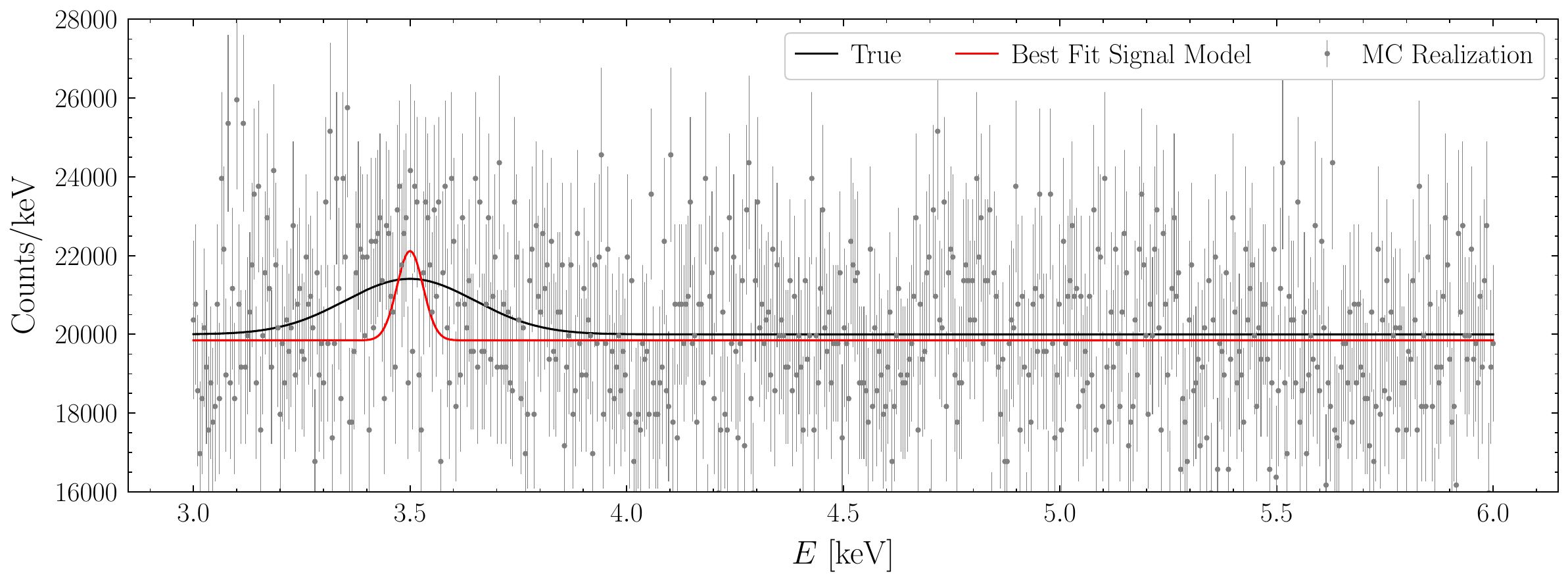}
\caption{An example simulated data set illustrating how mismodeling of the continuum background may generate artificial evidence for a narrow spectral signal.  The mock data is constructed from a background model consisting of an energy-independent contribution with 100 cts per energy bin on average, with the energy bins being 5 eV wide, in addition to a wide spectral feature centered at 3.5 keV that is described by a Gaussian with standard deviation of 150 eV and contributing on average 530 cts over all energy bins.  We analyze the simulated data under the hypothesis where the background model only consists of the energy independent contribution, with a nuisance parameter controlling the normalization, and a signal component being a narrow spectral feature centered at 3.5 keV with a standard deviation of 30 eV, also with a model parameter that controls the normalization of this feature.  While the mock data does not include a narrow spectral feature, the signal plus background model prefers a non-zero signal amplitude at over 3$\sigma$ significance because of the mismodeled broad spectral feature.  We illustrate the simulated data as constructed from the true model (black), along with the best-fit narrow signal plus flat background model (red).  One way of helping to mitigate mismodeling is to narrow the analysis energy range in order to allow the background model more freedom, as illustrated in Fig.~\ref{fig:prof_LL_ex}.
}
\label{fig:Toy_Model_Illustration}
\end{center}
\end{figure*}

In this section we construct simulated data that highlights the effects of mismodeling and possible mitigation strategies.  Our simulated data are inspired loosely by the XMM-Newton MOS data sets from clusters analyzed in~\cite{Bulbul:2014sua}, though the data are greatly simplified in this section.  In particular, we generate simulated counts in a mock detector across the energy range from $3.0$ to $6.0$ keV with energy bins of $5$ eV width.  The true background model, in counts, consists of an energy-independent background model that contributes on average 100 counts per bin.  On top of this flat background we add a broad spectral feature centered at 3.5 keV, described by a Gaussian distribution with standard deviation of $150$ eV and a total, expected number of counts of $530$.  We refer to data sets drawn from this model as the {\bf{Mismodeling  Data Sets}}.  An example simulated data set from this model, constructed by drawing Poisson counts from the total true model, is illustrated in Fig.~\ref{fig:Toy_Model_Illustration}.  The black curve, labeled ``True," is the underlying model while the data points represent the Monte Carlo (MC) realization. 

We analyze the mock data under the signal plus background hypothesis, with the signal model consisting of a narrow Gaussian feature centered at 3.5 keV with standard deviation of $30$ eV and the background consisting only of an energy-independent component.  The signal component is meant to approximate an XMM-Newton MOS response to an ultra-narrow spectral feature that is broadened by the detector energy resolution.  Importantly, since the background model does not include a broad spectral feature at $3.5$ keV, we are introducing {\it mismodeling} into the analysis. In particular, we are interested in how such mismodeling can give rise to artificial evidence for a narrow spectral feature at 3.5 keV.  We fit the signal plus background model to the MC data using a Gaussian likelihood, with Poisson counting errors:
\begin{equation}
\label{eq:LL}
    \mathcal{L}(\bm{d} | \mathcal{M}; \bm \theta) = \prod_i \mathcal{N}(\bm d_i | \mu_i = \mu_i( \bm \theta)) \,,
\end{equation}
where $\bm d$ is the data vector of counts in each bin, ${\bm \theta}$ are the model parameters for the combined signal and background model $\mathcal{M}$, $i$ is an index over energy bins, and $\mathcal{N}({\bf d}_i | \mu_i)$ is the Poisson probability, in the large-count (Gaussian) limit, of observing ${\bf d}_i$ counts for an expected value $\mu_i$.  We further divide the model parameter vector into ${\bm \theta} = \{ A, {\bm \theta}_{\rm nuis}\}$, with $A$ the signal amplitude parameter and ${\bm \theta}_{\rm nuis}$ the set of nuisance parameters that parameterize the background model.  In this case ${\bm \theta}_{\rm nuis}$ has a single parameter that controls the normalization of the flat background.
The best-fit signal plus background model is illustrated in red in Fig.~\ref{fig:Toy_Model_Illustration}. 

As a side-note, we emphasize that only positive $A > 0$ are physical. On the other hand, in the frequentist statistical analysis framework it is important to also consider negative $A$, which corresponds to negative flux, since the likelihood maximum could formally appear at negative fluxes (see~\cite{Safdi:2022xkm} for an extended discussion). Throughout this work we thus consider both positive and negative signal strengths, though when {\it e.g.} computing evidence in favor of the signal model we perform a one-sided test and require the best-fit signal amplitude to be positive.  

We may compute the evidence in favor of the signal model by computing the discovery test statistic (TS) $t$, which is defined as 
\es{eq:TS}{
    t = 2 \big[ &\mathrm{max}_{ {\bm \theta}}  \log {\mathcal L}(\mathbf{d} | {\mathcal M}, {\bm \theta}) - \\
    &\mathrm{max}_{ {\bm \theta_{\rm nuis}}}  \log {\mathcal L}(\mathbf{d} | {\mathcal M}_{\rm null}, {\bm \theta_{\rm nuis}}) \big].
}
Above, ${\mathcal M}_{\rm null}$ is the null model which is described by the background-only hypothesis.  In this case the null hypothesis only has the flat background component with a single normalization parameter.  The TS may be computed for one- or two-sided tests; in this work are primarily interested in the one-sided test, where $t$ is set to zero if the best-fit signal amplitude is negative.  In this case, and assuming Wilks' theorem, the relation between $t$ and the $p$-value of the data originating from the null hypothesis is $\sqrt{t} = \Phi^{-1}(1-p)$, where $\Phi^{-1}$ is the inverse cumulative distribution of the standard normal distribution (see, {\it e.g.},~\cite{Cowan:2010js}).  Note that significance is commonly quoted as $\sqrt{t}$, with $\sqrt{t}$ being the significance in ``$\sigma$," assuming Wilks' theorem.  For the purpose of setting upper limits on $A$ it is useful to define the profile likelihood
\es{eq:PL}{
    q(A) = 2 \big[ &\mathrm{max}_{ {\bm \theta}}  \log {\mathcal L}(\mathbf{d} | {\mathcal M}, {\bm \theta}) - \\
    &\mathrm{max}_{ {\bm \theta_{\rm nuis}}}  \log {\mathcal L}(\mathbf{d} | {\mathcal M}, \{A,\bm \theta_{\rm nuis}\}) \big] \,,
}
such that $t = q(0)$.  Again assuming Wilks' theorem, the 95\% upper limit on $A$ is given by the $A > \hat A$ where $q(A) \approx 2.71$, with $\hat A$ the value that maximizes the likelihood~\cite{Cowan:2010js}.  See~\cite{Safdi:2022xkm} for a review of the frequentist statistical procedures used in this work. 

\begin{figure}[t]  
\hspace{0pt}
\vspace{-0.2in}
\begin{center}
\includegraphics[width=0.49\textwidth]{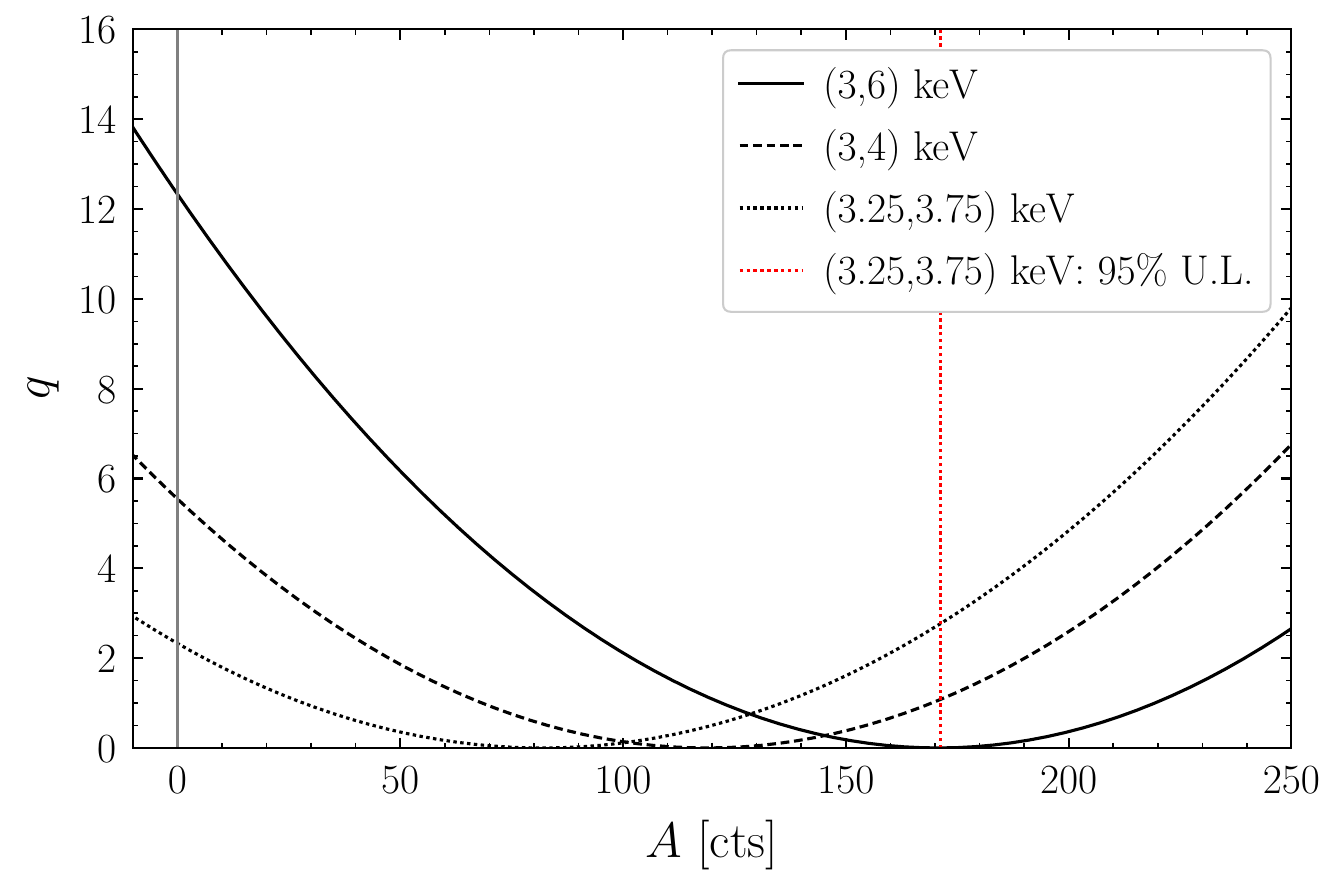}
\caption{ The profile likelihood for the signal parameter $A$ for the analyses of the mock data set illustrated in Fig.~\ref{fig:Toy_Model_Illustration}.  We show results for three different analyses, using increasing more restrictive energy ranges as indicated.  As the energy range decreases, the best-fit signal parameter (the value that minimizes $q$) moves towards the true value of zero, while the discovery TS ($q(A=0)$) decreases.  This trend is an indication of mismodeling, as the underlying model used to construct the simulated data does not have a real signal but rather a broad spectral feature centered around 3.5 keV.  As a further indication of the discrepancy between the different energy range analyses, in vertical dashed red we show the 95\% upper limit on $A$ found from the narrowest energy range analysis, which rules out the best-fit point from the largest energy range analysis.     }
\label{fig:prof_LL_ex}
\end{center}
\end{figure}

For the example shown in Fig.~\ref{fig:Toy_Model_Illustration}, the evidence in favor of the signal model over the null hypothesis is computed to be $t \approx 12.3$.  Note that the relative normalization of the flat background relative to the broad spectral feature at 3.5 keV in the simulated data is tuned to achieve a TS $t \sim 10$ (corresponding to around $3$$\sigma$), for a typical MC realization, under the test as described above.  More precisely, by performing this test on $10^4$ independent MC realizations, of which that shown in Fig.~\ref{fig:Toy_Model_Illustration} is just one representative member, we determine that the expected 68\% containment interval for $t$ is $\sim$$(3,14)$, with a median expectation of $8$.  We chose this target TS range for the example because it is typical of the significances found in {\it e.g.}~\cite{Bulbul:2014sua} for evidence of a 3.5 keV line.  The profile likelihood associated for the signal model parameter $A$, for our representative MC realization, is illustrated in Fig.~\ref{fig:prof_LL_ex} (labeled ``$(3,6)$ keV").  In this figure the signal normalization parameter is shown in units of the total number of counts, integrated over all energies.  The best-fit signal parameter is that which minimizes $q$ (maximizes the log likelihood).

In studies of the 3.5 keV line the possibility of mismodeling has often been assessed through the chi-square per degree of freedom:
\es{}{
\chi^2_\nu \equiv {1 \over \nu} \sum_i \left( {{\bf d}_i - \mu_i(\hat{ \bm \theta}) \over \sqrt{{\bf d}_i}} \right)^2 \,,
}
where $\nu$ is the number of degrees of freedom, equal to the number of data points minus the number of model parameters.  The quantity $\mu_i(\hat{ \bm  \theta})$ denotes the best-fit model prediction in counts in energy bin $i$.  A value of $\chi^2_\nu$ near unity implies that the hypothesis is a good description of the data, up to the expected statistical noise.  More precisely, for a given $\nu$ we may calculate the expected containment interval for $\chi^2_\nu$, at a given confidence, and then assess whether the observed value of $\chi^2_\nu$ appears consistent with that expected from statistical uncertainties alone. Interestingly, for the example discussed in this section (illustrated in Fig.~\ref{fig:prof_LL_ex}) we calculate $\chi^2_\nu \approx 0.98$, despite the model not including the broad spectral feature near 3.5 keV that went into constructing the simulated data.  Under the signal hypothesis, given $\nu = 598$, we expect $\chi^2_\nu \in (0.94,1.06)$ at 68\% containment.  Thus, we see that in this example the $\chi^2_\nu$ test is not an adequate test to indicate that mismodeling is present in the data.  This is because the mismodeling is localized over a relatively small region of energy, relative to the full analysis range, and thus the effect of the mismodeling on $\chi^2_\nu$ is ``washed out" by the statistical fluctuations at other energies, where the model is a reasonable description of the data.  Furthermore, performing this test over all $10^4$ MC realizations we determine that the expected range for $\chi^2_\nu$ in our analysis with mismodeling is $(0.97,1.09)$ at 68\% confidence, with a mean of $\sim$1.03. While the $\chi^2_\nu$ distribution is shifted towards higher values compared to that expected from statistical uncertainties only, the difference is minor and for an individual realization it is not possible to identify the mismodeling by using $\chi^2_\nu$ only.

With the above discussion in mind, in our reanalyses of $X$-ray data described in the remainder of this article we do calculate the chi-square per degree of freedom for each test as a possible diagnostic --- just not a definitive diagnostic -- of mismodeling.   When discussing $\chi_\nu^2$ in the context of mismodeling, it is instructive to calculate the $p$-value
\es{eq:p}{
p \equiv S_\nu(\chi_\nu^2) \,,
}
with $S_\nu(\chi_\nu^2)$ denoting the survival function of the chi-square distribution with $\nu$ degrees of freedom and with  $\chi_\nu^2$ representing the observed value. This $p$-value is interpreted as the probability of observing a chi-square per degree of freedom value as larger or larger than that observed in the data under the null hypothesis. Smaller $p$-values suggest increasing tension between the model and the data.  

Before discussing ways of mitigating mismodeling, we introduce one additional ensemble of mock data sets that we use in the following subsection, which we refer to as the {\bf Signal Data Sets}.
The Signal Data Sets have the same energy-independent background flux as the Mismodeling Data Sets but with the addition of a narrow spectral feature (standard deviation of 30 eV centered at 3.5 keV) instead of the broad spectral feature at 3.5 keV.  We chose the expected number of counts from the narrow spectral feature to be 130; this number is chosen such that the expected discovery TS in favor of the signal model at 68\% containment is in the range $\sim$$(3,15)$, roughly matching what we find from the signal model analyses of the Mismodeling Data Sets.  Importantly, however, the signal model is able to accurately describe the mock Signal Data Set data, with no mismodeling.

\subsection{Reducing the analysis energy range to mitigate mismodeling}
\label{sec:energy_range}

One of the central methods that we use in this work to assess for and help mitigate mismodeling is to decrease the width of the energy range used in the analysis.  In the example just discussed (illustrated in Fig.~\ref{fig:Toy_Model_Illustration}) the search is performed from 3.0 to 6.0 keV on the representative Mismodeling Data Set, but the signal itself has a full width half max (FWHM) of just $\sim$70 eV.  The relevant data for informing whether or not the signal hypothesis is preferred over the null hypothesis is the data in the immediate vicinity of 3.5 keV.  Increasing the width of the energy range will simply better determine the nuisance parameters associated with the background components that have support over this larger range.  If the background components accurately describe the data, then going to a larger energy range increases the sensitivity to a putative signal, since by better determining the nuisance parameters there is less potential degeneracy between the signal parameter and the nuisance parameters.  However, the danger is that if there is mismodeling and the background model does not accurately describe the data, then shrinking the statistical uncertainties on the nuisance parameters makes the analysis more susceptible to systematic uncertainties associated with the mismodeling.    

To illustrate this point, we repeat the analysis described in Sec.~\ref{sec:toy} on the same mock Mismodeling Data Set show in Figs.~\ref{fig:Toy_Model_Illustration} and~\ref{fig:prof_LL_ex}
but with increasingly narrow analysis energy ranges.  The profile likelihoods associated with $(3,4)$ keV and $(3.25,3.75)$ keV analysis ranges are illustrated in Fig.~\ref{fig:prof_LL_ex}.  As the energy range becomes more narrow there are two important trends to note: (i) the best-fit point in $A$ (the value that minimizes $q$) moves towards zero, which is the true value in this case since the simulated data does not contain a signal; and (ii) the detection TS $t$ ({\it i.e.}, $q(A = 0)$) decreases.  Indeed, while the significance for the signal model over the null hypothesis is $\sim$3.7$\sigma$ in the $(3,6)$ keV analysis, this significance drops to $\sim$$1.7$$\sigma$ in the $(3.25,3.75)$ keV search.  While some drop in significance is expected when going to a narrower energy range, since the background nuisance parameter is less well constrained, the combination of significance drop and best-fit model parameter moving towards zero is an indicator of mismodeling.  To emphasize this point, in Fig.~\ref{fig:prof_LL_ex} we show in vertical dotted red the 95\% one-sided upper limit on $A$ as computed from the $(3.25,3.75)$ keV analysis. Note that the best-fit from the $(3,6)$ keV analysis is excluded at almost precisely 95\% confidence from the $(3.25,3.75)$ keV analysis; this inconsistency is an indicator that mismodeling is present.

The example illustrated in Fig.~\ref{fig:prof_LL_ex} is one representative example from the $10^4$ MC realizations we analyze, but it illustrates the trends observed over the full ensemble.  In particular, while the 68\% containment interval over MC realizations for the $(3,6)$ keV analysis best-fit point $\hat A$ is $(85,178)$ cts, the equivalent containment intervals for the $(3,4)$ keV and $(3.25,3.75)$ keV analyses are $(52,151)$ cts and $(4,111)$ cts, respectively.  Thus, as the analysis energy range shrinks around 3.5 keV, the best-fit signal amplitude $\hat A$ moves towards the true value of zero.  The distribution of expected TSs $t$ also decreases as the energy range shrinks, with the 68\% containment interval becoming $(1.1,8.9)$ and $(0.1,4.3)$ for the 1 keV and 0.5 keV window analyses, respectively. 

It is interesting to contrast the examples above with ones over the data sets that have no mismodeling (the Signal Data Sets), as described in Sec.~\ref{sec:toy}.   Analyzing these $10^4$ mock data sets in the reduced energy ranges that are 1 keV and 0.5 keV wide, we find that the discovery TSs are, at 68\% confidence, in the ranges $(3,14)$ and $(2,13)$, while the best-fit signal amplitudes are in the associated ranges $(84,185)$ and $(80,188)$, respectively.  Without mismodeling, the discovery TSs are only mildly reduced, even when going to the narrowest 0.5 keV wide energy range, unlike in the cases where mismodeling is present.  Similarly, without mismodeling the best-fit signal amplitudes remain centered around the true value, while with mismodeling the best-fit signal amplitude ranges approach zero as the analysis energy range shrinks.  As discussed later in this article, some of the analyses for the 3.5 keV line on real $X$-ray data behave similarly to the mismodeling examples presented here when the analysis energy range is reduced.

\subsection{Global likelihood optimization}

The example discussed throughout this section is relatively simple in that there is a single signal parameter and a single background model parameter.  In contrast, the analyses that search of UXLs, such as those in~\cite{Bulbul:2014sua,Boyarsky:2014jta,Tamura:2014mta,Jeltema:2014qfa,Cappelluti:2017ywp}, tend to have dozens of model parameters to account for uncertainties related to the continuum background and also the locations and magnitudes of astrophysical emission lines that appear within the analysis energy ranges.  The presence of multiple model parameters generically leads to the presence of multiple local maxima in the likelihood.  The principle of likelihood maximization states that the best-fit model parameters are those which globally maximize the likelihood, but numerically it may be challenging to optimize the likelihood if there are a large number of model parameters.  Most prior analyses use the default likelihood optimization capabilities in the $X$-ray spectral fitting package \texttt{XSPEC}~\cite{1996ASPC..101...17A}.  \texttt{XSPEC} runs a Levenberg–Marquardt algorithm by default to maximize the likelihood, though minimization through \texttt{minuit}~\cite{James:1975dr,James:1994vla,iminuit} is also possible.  However, it is important to note that all of these algorithms only find the nearest local maximum of the likelihood from the starting parameters and not the global maximum.  In our analyses on the real $X$-ray data we find that in almost all cases running a local optimizer even with carefully chosen initial parameters will not converge to the global likelihood maximum; instead, global optimizers are necessary.

\begin{figure}[t]  
\hspace{0pt}
\vspace{-0.2in}
\begin{center}
\includegraphics[width=0.49\textwidth]{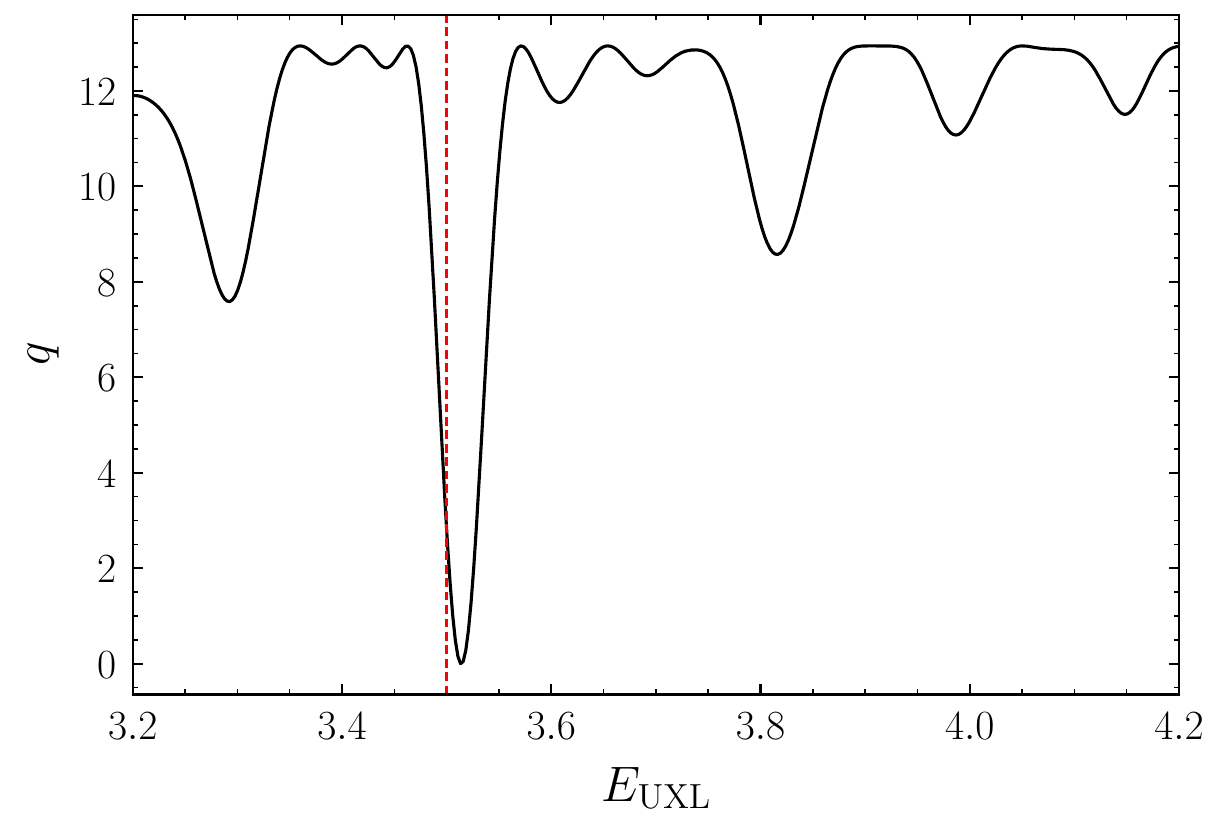}
\caption{ The profile likelihood as a function of the UXL energy $E_{\rm UXL}$, with all other model parameters profiled over at a given, fixed UXL energy.  This analysis is illustrated for a representative Signal Data Set that has a true line signal injected at 3.5 keV, as indicated in vertical dashed red.  When the line energy is allowed to float the true line energy is recovered within the expected statistical uncertainty, but the locations of local likelihood maxima are also clearly visible.  Local minimization algorithms need to start sufficiently close to the true minimum to avoid converging around local minima instead of the global minimum.    }
\label{fig:minimizer_error}
\end{center}
\end{figure}

In this section, we provide a simple example that illustrates how local optimization algorithms may miss the global likelihood maximum even in analyses with few model parameters.  In particular, we analyze the Signal Data Sets introduced in Sec.~\ref{sec:toy} with an energy independent background component and a signal model, but we allow the location of the signal model line, $E_{\rm UXL}$, to float as an additional model parameter instead of fixing it to the true value of 3.5 keV.  That is, our model has three model parameters in this example: the signal energy and normalization, in addition to the normalization of the flat background component. In Fig.~\ref{fig:minimizer_error} we illustrate an analysis of a representative Signal Data Set, where we construct the profile likelihood $q(E_{\rm sig})$ by profiling over the other two model parameters at fixed $E_{\rm sig}$.  The global minimum of $q$ is clearly consistent with the true signal energy of 3.5 keV, as indicated in vertical, dashed red.  However, the profile likelihood also shows local minima at energies away from the global minimum.  These minima are relatively easy to interpret; they arise from statistical fluctuations that create low-significance line-like features.

It is relatively straightforward to identify the global minimum for the example illustrated in Fig.~\ref{fig:minimizer_error}, but it is challenging when using a local minimizer.  As an example, we analyze this simulated data set using \texttt{minuit} with the starting point where the signal normalization and background normalization are taken at their global best-fit values but where $E_{\rm UXL} = 3.65$ keV.  In this case, the model converges to $E_{\rm UXL} \approx 3.608$ keV, which is the local minimum clearly visible in Fig.~\ref{fig:minimizer_error} directly to the right of the true minimum.  The TS difference between this local minimum and the true local minimum, which is at  $E_{\rm UXL} \approx 3.514$ keV, is $\sim$11.7.  Moreover, at the local minimum near $E_{\rm UXL} \approx 3.608$ the best-fit signal amplitude is in fact negative at slightly over 1$\sigma$ significance.  This simple example illustrates the important point that it is crucial to obtain the global likelihood maximum when performing profile likelihood in order to make self-consistent statements about the significance of a putative UXL.  The difficulty in achieving the global maximum, however, only increases as the number of model parameters increase.  

\section{Galaxy Cluster Data From XMM-Newton MOS}
\label{sec:xmm-cluster}

\begin{table*}[!]{
    \ra{1.3}
    \begin{center}
    \begin{tabular}{@{\extracolsep{4pt}}L{0.11\textwidth}rP{0.15\textwidth}*{4}{P{0.15\textwidth}}@{}}
    \hlinewd{1pt}
    & & \textbf{Original} & \multicolumn{4}{c}{\textbf{This work}} \\
    \cline{3-3}\cline{4-7} \multicolumn{2}{l}{\textbf{Analysis Range}} & Full & Full & 3-6 keV & 1 keV & 0.5 keV \\ \hlinewd{1pt}
    \multirow{5}{*}{\parbox{0.11\textwidth}
    {\raggedright\textbf{XMM\\Perseus}}} & $\chi_\nu^2$ & $613.8/574$ & \multirow{5}{*}{\textemdash} & 593.9 / 564 & 199.2/176& 88.4/85 \\ 
    & $p$ & $0.12$ & & 0.19 & 0.11 & 0.38 \\ 
    & $\hat{A}$ & $52.0_{-15.2}^{+24.1}$ & & $4.0^{+8.3}_{-8.7}$ & $16.3^{+12.9}_{-13.6}$& ${-3.3}^{+17.5}_{-26.3}$\\ 
    & $t$ & $17.1$ &  &  $0.2$ & $1.6$ & $0$ \\ 
    & $A^{95}$ & \textemdash & & 18.0 & 37.1 & 22.5 \\ \hline
    \multirow{5}{*}{\parbox{0.11\textwidth}
    {\raggedright\textbf{XMM\\Perseus, Cored}}} & $\chi_\nu^2$ & $596.1/574$ & \multirow{5}{*}{\textemdash} & 602.8 / 567 & 184.7/175& 86.4/91 \\ 
    & $p$ & $0.25$ & & 0.14 & 0.29 & 0.62 \\ 
    & $\hat{A}$ & $21.4_{-6.3}^{+7.0}$ & & $1.6^{+8.1}_{-8.7}$ & $-5.4^{+11.3}_{-12.2}$& ${-14.6}^{+12.4}_{-13.6}$\\ 
    & $t$ & $12.8$ &  &  $0.02$ & $0$ & $0$ \\ 
    & $A^{95}$ & \textemdash & & 18.0 & 37.1 & 22.5 \\ \hline
    \multirow{5}{*}{\parbox{0.11\textwidth}
    {\raggedright\textbf{XMM\\ Joint CCO}}} & $\chi_\nu^2$ & $562.3/569$ & \multirow{5}{*}{\textemdash} & 1759.9/1715 & 590.2/551 & 320.3/277\\ 
    & $p$ & $0.57$ & & $0.22$ & $0.12$ & $0.04$\\ 
    & $\hat{A}^{*}$ & $1.8_{-0.7}^{+0.8}$ & & $1.6^{+0.4}_{-0.5}$ & $0.6^{+0.5}_{-0.5}$ & $-0.5^{+0.5}_{-0.5}$ \\ 
    & $t$ & $15.7$ & & $12.2$ & $1.4$ & $0$\\ 
    & $A^{95*}$ & \textemdash & & $2.3$ & $1.4$ & $0.4$ \\ \hline
    \multirow{5}{*}{\parbox{0.11\textwidth}
    {\raggedright\textbf{XMM\\M31}}}& $\chi_\nu^2$ & $97.8/74$ &1225.3/1166  & 583.6/588 & 203.1/198 & 98.1/98\\ 
    & $p$ & $0.036$ &$0.11$ & $0.54$ & $0.39$ & $0.48$ \\ 
    & $\hat{A}$ & $4.9_{-1.3}^{+1.6}$ & $2.1^{+0.9}_{-0.9}$& $1.3^{+1.0}_{-1.1}$ & $1.2^{+0.9}_{-0.8}$ & $11^{+1.0}_{-1.0}$\\ 
    & $t$ & $13.0$ & $5.5$ & $1.7$ & $2.1$ & $1.4$\\ 
    & $A^{95}$ & \textemdash & $3.6$ & $3.0$ & $2.7$ & $2.8$\\ \hline
    \multirow{5}{*}{\parbox{0.11\textwidth}
    {\raggedright\textbf{Chandra\\Perseus}}} & $\chi_\nu^2$ & $158.7/152$ & $216.1/211$ & $189.9/180$ & $47.7/50$ & $24.5/22$ \\ 
    & $p$ & $0.45$ & $0.39$ & $0.29$ & $0.57$ & $0.32$ \\ 
    & $\hat{A}$ & $18.6_{-8.0}^{+7.8}$ & $-0.2_{-8.7}^{+8.9} $ & $-3.6_{-9.7}^{+11.7}$ & $-15.0_{-12.7}^{+15.3}$ & $-15.8_{-18.0}^{+17.0}$ \\ 
    & $t$ & $6.2$ & $0$ & $0$ & $0$ & $0$ \\ 
    & $A^{95}$ & \textemdash & $14.9$ & $16.3$ & $16.3$ & $12.3$ \\ \hline
    \multirow{5}{*}{\parbox{0.11\textwidth}
    {\raggedright\textbf{Chandra\\Deep Field}}} & $\chi_\nu^2$ & \textemdash & 614.6/617 & 375.6/403 & 126.4/131 & 47.7/63 \\ 
    & $p$ & \textemdash & $0.52$& $0.83$ &  0.59 & $0.92$ \\ 
    & $\hat{A}$ & $0.39_{-0.25}^{+0.21}$ & $-0.30^{+0.28}_{-0.27}$ & $-0.33^{+0.27}_{-0.28}$ & $-0.33^{+0.31}_{-0.34}$ & $0.11^{+0.34}_{-0.34}$ \\ 
    & $t$ & $6.3$ & $0$ & $0$ & $0$ & 0.1 \\ 
    & $A^{95}$ & \textemdash & $0.15$ & $0.12$ & $0.17$ & $0.68$ \\ \hlinewd{1pt}  
    \end{tabular}
    \end{center}}\caption{\label{tab:Results} A compilation of the results derived in this work for each of our analyses along with those of the original analyses. $\chi_\nu^2$ refers to the reduced $\chi^2$ of the null fit, and the corresponding $p$-value is also reported. $\hat{A}$ is the best-fit flux for the 3.5 keV line reported in units of $10^{-6}$ cts/cm$^2$/s with associated 1$\sigma$ uncertainties ($^*$except for the XMM Joint CCO analysis, where it has units of $10^{10} \sin^2(2\theta)$, see text for details) and $t$ is the discovery TS. Note that $t$, defined in~\eqref{eq:TS}, is sometimes referred to as $\Delta\chi^2$, though it is distinguished by being a one-sided test statistic and explicitly set to zero if the best-fit signal strength is negative. The 95\% one-sided upper limit on the 3.5 keV line flux is $A^{95}$ in the same units as $\hat{A}$. For XMM-Newton MOS Perseus, the fit is performed on the data realization with the median $\chi_\nu^2$ (see App.~\ref{app:Randomization}); for the others, it is performed on the single realization generated in this work. If the original analysis range was 3-6 keV, the ``Full" column is not populated. The original energy range for XMM-Newton MOS M31 was 2--8 keV; for Chandra Perseus, 2.5--6 keV; for Chandra Deep Field, 2.4--7 keV. For the Chandra Deep Field analysis, no $\chi_\nu^2$ is reported because we show results for the modeled background scenario. See text and App.~\ref{app:DeepFieldBkgSub} for details.} 
\end{table*}

Having illustrated examples of mismodeling and optimization confusion using local optimizers on toy data sets, we now turn our attention to the reanalyses of the original data sets that produced evidence for the 3.5 keV line.  A summary of all our results is provided in Tab.~\ref{tab:Results}.
The strongest claimed evidence for a line-like excess at rest energy of 3.5 keV comes from observations of galaxy clusters using the MOS instrument onboard the XMM-Newton $X$-ray observatory analyzed in \cite{Bulbul:2014sua}, with somewhat weaker evidence for a corresponding excess found in corresponding data collected by the PN instrument.  In this section we reanalyze the key XMM-Newton data sets from~\cite{Bulbul:2014sua}, though in Sec.\ref{sec:chandra-perseus} we also revisit their Chandra Perseus analysis.  Note that when possible we implement data reduction and modeling procedures identical to those used in \cite{Bulbul:2014sua}.

Ref.~\cite{Bulbul:2014sua} claims evidence for a 3.5 keV line in a number of XMM-Newton MOS analyses.  In particular, they stack data from 
73 galaxy clusters out to redshifts $z\sim 0.35$, finding evidence for a UXL at 3.57 keV at approximately 5$\sigma$ local significance. On the other hand, their evidence for the 3.5 keV line (more precisely the 3.57 keV line) is driven primarily by four bright objects: the clusters Perseus, Coma, Ophiuchus, and Centaurus.  In an analysis of the Perseus cluster alone they fix the UXL line energy (in the cluster frame) to be 3.57 keV and find approximately 4$\sigma$ evidence for the signal model over the null hypothesis; below, we repeat this analysis following as closely as possible the procedure in~\cite{Bulbul:2014sua} and find no preference ($< 1 \sigma$) for the signal model, in strong tension with the claim in~\cite{Bulbul:2014sua}.   Ref.~\cite{Bulbul:2014sua} also considers a core-masked analysis variant of the Perseus data in order to isolate the DM-abundant  outer regions of the cluster from the active core; they claim $\sim$$3.6$$\sigma$ evidence in favor of the UXL from that analysis. In contrast, we find no evidence ($t \sim 0$) for the 3.57 keV UXL in our same analysis of the same data set. 

Ref.~\cite{Bulbul:2014sua}  then considers a stacked analysis of the data from the next three brightest clusters: Coma, Ophiuchus, and Centaurus, again fixing the rest-frame UXL energy to 3.57 keV.  
We avoid blueshifting the XMM-Newton data, which is necessary for stacking data from different targets in the source frame, because rebinning the data during the blueshifting procedure introduces additional stochasticity and/or correlations along the lines of those discussed in App.~\ref{app:Randomization}.  
Instead, we analyze the data from each of these clusters individually and then join the results together in the context of a joint likelihood, assuming a decaying DM model. Ref.~\cite{Bulbul:2014sua} found $\sim$$4\sigma$ evidence from these three clusters in favor of the signal model.  We find comparable evidence for a 3.5 keV line in Centaurus, though that evidence disappears when analyzing the data over more restrictive energy windows.  The other two clusters show no evidence for a UXL, with the joint analyses finding no evidence for a line in the 1 keV and the 0.5 keV analysis windows.

\subsection{Data Reduction}
\label{sec:xmm-cluster-reduction}

For each of the observations under consideration, we retrieve the raw data products from the XMM-Newton Science Archive. To reduce the data we use the Science Analysis System~\cite{XMM-SAS} (SAS) version 14.0 Extended Source Analysis Software (ESAS) subpackage, which is used for modeling sources covering the full XMM-Newton field-of-view and diffuse backgrounds. We reduce the data following the same procedure on individual exposures as in~\cite{Foster:2021ngm}, except that we instead use the CIAO~\cite{2006SPIE.6270E..1VF} version 4.14 and CALDB version 4.9.8 task \texttt{wavdetect} to identify point sources in the 0.4-7 keV range to match the data reduction procedures in~\cite{Bulbul:2014sua}. 

We briefly summarize the process here, although we follow the standard ESAS pipeline. For each observation ID, we obtain the associated list of science exposures taken by the MOS instrument, which are those data sets which were collected with the spacecraft in Science mode. From these exposures, we filter the event list so that it only includes events taken in periods of low background, to reduce soft-proton contamination of the spectrum. We mask all point sources in the field of view, as explained more below, and CCDs operating in anomalous states. We use the resulting data products to generate the photon-count data, the ancillary response file (ARF), and the redistribution matrix file (RMF), over the full field of view. 

To construct the stacked data for each target, we sum the photon-count data while we average the detector response, composed of the ARF and RMF, weighted by the total counts between 2 and 10 keV. 
Following the literature ({\it e.g.}, Ref.~\cite{Bulbul:2014sua}), errors are treated in the Gaussian approximation to the Poisson distribution. We additionally generate the quiescent particle backgrounds (QPB) and associated statistical uncertainties. The QPB data are subtracted from the counts data at each source with the statistical uncertainties added to the counts uncertainties in quadrature.

Our point-source masking procedure is as follows. As mentioned previously, we use the CIAO task \texttt{wavdetect}, which is a Mexican Hat wavelet source detection algorithm that correlates the image at each pixel with wavelets at different scales and produces a file containing the sky regions to exclude. We choose the correlation scales 4,8,16,32,64 in units of pixels and set the detection threshold such that on average we detect one fake source per image. We then feed the SAS task \texttt{mos-spectra} the exclusion regions. The \texttt{mos-spectra} task outputs the summed spectra over the region of interest along with the associated ARF and RMF averaged over the region.  Note that for the Perseus core-masked analysis only we also mask the inner $1'$ around the core center, which we define as the point of maximum counts in the image. 

A particular challenge for comparing analyses of nominally identical $X$-ray data sets 
is that the data reduction tools for $X$-ray telescopes typically involve randomization. For data collected with the MOS instrument on XMM-Newton, the \texttt{emchain} data reduction task randomizes events between adjacent sky pixels, time frames, and analog-digital-unit (ADU) energy bins. This randomization is implemented to mitigate various undesirable signal processing and instrumental  effects, \textit{i.e.}, aliasing and interference. As a result, since neither the data products nor the random seed used in prior works is publicly available, we are unable to produce exactly identical data sets for our analysis. We study the significance of the randomization intrinsic to the data reduction in App.~\ref{app:Randomization}, finding that it results in considerable variance in the chi-square per degree of freedom $\chi_\nu^2$. As a result, we find that it is not meaningful to compare our $\chi_\nu^2$ values to those in the literature, given we are using slightly different data sets. On the other hand, the $p$-values defined in~\eqref{eq:p} are still useful measures of mismodeling. 

In this work, we also elect to analyze the data sets from each cluster individually rather than stacking the cluster data after blueshifting each data set to the source frame, for the reasons already given.
However, in an attempt to use as similar a region of interest as possible to \cite{Bulbul:2014sua}, we analyze an energy range which is 3 keV wide in the source frame and is defined by selecting all detector-frame bins with central energy $E_\mathrm{center}$ such that $3 \leq (1+z_\mathrm{source})E_\mathrm{center} \leq 6$. We then consider two alternate energy ranges similarly defined in the source frame: a 1 keV interval (3.07 -- 4.07 keV) and a 500 eV interval (3.32 -- 3.82 keV) centered on the claimed 3.57 keV excess.  

\subsection{Model Components and Likelihood }
\label{sec:xmm-cluster-models}

To the extent possible, we attempt to construct identical background models to those used in the original analyses of \cite{Bulbul:2014sua}. A key component of these analyses is that out of a set of possible model components, only those which improve the goodness-of-fit above some prescribed threshold are kept in the full background model then used in searches for line-like excesses. We begin by itemizing the set of all possible components that could be included in our background model and then we describe the procedure by which candidate background components are either included or excluded from our final background model.

We allow continuum backgrounds to be described by up to four possible components. In particular, the continuum background model is partially composed of up to two \texttt{nlapec} models (referred to as {\it line-free apec} in Ref.~\cite{Bulbul:2014sua}) with abundances of trace elements relative to the solar abundances fixed at $0.3$. These models are parameterized by a temperature parameter $T$ and an intensity parameter $I$. The \texttt{nlapec} model is the $X$-ray spectrum of a collisionally-dominated optically-thin plasma at a fixed temperature accounting for bremsstrahlung, radiative recombination continuum, and two-photon emission, so that the line emission is subtracted out. Though up to four \texttt{nlapec} components were used in some analyses in \cite{Bulbul:2014sua} (in particular, they use two \texttt{nlapec} models for Perseus and four for the stacked clusters analysis), we find two are sufficient in the sense that including more than two models results in chi-square differences less than two. As in \cite{Bulbul:2014sua}, we also allow for the possibility of two power-law components characterized by an intensity and a power-law index, $I$ and $k$, respectively, to describe non-thermal $X$-ray backgrounds. One power-law component is folded through the instrumental response of the telescope while the other is not. Unlike in~\cite{Bulbul:2014sua}, 
we allow these power-law intensities and indices to be freely fit to the data.\footnote{Note that instead~\cite{Bulbul:2014sua} fixes the power-law model parameters through external data sets and analyses but does not provide sufficient information to determine the values they use.  On the other hand, by allowing these model parameters to float freely we are being conservative, since this can only lead to more degeneracy between the continuum model components and our signal parameter of interest.}

In addition to the continuum components, we include 13 background lines of astrophysical origin, as described further below,
in the 3-6 keV range using the \texttt{gauss} and \texttt{zgauss} line profiles in \texttt{XSPEC} for unshifted and redshifted lines, respectively. Astrophysical lines are appropriately redshifted according to the best-fit redshifts in \cite{Bulbul:2014sua}, and each line is characterized by three parameters: a rest energy $E$, a width $\Delta E$, and an intensity $I$. In Tab.~\ref{tab:MOS_Lines}, we provide the list of those 13 lines with associated expected rest energies. Following~\cite{Bulbul:2014sua} we do not consider instrumental lines in our XMM-Newton cluster analyses.

In our analysis procedure, we allow the inferred rest energies to vary by up to $\delta E = 5 \, \mathrm{eV}$ from their expected rest energy.  The line widths are allowed to freely float in the range $\Delta E / E \in (10^{-4}, 10^{-2})$, where $E$ is the rest-energy of the line and $\Delta E$ is the line width. Finally, for most lines, the intensities are allowed to take any non-negative value, whereas for the five lines near in energy to the purported 3.5 keV excess, an upper bound is placed on their estimated intensity. For both Perseus and CCO, the intensity bounds, if relevant, are provided in Tab.~\ref{tab:MOS_Lines}. We emphasize that all of these choices and bounds come directly from the original work~\cite{Bulbul:2014sua}. 

Finally, we account for the possible attenuation of $X$-ray flux due to the optical depth of hydrogen along the line-of-sight with the \texttt{wabs} absorption model in \texttt{XSPEC}. This absorption is applied to the two \texttt{nlapec} components and one power-law component. The remaining power-law component is unabsorbed in order to describe continuum instrumental backgrounds.
We use the \texttt{XSPEC} default abundances such that the absorption is characterized by a single parameter $\eta_H$, the hydrogen column depth. Though the expected hydrogen depth along the line of sight for the various observational targets can be obtained from \texttt{HEASoft}, this hydrogen depth accounts for only the Milky Way contribution and not the contribution of the clusters themselves. Hence, we allow $\eta_H$ to take on arbitrary non-negative values. 

In total, the nuisance parameter vector which determines the background model is given by
\mbox{$\bm{\theta} = \{\bm{\theta}_\mathrm{nlapec}, \bm{\theta}_\mathrm{pl}, \bm{\theta}_\mathrm{line}, \eta_H\}$}, which are corresponding defined by 
\es{}{
\bm{\theta}_\mathrm{nlapec} &= \{\{I_i, T_i\}_{i=1}^{N_{\rm nlapec}}\} \,, \\ \bm{\theta}_\mathrm{pl} &=\{\{I_\mathrm{pl, 1}, k_\mathrm{pl, 1}\}, \{I_\mathrm{pl, 2}, k_\mathrm{pl, 2}\}\}, \\
 \bm{\theta}_\mathrm{line} &= \{\{E_i, \Delta E_i, I_i\}_{i=1}^{N_\mathrm{astro.}}\} \,.
 }
Given a nuisance parameter vector $\bm \theta$ and the signal normalization parameter $A$, the total mean model prediction per energy bin $\mu(A, \bm{\theta})$, which we treat as a vector over energy bins, is constructed in the following way: 
\begin{widetext}
\begin{align}
\begin{split}
    \mu_\mathrm{nlapec}(\bm{\theta}_\mathrm{nlapec},\eta_H) &= \mathrm{RSP} \star \mathbf{wabs}(\eta_H) \sum_i^{N_\mathrm{nlapec}} \mathbf{nlapec}(I_i, T_i) \\ 
    \mu_\mathrm{pl}(\bm{\theta}_\mathrm{pl},\eta_H) &=\mathrm{RSP} \star \mathbf{wabs}(\eta_H) \mathbf{powerlaw}(I_\mathrm{pl,1}, k_\mathrm{pl, 1}) + \mathbf{powerlaw}(I_\mathrm{pl,2}, k_\mathrm{pl, 2}) \\
    \mu_\mathrm{line}(\bm{\theta}_\mathrm{line}, \eta_H) &= \mathrm{RSP} \star \sum_i^{N_\mathrm{astro.}} \mathbf{zgauss}(E_i, \Delta E_i, I_i, z) \\
    \mu_\mathrm{bkg.}(\bm{\theta}) &=  \mu_\mathrm{nlapec} + \mu_\mathrm{powerlaw} + \mu_\mathrm{line} \\
    \mu(A, \bm{\theta}) &= \mu_\mathrm{bkg.}(\bm{\theta}) + \mathrm{RSP} \star  \mathbf{zgauss}(3.57, 0, A, z) \,,
\end{split}
\end{align}
\end{widetext}
where $z$ is the redshift of the observational target and $\star$ indicates that we have folded the predicted spectrum with the instrumental response (RSP). Explicitly, the folding operation $\star$ on a spectral model $\mathcal{S}(E)$ that is a function of input energy $E$ is defined by 
\begin{equation}
    \mathrm{RSP} \star \mathcal{S} = \int dE^\prime \mathrm{RMF}_i(E^\prime) \mathrm{ARF}(E^\prime) \mathcal{S}(E^\prime) \,,
\end{equation}
so that the output $\mathrm{RSP} \star \mathcal{S}$ (up to a dimensionful constant) is the number of expected counts in the detector. The RMF accounts for the energy resolution while the ARF accounts for the effective area of the instrument.
Note that we do not convolve one of the power laws with the RSP; this is equivalent to using the diagonal response matrices as in Ref.~\cite{Bulbul:2014sua}. We extend our null background-only model to the signal model hypothesis by including an additional appropriately redshifted, zero-width line with intensity determined by the signal parameter $A$ at a rest-energy of exactly 3.57 keV. 

Equipped with our model prediction for the signal-plus-background model ${\mathcal M}$ the likelihood for observed data $\mathbf{d}$ in the Gaussian limit is given by
\begin{equation}
\label{eq:like}
    \mathcal{L}({\bf d} | {\mathcal M}, \{A,\bm{\theta} \}) = \prod_i\mathcal{N}(\mathbf{d}_i | \mu = \mu_i(\bm{\theta})) \,,
\end{equation}
where $\mathbf{d}_i$ is the observed number of counts in bin $i$. 

We also briefly comment on the procedure used to maximize the likelihood. Given the large number of parameters (up to 48 in the model prior to dropping) and the high degree of degeneracy between the different model components, local maximization is particularly unreliable in terms of identifying the global maximum relevant for frequentist maximum-likelihood-estimate (MLE)-based analyses. We instead use {\it differential evolution}, implemented in \texttt{SciPy}~\cite{2020SciPy-NMeth}, 
using a population size which is 100 times larger than the number of model parameters, enforcing an absolute tolerance of $10^{-2}$ and a relative tolerance of $10^{-4}$. (Note that we minimize minus twice the log likelihood instead of directly maximizing the likelihood.) Since a well-fit log-likelihood is typically $\mathcal{O}(100-1000)$, this ensures sufficient precision for limit-setting and component-dropping. After optimizing with differential evolution, we polish the fit with a local optimization using the \texttt{MIGRAD} algorithm implemented in \texttt{minuit}~\cite{James:1975dr,James:1994vla,iminuit}. For optimizer stability, we modify the \texttt{XSPEC} implementations of \texttt{wabs} and \texttt{nlapec} to use cubic-spline interpolation rather than linear interpolation in order to avoid the possibility of spurious convergence of parameters at interpolation nodes where the first derivative of the model prediction with respect to the model parameters is discontinuous under linear interpolation.

\subsection{Likelihood Maximization and Model Component Dropping}

Following \cite{Bulbul:2014sua}, we only keep lines that improve the goodness-of-fit of our background model to the data by $\Delta\chi^2 \geq 3$.
We begin by fitting the background model including all candidate model components to the data within the 3-6 keV range. We then independently remove each line, refit the reduced model, and evaluate the $\Delta \chi^2$ associated with excluding the candidate line from the background model. The line associated with the smallest $\Delta \chi^2 \leq 3$ is removed from the model. This procedure is repeated until all remaining lines are associated with $\Delta \chi^2 > 3$ when removed. A similar procedure is applied for the continuum model components, though now we use a threshold of $\Delta \chi^2 \leq 2$ as each continuum model component is described by only two parameters. Once no more components can be dropped subject to our criteria, we take the collection of remaining model components to be our 3-6 keV background model. For simplicity, we also fix the hydrogen depth parameter to its best-fit value for all subsequent likelihood evaluations.

A central aspect of this work is to examine the robustness of the 3.5 keV signal as the energy window of the analysis is shrunk. However, given that~\cite{Bulbul:2014sua} only uses a wide analysis window, we must make choices -- described below -- in how to modify the model when performing narrow-energy-range analyses.

We develop a background model for fitting in a 1 keV interval centered around the purported 3.57 keV excess by lightly modifying the background model developed for the 3-6 keV interval. First, independent of the continuum components that were included in the 3-6 keV background we use only a single folded power law, as for narrower ranges in energy we find that additional components would always change $\Delta \chi^2$ by an amount less than two. 
We then repeat our line-dropping procedure, using the lines that were included in the 3-6 keV background as candidates for fitting in the 1 keV interval. Similarly, we develop a background model for a 500 eV interval centered at 3.57 keV by using the 1 keV interval lines as the initial line candidate list along with a single folded power law.\footnote{Recall that the spirit of this work is to, as much as possible, avoid developing our own analysis strategies but rather to self-consistently apply those from the original works. With that said, some choices need to be made when going to narrower energy windows, since the original works did not perform these analysis variations. Our choices described here are made in an attempt to modify the models as little as possible when shrinking the energy windows, except for the exclusion of unimportant model components.  }

\begin{table*}[htb]{
    \ra{1.3}
    \begin{center}
    \begin{tabular}{c*{16}{C{0.055\textwidth}}}
    \hlinewd{1pt} 
    \textbf{Element} &  \textbf{Ar} & \textbf{Ar} & \textbf{K} & \textbf{K} & \textbf{Ar} & \textbf{Ar} & \textbf{K} & \textbf{Ca} & \textbf{Ca} & \textbf{Ar} & \textbf{Ca} & \textbf{Ca}  & \textbf{Cr} \\ 
    Energy [keV] & 3.124 & 3.32 & 3.472 & 3.511 & 3.617 & 3.685 & 3.705 & 3.861 & 3.902 & 3.936 & 4.107 & 4.584 & 5.682 \\ \hlinewd{1pt} 
    \textbf{Perseus Bound} & -- & -- & $5.55$ & $13.7$ & $1.92$ & $45.3$ & $34.8$ & -- & -- & -- & -- & --  \\ 
    3 keV Fit   & $194^{+12}_{-12}$ & $212^{+11}_{-11}$ & -- & $13.7$ & -- & -- & $34.8$ & -- & $165^{+8}_{-8}$ & -- & $111^{+7}_{-7}$ & -- & $13^{+6}_{-6}$ \\ 
    1 keV Fit   & $201^{+18}_{-18}$ & $214^{+14}_{-14}$ & -- & -- & -- & -- & $30$ & -- & $152^{+12}_{-12}$ & -- & $73^{+27}_{-27}$ & -- & -- \\ 
    500 eV Fit& -- & $213^{+16}_{-16}$ & -- & -- & -- & -- & 31 & -- & $113^{+58}_{-58}$ & -- & -- & -- & -- \\ \hline
    \textbf{Centaurus Bound} & -- & -- & $0.81$ & $2.46$ & $2.10$ & $7.5$ & $15.6$ & -- & -- & -- & -- & -- & -- \\ 
    3 keV Fit  & $88^{+6}_{-6}$ & $67^{+5}_{-5}$ & -- & 2.46 & -- & -- & 15.6 & -- & $71^{+4}_{-4}$ & -- & $26^{+3}_{-3}$ & -- & -- \\ 
    1 keV Fit   & $87^{+8}_{-8}$ & $56^{+6}_{-6}$ & -- & -- & -- & -- & -- & -- & $60^{+4}_{-4}$ & -- & -- & -- & -- \\ 
    500 eV Fit   & -- & $54^{+8}_{-8}$ & -- & -- & -- & -- & -- & -- & -- & -- & -- & -- & -- \\ \hline
    \textbf{Coma Bound}  & -- & -- & $0.81$ & $2.46$ & $2.10$ & $7.5$ & $15.6$ & -- & -- & -- & -- & -- & -- \\ 
    3 keV Fit   & $35^{+10}_{-10}$ & 36$^{+10}_{-10}$ & -- & -- & -- & -- & -- & -- & 23$^{+8}_{-8}$ & -- & $19^{+6}_{-6}$ & -- & -- \\ 
    1 keV Fit  & $37^{+18}_{-18}$ & $34^{+15}_{-15}$ & -- & -- & -- & -- & -- & -- & 21$^{+7}_{-7}$& -- & -- & -- & -- \\ 
    500 eV Fit  & -- & -- & -- & -- & -- & -- & -- & -- & -- & -- & -- & -- & -- \\ \hline
    \textbf{Ophiuchus Bound}  & -- & -- & $0.81$ & $2.46$ & $2.10$ & $7.5$ & $15.6$ & -- & -- & -- & -- & -- & -- \\ 
    3 keV Fit   & -- & -- & -- & -- & -- & -- & -- & -- & -- & -- & -- & -- & -- \\ 
    1 keV Fit & -- & -- & -- & -- & -- & -- & -- & -- & -- & -- & -- & -- & -- \\ 
    500 eV Fit & -- & -- & -- & -- & -- & -- & -- & -- & -- & -- & -- & -- & -- \\ \hlinewd{1pt} 
    \end{tabular}\end{center}}
\caption{\label{tab:MOS_Lines} The list of spectral lines which are included in our background model for the four XMM-Newton MOS galaxy cluster analyses we consider in this work. The line intensity nuisance parameter for the five lines near in energy to the purported 3.5 keV line are bounded from above by the corresponding values in the table. These upper bounds adopted from \cite{Bulbul:2014sua} vary between the Perseus and CCO data sets. We additionally provide the bounds on the line intensities where relevant and the best-fit line intensities for the two analyses. Lines without an associated best-fit line intensity are those which are not included in the final background model after our line-dropping procedure, which is taken directly from~\cite{Bulbul:2014sua}. Both line intensities and intensity bounds are provided in units of $10^{-6}$ photons/cm$^2$/s. Note that lines which are included in the Perseus background model may be excluded in the CCO background model and vice-versa.}
\end{table*}

\subsection{Data Analysis}
\label{sec:xmm-cluster-individual}

We now apply the methodology discussed above to the Perseus, Centaurus, Coma, and Ophiuchus XMM-Newton MOS data sets. We begin by discussing Perseus before turning to the other clusters. 

\subsubsection{Perseus Cluster}

\begin{figure*}[htb]  
\hspace{0pt}
\vspace{-0.2in}
\begin{center}
\includegraphics[width=0.99\textwidth]{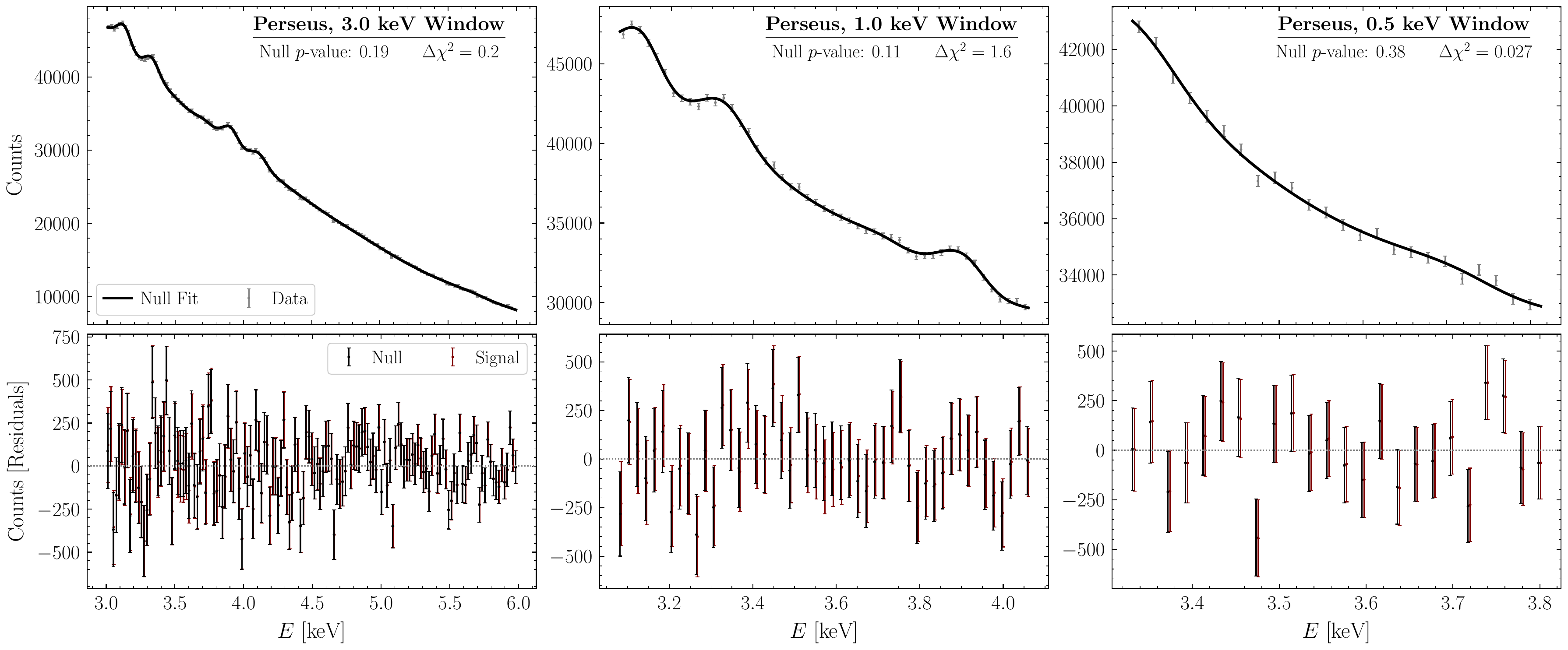}
\includegraphics[width=0.99\textwidth]{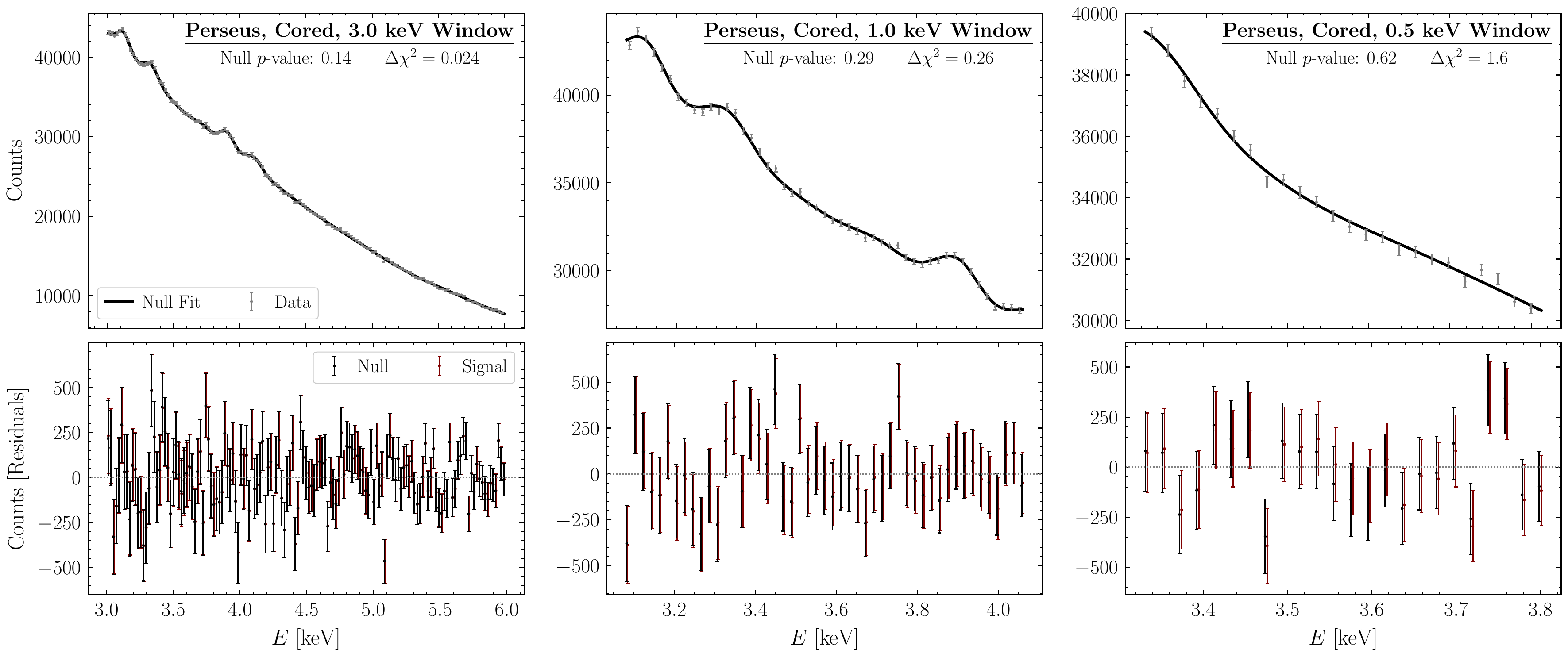}
\caption{(Top panels) The stacked XMM-Newton MOS data of the Perseus cluster (gray points with 1$\sigma$ statistical error bars) along with the best fit null model (black)
in each of our analysis energy windows. On the left is the 3 keV window of Ref.~\cite{Bulbul:2014sua}, middle 1 keV, and right 0.5 keV.  The bottom panels illustrate the residuals after subtracting the best-fit null and signal models. Note that we down-bin the data by a factor of 4 for presentation purposes only. (Bottom panels) As in the top panels, but with the core of the Perseus cluster masked.}
\label{fig:Perseus_Data}
\end{center}
\end{figure*}

\begin{figure}[htb]  
\hspace{0pt}
\vspace{-0.2in}
\begin{center}
\includegraphics[width=0.49\textwidth]{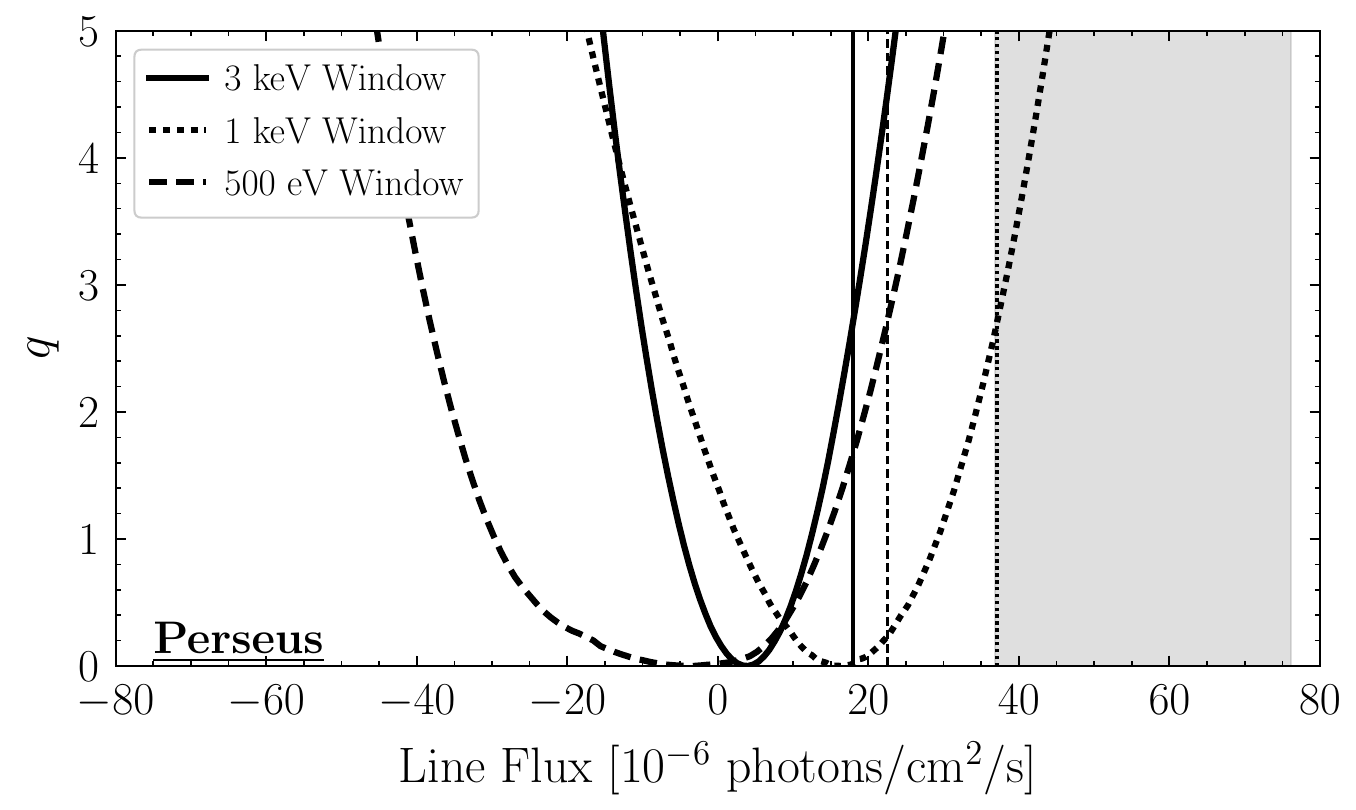}
\includegraphics[width=0.49\textwidth]{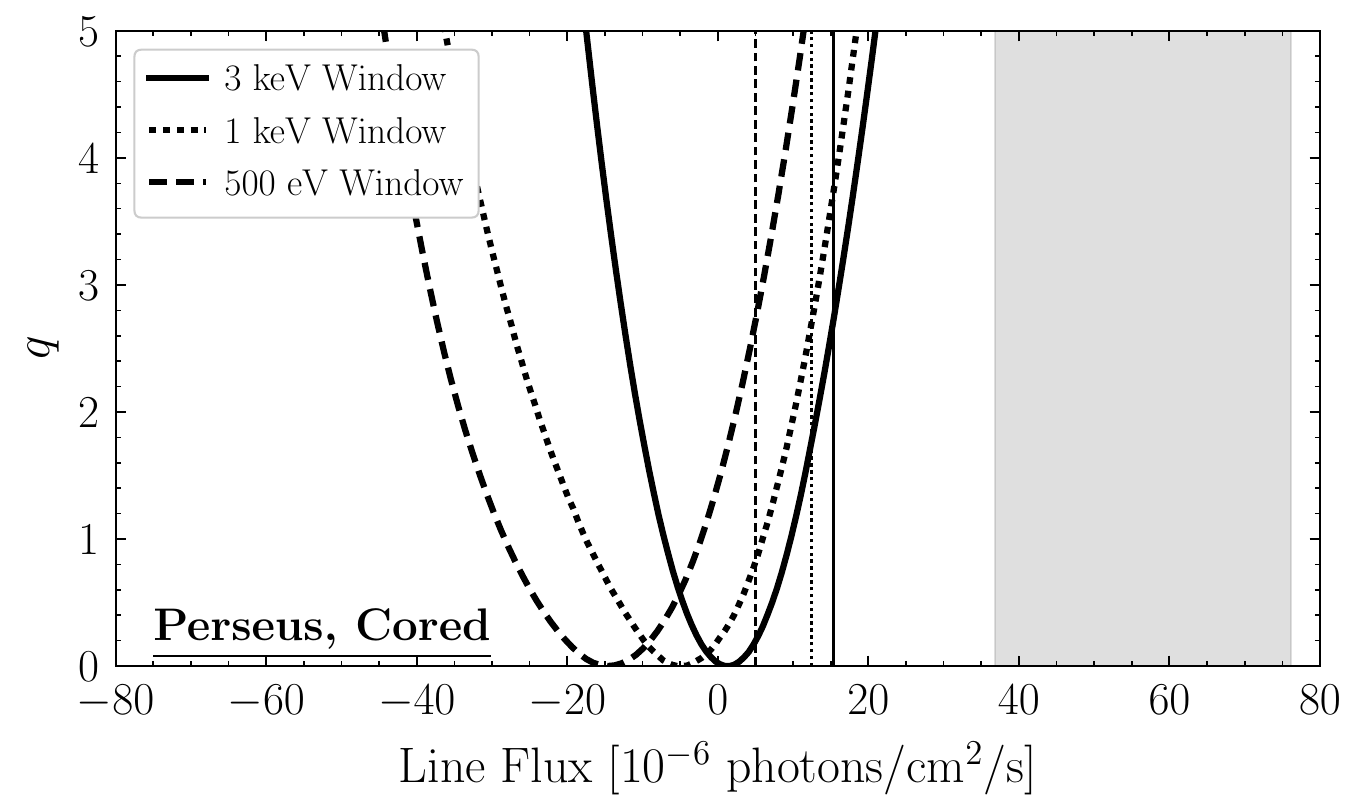}
\caption{(\textit{Above}) The profile likelihoods for the Perseus cluster analyses in each of the three analysis energy windows: 3 keV (solid), 1 keV (dotted), and 0.5 keV (dashed). The 95\% upper limits from each fit are shown as horizontal lines with corresponding styles. The 1$\sigma$ best fit region for the 3.5 keV line flux in Ref.~\cite{Bulbul:2014sua} is in shaded gray. (\textit{Below}) As in the top panel, but with the core of the Perseus cluster masked.}
\label{fig:Perseus_Likelihood}
\end{center}
\end{figure}

In this section we discuss our re-analysis of the Perseus cluster data (without and with the core mask) taken with the MOS camera on-board XMM-Newton. Ref.~\cite{Bulbul:2014sua} found $\sim$4$\sigma$ evidence for an additional emission line at 3.57 keV in the core-unmasked data, with a much larger flux 52.0$_{-15.2}^{+24.1} \times 10^{-6}$ cts/cm$^2$/s than in any other cluster analyzed. The null model fit gave $\chi_\nu^2 = 613.8/574$, corresponding to a $p$-value $p=0.12$. Note the other camera onboard XMM-Newton, PN, did not detect a line and placed an upper limit on the line strength about three times smaller than the MOS central value.  We fix the Perseus redshift at the value $z =  0.016$ taken in~\cite{Bulbul:2014sua}.

We first reanalyze the core-unmasked MOS data for evidence of a 3.57 keV UXL in the original analysis window of~\cite{Bulbul:2014sua}.
The best fit null model
is shown in the upper left panel of Fig.~\ref{fig:Perseus_Data} alongside the Perseus data.  The sub-panel below illustrates the residual counts, down-binned by a factor of four for illustrative purposes, for both the null and signal model. The profile likelihood for this analysis is illustrated in Fig.~\ref{fig:Perseus_Likelihood} (3 keV window). We find no evidence ($t < 1$) for a UXL, recovering the best fit flux $4.0_{-8.7}^{+8.3} \times 10^{-6}$ cts/cm$^2$/s, with null $\chi_{\nu}^2 = 593.9 / 564$ ($p=0.2$). 
Moreover, we place a 95\% one-sided upper limit on the UXL flux of $17\times 10^{-6}$ cts/cm$^2$/s, which excludes the best-fit flux from Ref.~\cite{Bulbul:2014sua} for their nearly identical analysis of the same data set.

We are not able to identify why~\cite{Bulbul:2014sua} finds 4$\sigma$ evidence for a UXL at 3.57 keV and we find no evidence for the line with a nearly identical analysis. One possibility is that Ref.~\cite{Bulbul:2014sua} did not converge to the global minimum; if we instead use local rather than global optimizers, we are able to, sometimes, artificially find modest evidence for a line, depending on the initial starting parameters for the local optimizers.
As a test, we use the XSPEC default Levenberg–Marquardt minimization algorithm instead of the global optimizer, on the analysis already after line dropping, with initial starting parameters randomly chosen within the parameter ranges used in our global optimization. Over 100 different realizations of this exercise, we find that the local optimizer converges to a minimum with a mean $\chi^2$ difference of 140 above the global optimizer result. Approximately 43\% of the samples find a best-fit signal flux consistent with that found in~\cite{Bulbul:2014sua} within the 1$\sigma$ flux uncertainties quoted in~\cite{Bulbul:2014sua}.
On the other hand, given the inherent stochasticity in the data reduction (see App.~\ref{app:Randomization}), it is not possible to compare our minimum with theirs, since we are working with slightly different randomized data sets, in order to make any definitive statement about whether they reached the global minimum.

The analysis described above uses the 3 keV energy window (from 3-6 keV) that was used in Ref.~\cite{Bulbul:2014sua}.
Next, we study how our results vary under reductions of the analysis window.  Using a smaller analysis window should make the analysis more robust to mismodeling, as discussed in Sec.~\ref{sec:methods}.  We repeat the analysis described above in two narrower energy windows: a 1 keV window from 3-4 keV and a 0.5 keV window centered on 3.57 keV, with energies quoted in the source frame. As we move to smaller windows, the number of model parameters decreases because emission lines may fall outside of the analysis range.  We also simplify the continuum model as previously described. The best-fit models in all three cases are shown in the top panels of Fig.~\ref{fig:Perseus_Data}, with profile likelihoods illustrated in Fig.~\ref{fig:Perseus_Likelihood}.\footnote{Note that the 500 eV-wide analysis gives a wider profile likelihood in Fig.~\ref{fig:Perseus_Likelihood}, which is clearly non-quadratic, because of large degeneracies associated with $X$-ray lines that are slightly outside of the analysis energy range but still contribute enough flux to be included in the analysis through our line-dropping procedure. The non-quadratic behavior arises because of the upper-limits on the line flux of these spectral features (see Tab.~\ref{tab:MOS_Lines}).}  In the smaller windows, we find statistically compatible results with zero line flux and with the 3 keV window result. 
In each case we can place a 95\% upper limit on the UXL flux that excludes the entirety of the $1\sigma$ containment interval recovered in~\cite{Bulbul:2014sua}.

We extend our analysis to the core-masked Perseus XMM-Newton MOS data set, which is constructed following~\cite{Bulbul:2014sua} as described previously. The fits to the data and residuals under the null and signal hypotheses are illustrated in the lower panels of Fig.~\ref{fig:Perseus_Data}. As in the core-unmasked analyses, the $p$-values associated with the $\chi^2_\nu$ values of the null-hypothesis fits are acceptable ($p > 0.05$), though as discussed in Sec.~\ref{sec:methods} this is not a definitive diagnostic of mismodeling for the purpose of searching for narrow spectral features. Still, as shown in the lower panel of Fig.~\ref{fig:Perseus_Likelihood} and in Tab.~\ref{tab:Results}, we find no evidence for a UXL in any of the analysis variations; in fact, in the two narrowest window analyses the best-fit fluxes are slightly negative, while the analysis in the original window size returns a best-fit flux nearly identical to zero ($t \approx 0.02$). 

Our results strongly suggest that there is no evidence for a 3.5 keV line in the Perseus XMM-Newton MOS data.
Our results over all analysis windows are summarized in Tab.~\ref{tab:Results}.

\subsubsection{Centaurus, Coma, and Ophiuchus Clusters}

\begin{table*}[!]{
    \ra{1.3}
    \begin{center}
    \begin{tabular}{@{\extracolsep{4pt}}L{0.11\textwidth}r*{3}{P{0.15\textwidth}}@{}}
    \hlinewd{1pt}
    \multicolumn{2}{l}{\textbf{Analysis Range}} & 3-6 keV & 1 keV & 0.5 keV \\ \hlinewd{1pt}
    \multirow{5}{*}{\parbox{0.11\textwidth}
    {\raggedright\textbf{XMM\\Centaurus}}} & $\chi_\nu^2$ & 590.3/570& 220.8/184&132.0/91\\ 
    & $p$ & $0.27$& $0.03$ & $0.003$ \\ 
    & $\hat{A}$ & $15.3^{+3.9}_{-3.9}$ & $8.0^{+5.0}_{-4.7}$& $4.6^{+5.0}_{-6.9}$ \\ 
    & $t$ & $13.0$ & $3.1$ & $0.43$ \\ 
    & $A^{95}$ & $21.7$ & $16.0 $& $12.7$\\ \hline  
    \multirow{5}{*}{\parbox{0.11\textwidth}
    {\raggedright\textbf{XMM\\Coma}}} & $\chi_\nu^2$ & 582.8/569 & 182.6 /181 & 97.5/94\\ 
    & $p$ & $0.34$ & $0.45$ & $0.38$\\ 
    & $\hat{A}$ & $6.9^{+7.2}_{-7.2}$& $-2.2^{+9.2}_{-8.1}$ & $-10.3^{+7.6}_{-7.3}$\\ 
    & $t$ & $0.97$ & $0$ & $0$\\ 
    & $A^{95}$ & 18.7 & 12.3 & 2.4 \\ \hline  
    \multirow{5}{*}{\parbox{0.11\textwidth}
    {\raggedright\textbf{XMM\\Ophiuchus}}} & $\chi_\nu^2$ & 586.8/578 & 186.8/188 & 90.8/94\\ 
    & $p$ & 0.39 & 0.51 & 0.57\\ 
    & $\hat{A}$ & $5.4^{+13.7}_{-13.6}$ & $-20.5^{+20.8}_{-16.8}$ & $-23.6^{+16.5}_{-16.6}$ \\ 
    & $t$ & $0.16$ & $0$ & $0$\\ 
    & $A^{95}$ & 27.9 & 15.0 & 3.5 \\ \hlinewd{1pt}
    \end{tabular}
    \end{center}}\caption{\label{tab:IndividualResults} The same as Tab.~\ref{tab:Results}, but for the individual clusters. We show no comparison to the original work Ref.~\cite{Bulbul:2014sua} because that work stacked the three clusters. Our results for the joint cluster analysis are shown in Tab.~\ref{tab:Results}.}
\end{table*}

\begin{figure*}[htbp]  
\hspace{0pt}
\vspace{-0.2in}
\begin{center}
\includegraphics[width=\textwidth]{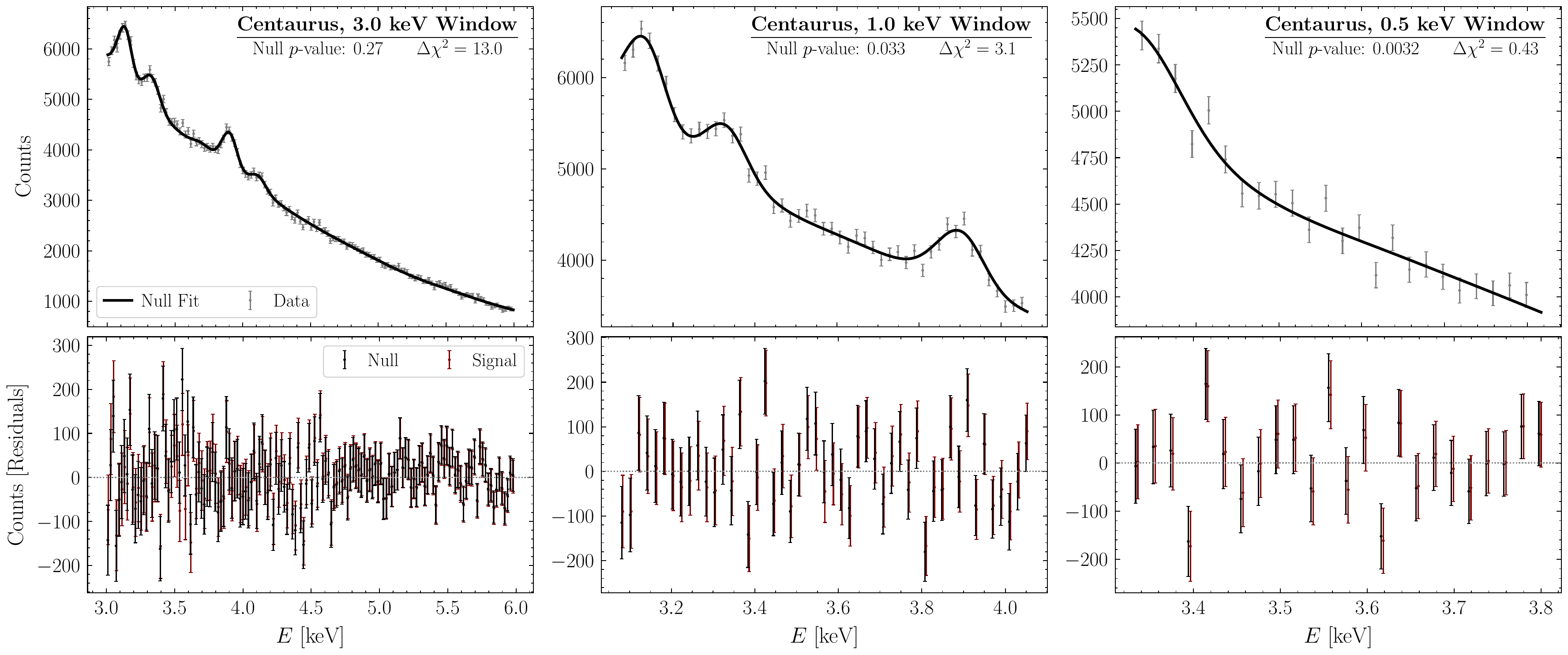}
\includegraphics[width=\textwidth]{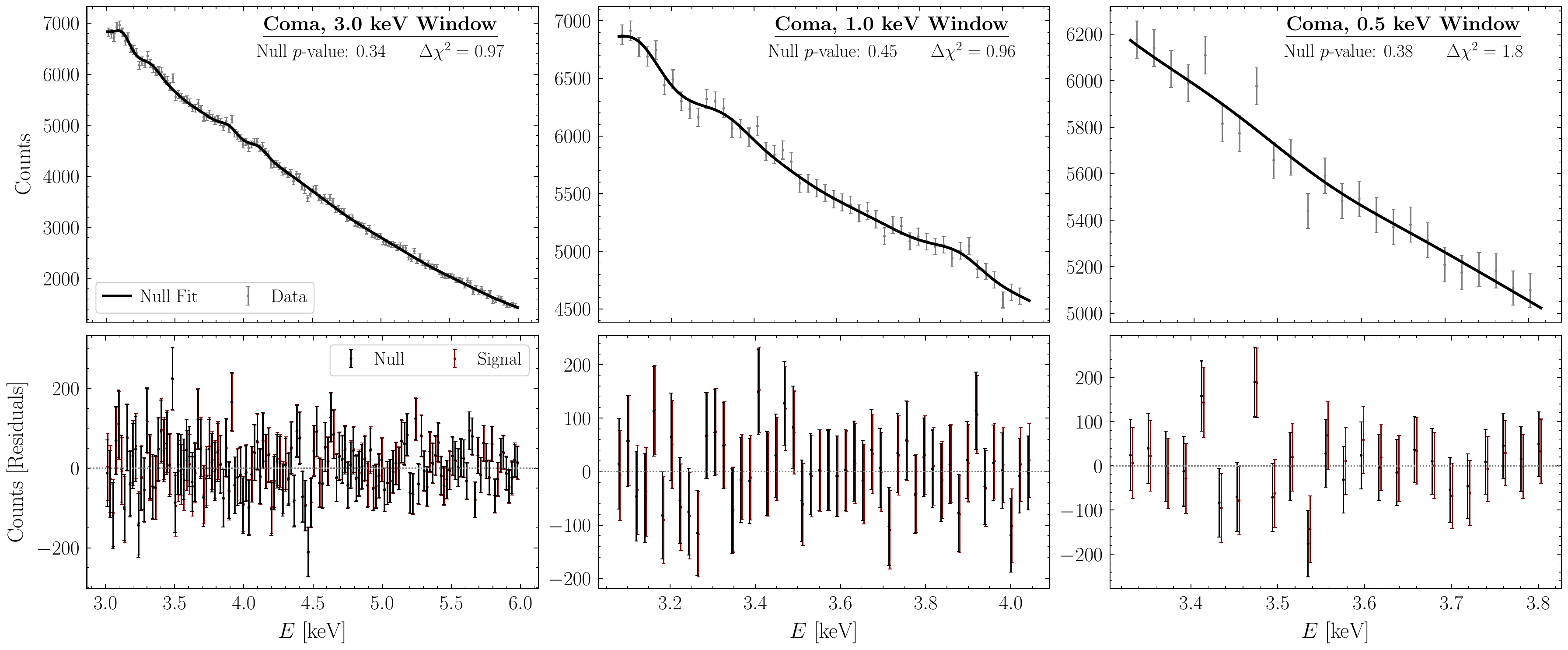}
\includegraphics[width=\textwidth]{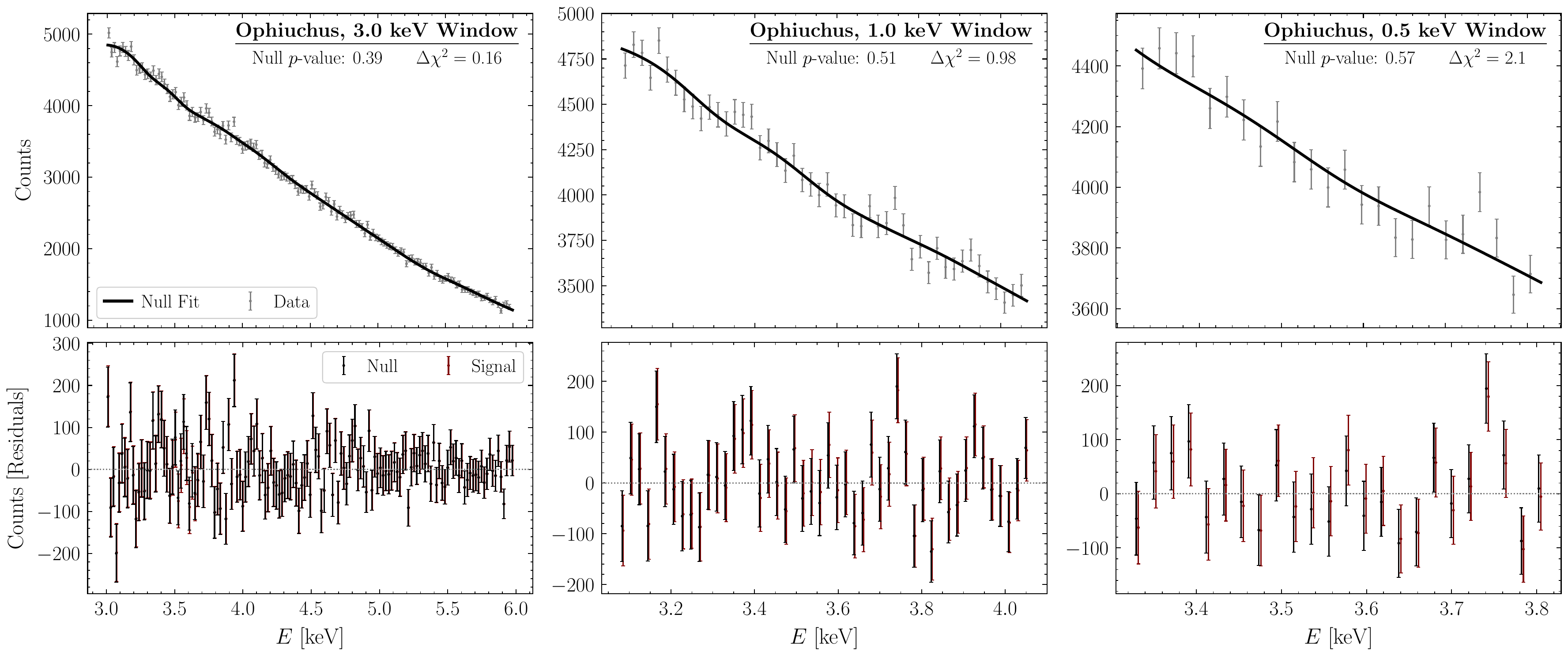}
\caption{The same as Fig.~\ref{fig:Perseus_Data} but for the Centaurus cluster (upper panels), the Coma cluster (middle panels), the Ophiuchus cluster (lower panels).}
\label{fig:CCO_Data}
\end{center}
\end{figure*}

\begin{figure}[htb]  
\hspace{0pt}
\vspace{-0.2in}
\begin{center}
\includegraphics[width=0.49\textwidth]{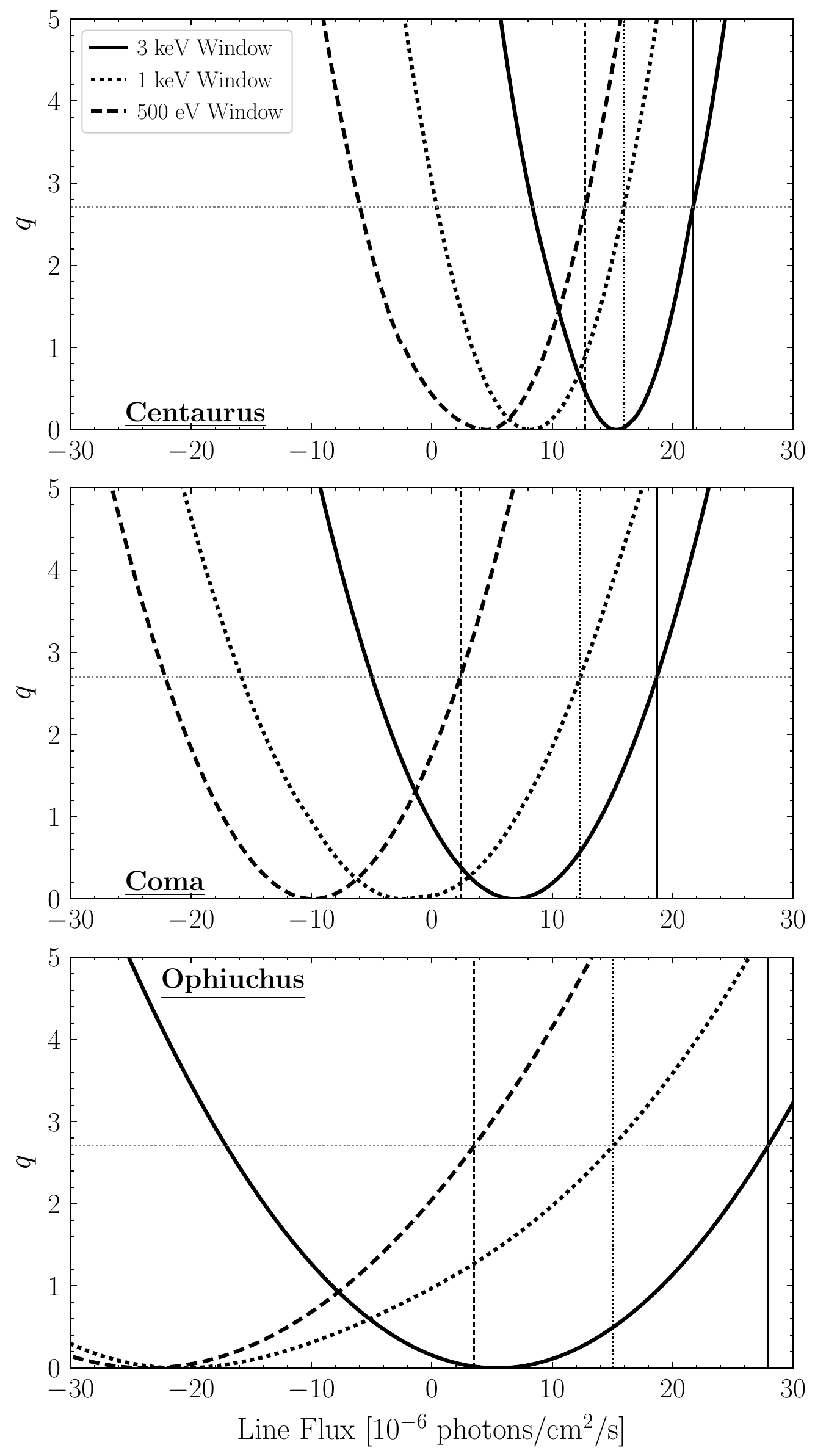}
\caption{The same as Fig.~\ref{fig:Perseus_Likelihood} but for the Centaurus cluster (upper panel), the Coma cluster  (middle panel), and the Ophiuchus cluster (lower panel).}
\label{fig:CCO_Likelihood}
\end{center}
\end{figure}

We now repeat the strategies employed in the previous section for Perseus on the next three brightest clusters in the sample: Centaurus, Coma, and Ophiuchus. While Ref.~\cite{Bulbul:2014sua} analyzed their stacked MOS spectra, here we analyze each cluster individually. In the stacked spectra, a 3.57 keV line was detected in~\cite{Bulbul:2014sua} at $\sim$4$\sigma$ with flux 15.9$_{-3.8}^{+3.4} \times 10^{-6}$ cts/cm$^2$/s. The null model fit gave $\chi_\nu^2 = 562.3/569$, corresponding to a $p$-value $p=0.57$. Again the PN camera did not detect a line and set an upper limit on the line flux smaller than the MOS detection.

We first reanalyze these data as individual clusters for evidence of a 3.57 keV line in the original analysis energy window of~\cite{Bulbul:2014sua}, replicating as closely as possible the original analysis but with global optimization. Following~\cite{Bulbul:2014sua}, we take the redshifts of Centaurus, Coma, and Ophiuchus to be $0.009$, $ 0.022$, and $0.028$, respectively.
Our best-fit models compared to the MOS data for the three clusters are shown in Fig.~\ref{fig:CCO_Data}.  The profile likelihoods for the signal strength normalization are illustrated in Fig.~\ref{fig:CCO_Likelihood}.  The results for the individual clusters are also summarized in Tab.~\ref{tab:IndividualResults}. Ophiuchus and Coma show no evidence for a 3.57 keV UXL ($t \lesssim 1$).  The Centaurus analysis, on the other hand, has evidence for a UXL ($ t \approx 13)$.  
The null-hypothesis fits converge to global minimums with $\chi_\nu^2 = 590.3/570, 582.8/569, 586.8/578$ ($p = 0.27, 0.34, 0.39$) for Centaurus, Coma, and Ophiuchus, respectively. 

When we analyze the spectra in successively smaller windows of 1 keV and 0.5 keV width, we find that in Centaurus the evidence for the UXL disappears entirely (see, {\it e.g.}, Fig.~\ref{fig:CCO_Likelihood}), as may be expected if the results are driven by mismodeling.  In particular, as given in Tab.~\ref{tab:IndividualResults}, in the 1 keV and 0.5 keV wide analysis windows we find $t = 3.1$ and $t = 0.43$, respectively. In both Coma and Ophiuchus the best-fit fluxes are negative in the two narrower analysis windows.  We conclude that the Centaurus, Coma, and Ophiuchus clusters do not show robust evidence for a 3.5 keV UXL.

\subsection{Joint Interpretation of Clusters}

\begin{figure}[htb]  
\hspace{0pt}
\vspace{-0.2in}
\begin{center}
\includegraphics[width=0.49\textwidth]{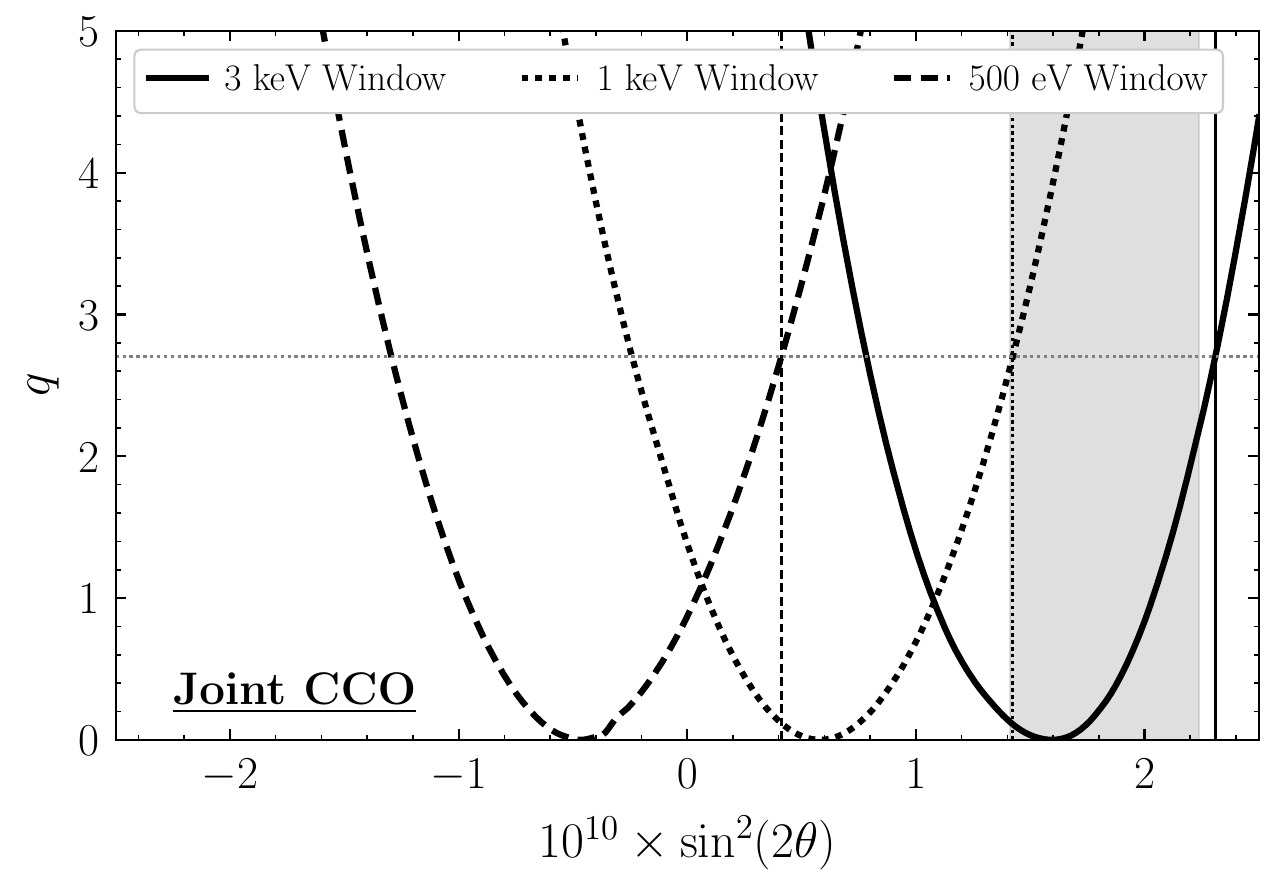}
\caption{The joint profile likelihood for a 3.57 keV UXL in Centaurus, Coma, and Ophiuchus under the assumption that the line arises from DM decay.  The DM decay assumptions allows us to join the profile likelihoods from the individual clusters, shown in Fig.~\ref{fig:CCO_Likelihood}, to constrain the common sterile-active mixing angle $\sin^2(2\theta)$ (see text for details).  The gray band is the best-fit mixing angle range at 1$\sigma$ from the analysis in~\cite{Bulbul:2014sua} to explain their result for the 3.57 keV UXL from their stacked Centaurus, Coma, and Ophiuchus analysis.  In contrast, we find no evidence for a UXL in the joint analysis, and our 95\% upper limits from our 1 keV and 0.5 keV window analyses rule out the full 1$\sigma$ best-fit parameter space from~\cite{Bulbul:2014sua} as indicated by the vertical dashed lines. }
\label{fig:JointCCO_Likelihood}
\end{center}
\end{figure}

As already discussed, we do not perform a stacked analysis of Centaurus, Coma, and Ophiuchus because of statistical complications related to the blueshifting procedure. Instead, having analyzed the clusters individually in the previous subsection we compute the joint profile likelihood by adding the individual profile likelihoods under the DM interpretation.  In particular, for sterile neutrino DM the decay rate to an active neutrino and an $X$-ray is
\es{}{
\Gamma \approx 1.4 \times 10^{-29} \, {\rm s}^{-1} \left( {\sin^2(2 \theta) \over 10^{-7}} \right) \left( {m_s \over 1 \, \, {\rm keV} }\right)^5 \,,
}
where $m_s$ is the DM mass and $\theta$ is the active-sterile neutrino mixing angle~\cite{Pal:1981rm}.  Note that the $X$-ray energy is $m_s / 2$. The flux incident on the detector is then (see, {\it e.g.},~\cite{Safdi:2022xkm})
\es{}{
\Phi = { \Gamma \over 4 \pi m_s} D \,, \qquad D \equiv \int ds\, d\Omega \,\rho_{\rm DM}(s,\Omega) \,,
}
where $s$ is the line-of-sight from the detector and the integral over $d\Omega$ covers the solid angle within the field of view.  In the spirit of following~\cite{Bulbul:2014sua} as closely as possible, we take their assumed values $D \approx 2.41 \times 10^{10}$ $M_\odot/{\rm Mpc}^2$ ($D \approx 2.78 \times 10^{10}$ $M_\odot/{\rm Mpc}^2$) ($D \approx 3.05 \times 10^{10}$ $M_\odot/{\rm Mpc}^2$) for Centaurus (Coma) (Ophiuchus).  Note that $\Phi$ has units of flux (${\rm cts}/{\rm cm}^2/{\rm s}$), which allows us to interpret the profile likelihoods in Fig.~\ref{fig:CCO_Likelihood} in terms of profile likelihoods for $\sin^2(2 \theta)$. We then join the profile likelihoods to construct the joint profile likelihood illustrated in Fig.~\ref{fig:JointCCO_Likelihood}. The gray band in that figure is the best-fit parameter space at 1$\sigma$ from~\cite{Bulbul:2014sua} to explain their stacked Centaurus, Coma, and Ophiuchus result using decaying DM. 
Our 3 keV window analysis finds $t \approx 12.2$ in favor of the signal model, though this is driven by Centaurus, with a best-fit mixing angle consistent with the 1$\sigma$ band recovered in~\cite{Bulbul:2014sua}. On the other hand, our narrower energy window analyses find $t \approx 1.4$ and $t = 0$, for 1 keV and 0.5 keV wide windows, respectively (see Tab.~\ref{tab:Results}).  In fact, our 95\% upper limits from the narrow window analyses exclude the parameter space to explain the UXL at 1$\sigma$ found in~\cite{Bulbul:2014sua}.

\section{M31 Data from XMM-Newton}
\label{sec:xmm-m31}

Ref.~\cite{Boyarsky:2014jta} used the MOS camera on XMM-Newton to detect the 3.5 keV line at $3.2\sigma$ significance in the M31 ``on-center" observations, which are defined in Ref.~\cite{Boyarsky:2014jta} as those within $1.5'$ of the M31 center. In this section we critically reanalyze these observations, following as closely as possible the analysis procedure described in~\cite{Boyarsky:2014jta}.

\subsection{Data Reduction} 
\label{sec:xmm-m31-reduction}

We reduce the XMM-Newton M31 observations identically to the XMM-Newton cluster observations, detailed in Sec.~\ref{sec:xmm-cluster-reduction}, except for two changes to reproduce Ref.~\cite{Boyarsky:2014jta}: (i) we use the ESAS point-source finding and masking task \texttt{cheese} instead of \texttt{wavdetect}, and (ii) we do not subtract the QPB from the data.\footnote{
Ref.~\cite{Boyarsky:2014jta} does not provide a description of their point-source removal, so we reduce the data using both the default ESAS point source removal task \texttt{cheese} and \texttt{wavdetect}. A visual inspection of the \texttt{cheese} results shows that it clearly fails to find many point sources, but it results in a power-law spectral index consistent with that shown in Ref.~\cite{Boyarsky:2014jta}. On the other hand, \texttt{wavdetect} finds more point sources, but results in a spectrum with significantly less counts than in Ref.~\cite{Boyarsky:2014jta}. We thus elect to analyze the results obtained with \texttt{cheese} to follow as closely as possible Ref.~\cite{Boyarsky:2014jta}.
}

\subsection{Likelihood and Model Components}
\label{sec:xmm-m31-models}

\begin{table*}[htb]{
    \ra{1.3}
    \begin{center}
    \begin{tabular}{c*{8}{C{0.1\textwidth}}}
    \hlinewd{1pt} 
    \textbf{Element} & \textbf{Si XIV} & \textbf{Al XIII} & \textbf{Si XII}  & \textbf{Si XII}  & \textbf{Si XII}  &\textbf{Si XV}  & \textbf{S XIV}  & \textbf{S XIV}  \\ 
    Energy [keV] & 2.01 & 2.05 & 2.18 & 2.29 & 2.34 & 2.45 & 2.51 & 2.62  \\ \hlinewd{1pt} 
    2-8 keV Fit & $8.6^{+4.0}_{-4.0}$ & $6.5^{+3.1}_{-3.1}$ & $8.6^{+1.6}_{-1.6}$ & $5.3^{+1.6}_{-1.6}$ & - & $7.9^{+1.4}_{-1.4}$ & -- & $4.4^{+1.1}_{-1.1}$  \\
    3 keV Fit & -- & -- & -- & -- & -- & -- & -- & -- \\
    1 keV Fit & -- & -- & -- & -- & -- & -- & -- & --  \\
    0.5 keV Fit & -- & -- & -- & -- & -- & -- & --  \\
    \hlinewd{1pt}
    \textbf{Element} & \textbf{S XV} 
    & \textbf{Ar XVII}  & \textbf{Ar XVIII} + \textbf{K} K$\alpha$ & \textbf{Ar XVII} & \textbf{Ar XVII} + \textbf{Ca} K$\alpha$ & \textbf{Ca XIX} & \textbf{Ca XIX} & \textbf{Ar XVIII}  \\ 
    Energy [keV] & 2.88 & 3.124 & 3.315 & 3.617 & 3.688 & 3.861 & 3.902 & 3.936 \\ \hlinewd{1pt} 
    2-8 keV Fit & -- & $2.1^{+0.9}_{-0.9}$ & $1.8^{+0.9}_{-0.9}$ & $2.8^{+0.8}_{-0.8}$ & -- & $1.8^{+0.8}_{-0.8}$ & -- & --  \\
    3 keV Fit & -- & -- & -- & $1.9^{+0.8}_{-0.8}$ & -- & $1.2^{+0.8}_{-0.8} $& -- & -- \\
    1 keV Fit & -- & -- & -- & -- & -- & -- & -- & --  \\
    0.5 keV Fit & -- & -- & -- & -- & -- & -- & --  \\
    \hlinewd{1pt}
    \textbf{Element} & \textbf{Cr} K$\alpha$ & \textbf{Mn} K$\alpha$ & \textbf{Fe} K$\alpha$ & \textbf{Fe} K$\beta$ & \textbf{Ni} (K$\alpha$) & \textbf{Ni} K$\alpha$ & \textbf{Cu} K$\alpha$ &  \\ 
    Energy [keV] &  5.41 & 5.893 & 6.398 & 7.058 & 7.461 & 7.489 & 8.028 &  \\ \hlinewd{1pt} 
    2-8 keV Fit & $22.0^{+1.0}_{-1.0}$ & $23.6^{+1.2}_{-1.2}$ & $42.9^{+1.5}_{-1.5}$ & $9.0^{+2.0}_{-2.0}$ & $32.4^{+2.7}_{-2.7}$ & -- & -- &   \\
    3 keV Fit & $22.6^{+1.1}_{-1.1}$ & $24.5^{+1.6}_{-1.6}$ & -- & -- & -- & -- & -- &  \\
    1 keV Fit & -- & -- & -- & -- & -- & -- & -- &   \\
    0.5 keV Fit & -- & -- & -- & -- & -- & -- & --  \\
    \hlinewd{1pt}
    \end{tabular}\end{center}}
\caption{\label{tab:M31_Lines} As in Tab.~\ref{tab:MOS_Lines}, but for the M31 XMM-Newton MOS data set. Following~\cite{Boyarsky:2014ska}, no bounds are put on line intensities.}
\end{table*}

As compared with the galaxy clusters discussed in Sec.~\ref{sec:xmm-cluster}, M31 is expected to be a relatively cleaner environment in $X$-rays, and so, as in \cite{Boyarsky:2014jta}, we do not include any plasma components in our continuum model. Instead the continuum is composed of one folded and one unfolded power-law, which represent the M31 $X$-ray emission and instrumental soft-proton contamination, respectively. The brightest line-like emission is expected to be produced by detector fluorescence; following Ref.~\cite{Boyarsky:2014jta}, we consider $K\alpha$ fluorescence lines associated with Cr, Mn, K, Fe, Ni, Ca, and Cu, as well as $K\beta$ lines associated with Fe. Ref.~\cite{Boyarsky:2014jta} also includes several astrophysical lines at low energies that were introduced to explain observed, large residuals. On the other hand, Ref.~\cite{Boyarsky:2014jta} is unspecific with regard to precisely what lines were introduced. To consider astrophysical lines in a principled manner, we include in our candidate line list all astrophysical lines with rest energies between 3-4 keV from \cite{Bulbul:2014sua}, excluding those within the range of 3.4-3.6 keV, as was done in~\cite{Boyarsky:2014ska}. Then, as in our analysis of the clusters, we globally optimize and iteratively  test the significance of lines, keeping only those which improve the $\chi^2$ by 3 or more. The complete list of lines we consider along with details regarding which are ultimately included in the final model is provided in Tab.~\ref{tab:M31_Lines}. In contrast with the cluster analyses, the M31 astrophysical lines do not need to be redshifted.

The nuisance parameter vector is given by \mbox{$\bm{\theta} = \{\bm{\theta}_\mathrm{pl}, \bm{\theta}_\mathrm{line}\}$}, which are correspondingly defined by 
\es{}{
\bm{\theta}_\mathrm{pl} &=\{\{I_\mathrm{pl, 1}, k_\mathrm{pl, 1}\}, \{I_\mathrm{pl, 2}, k_\mathrm{pl, 2}\}\} \,, \\
\bm{\theta}_\mathrm{line} &= \{\{E_i, \Delta E_i, I_i\}_{i=1}^{N_\mathrm{astro.}}, \{E_i, \Delta E_i, I_i\}_{i=1}^{N_\mathrm{inst.}}\} \,.
}
Note that the number of astrophysical and instrumental lines, $N_\mathrm{astro.}$ and $N_\mathrm{inst.}$ respectively, differ between the cluster analysis and this analysis, and no hydrogen absorption is applied.
The signal plus background model prediction per energy bin, $\mu(A,\bm{\theta})$, is given by
\begin{align}
\begin{split}
    \mu_\mathrm{pl}(\bm{\theta}_\mathrm{pl}) =&\mathrm{RSP} \star \mathbf{powerlaw}(I_\mathrm{pl,1}, k_\mathrm{pl, 1})\\
    &+ \mathbf{powerlaw}(I_\mathrm{pl,2}, k_\mathrm{pl, 2}) \\
    \mu_\mathrm{line}(\bm{\theta}_\mathrm{line}) &= \mathrm{RSP} \star \sum_i^{N_\mathrm{lines}} \mathbf{gauss}(E_i, \Delta E_i, I_i) \\
    \mu_\mathrm{bkg.} &= \mu_\mathrm{pl} + \mu_\mathrm{line}.\\
    \mu(A,\bm{\theta}) &= \mu_\mathrm{bkg.}(\theta) + \mathrm{RSP} \star \mathbf{gauss}(3.53, 0, A).
\end{split}
\end{align}
The total model therefore consists of the background model plus a zero-width signal line at 3.53 keV with intensity parameter $A$ (the signal line best-fit energy in Ref.~\cite{Boyarsky:2014jta} was 3.53 keV). We then use the identical likelihood as for the cluster analysis \eqref{eq:like} but with the model prediction given above. 

\subsection{Data Analysis}
\label{sec:xmm-m31-analysis}

Ref.~\cite{Boyarsky:2014jta} found evidence at the level of $t = 13$ (3.2$\sigma$) for a line at 3.53 keV with flux $4.9_{-1.3}^{+1.6} \times 10^{-6}$ cts/cm$^2$/s. Their analysis covered the energy range 2 - 8 keV and down-binned the data to bins of width 60 eV, so that the null model had $\chi_\nu^2 = 97.8/74$, corresponding to a $p$-value $p=0.036$. We reanalyze these data using the model described in the previous subsection, starting with the same analysis energy window. In contrast with~\cite{Boyarsky:2014jta}, however, we do not downbin the data in energy. 

\begin{figure*}[htb]  
\hspace{0pt}
\vspace{-0.2in}
\begin{center}
\includegraphics[width=0.99\textwidth]{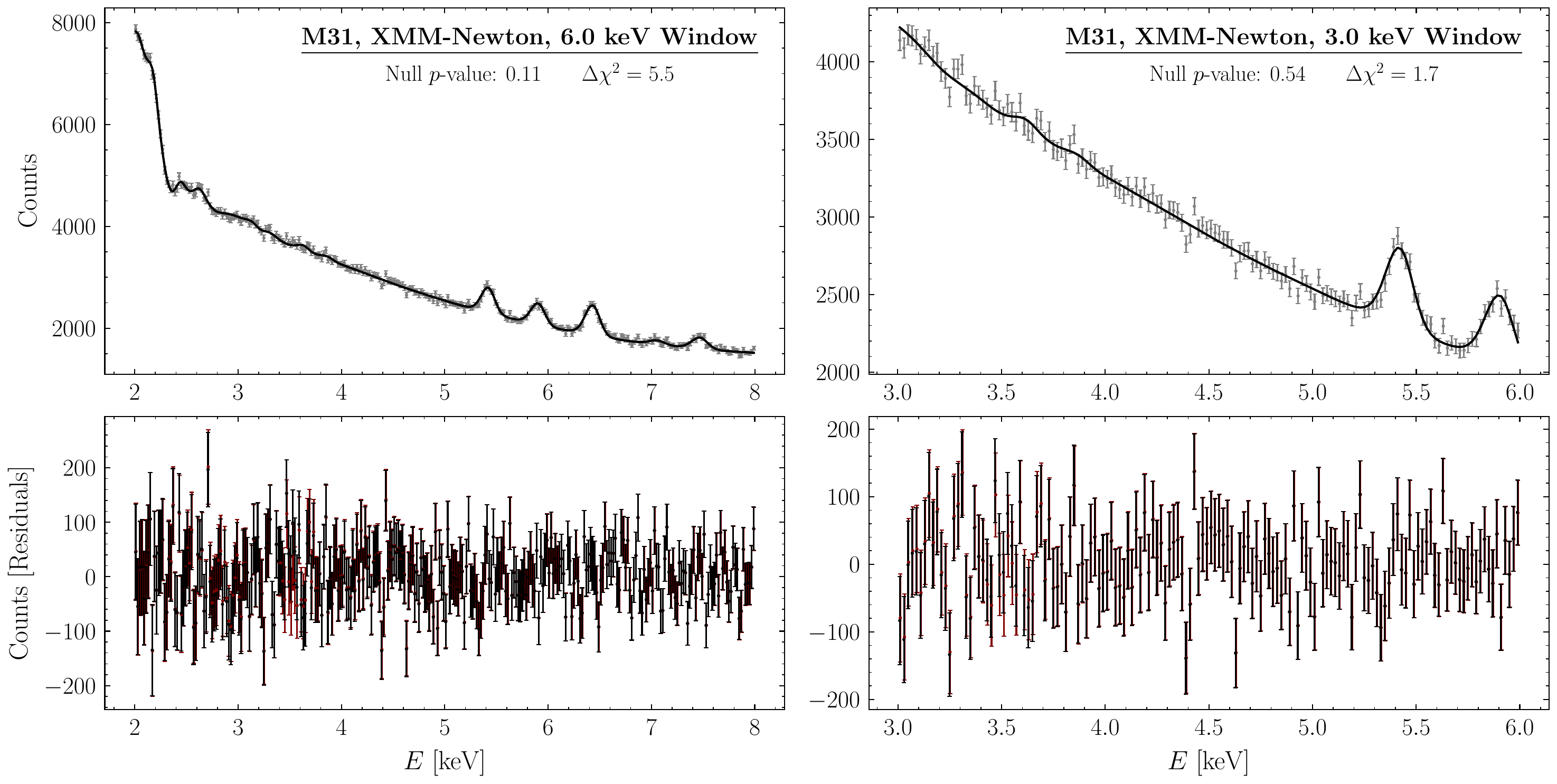}
\includegraphics[width=0.99\textwidth]{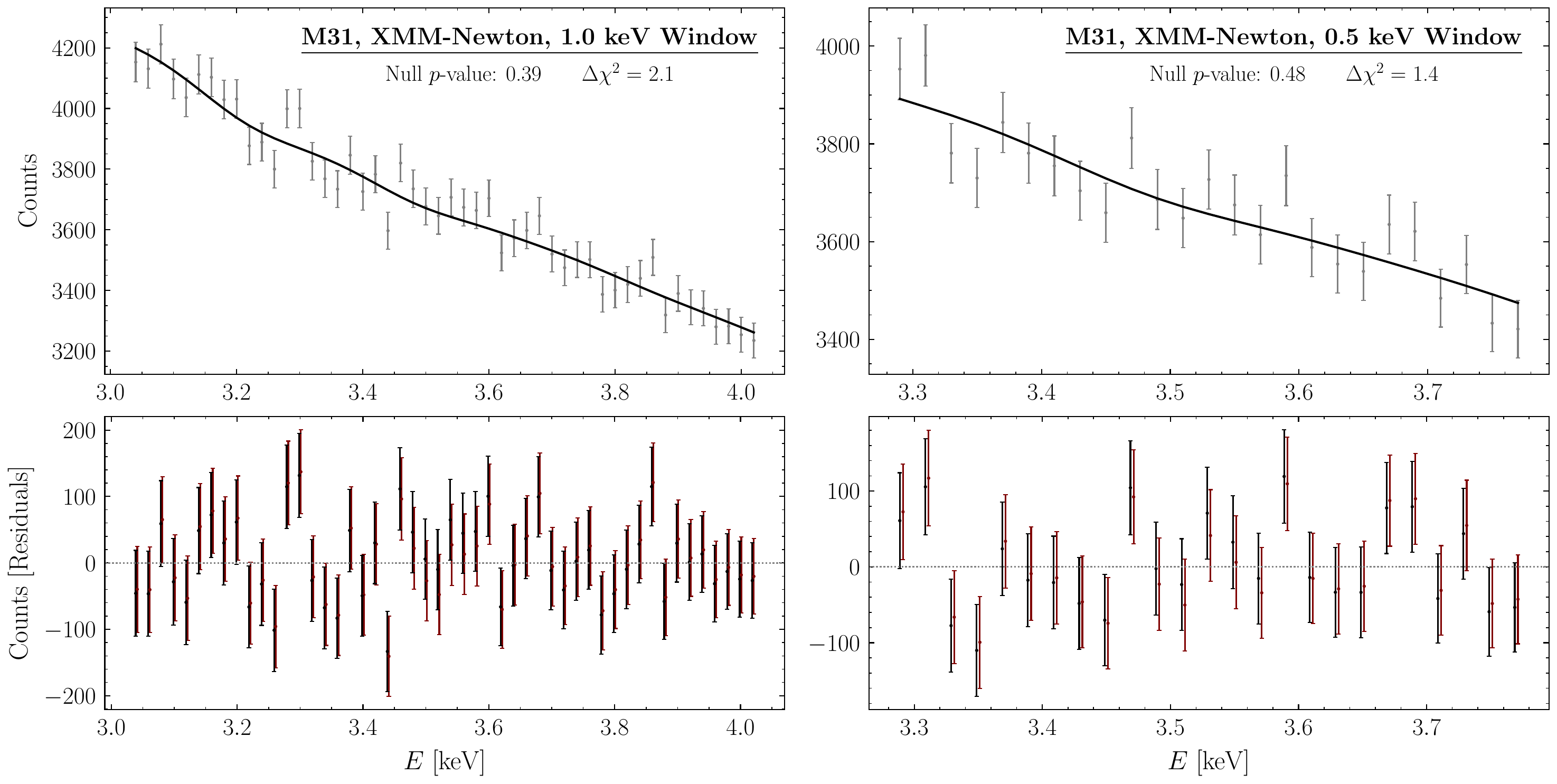}

\caption{The stacked XMM-Newton MOS data of M31 (gray points with error bars) along with the best fit null model in each of our analysis windows. In the upper left is the 2\textemdash8 keV window of Ref.~\cite{Boyarsky:2014jta}, upper right is the 3.0 keV window, lower left 1.0 keV, and lower right 0.5 keV. The bottom panels illustrate the residuals after subtracting the best-fit null and signal models.}
\label{fig:M31_Data}
\end{center}
\end{figure*}

\begin{figure}[htb]  
\hspace{0pt}
\vspace{-0.2in}
\begin{center}
\includegraphics[width=0.49\textwidth]{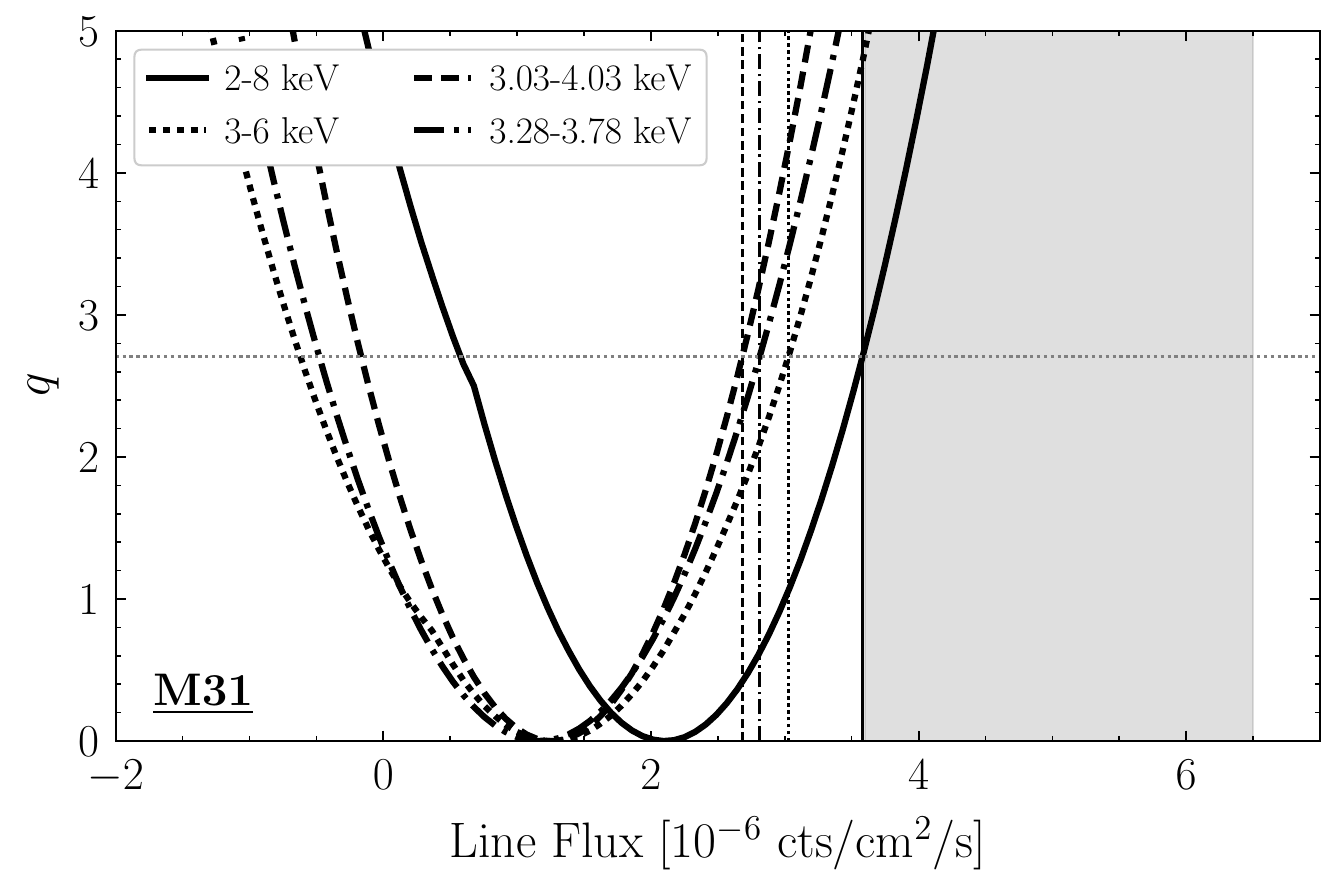}
\caption{The profile likelihood in each of the four M31 XMM-Newton MOS analysis windows: $2 - 8$ keV (solid), 3.0 keV (dotted), 1.0 keV (dashed), and 0.5 keV (dash-dotted). The 95\% upper limits from each fit are shown as vertical lines with corresponding styles. The 1$\sigma$ best fit region for the 3.5 keV line flux in Ref.~\cite{Boyarsky:2014jta} is in shaded gray.}
\label{fig:M31_Likelihood}
\end{center}
\end{figure}

Using the same machinery as in Sec.~\ref{sec:xmm-cluster}, we construct the profile likelihood for the UXL flux $A$ in the original analysis energy window as~\cite{Boyarsky:2014jta}, though we use the global minimizer {\it differential evolution} as previously described. (Ref.~\cite{Boyarsky:2014jta} instead uses local optimization.) We show the best fit models in the upper left panel of Fig.~\ref{fig:M31_Data}, and the profile likelihood in Fig.~\ref{fig:M31_Likelihood}. We find $t = 5.5$ evidence for the UXL with best fit flux $2.1_{-0.9}^{+0.9} \times 10^{-6}$ cts/cm$^2$/s, with null $\chi_\nu^2 = 1225.3/1166$ ($p=0.11$).  As in the MOS cluster analyses described in Sec.~\ref{sec:xmm-cluster}, we are unable to reproduce a UXL line consistent with that found in the original work~\cite{Boyarsky:2014jta}. For example, the best-fit UXL line flux from~\cite{Boyarsky:2014jta} at 1$\sigma$ confidence is shaded gray in Fig.~\ref{fig:M31_Likelihood}; our upper one-sided 95\% upper limit on the flux from our full-energy-range analysis excludes the entire 1$\sigma$ interval from~\cite{Boyarsky:2014jta}.  

Next, we study how our results vary under reductions of the analysis energy window. We perform the analysis in three additional windows: a 3-6 keV window, a 1 keV wide window (centered around 3.53 keV), and a 0.5 keV wide window (centered around 3.53 keV). The best-fit models in all four cases are shown in Fig.~\ref{fig:M31_Data}. If there is no mismodeling present in the data, we expect  
that the recovered line flux in all four cases to be compatible.  
On the other hand, we observe that the best fit UXL flux decreases with window size (see Fig.~\ref{fig:M31_Likelihood}), though the three narrower-window analyses all have consistent best-fit fluxes.  On the other hand, all three of the narrower-window analyses find $t < 2$, indicating no evidence for a UXL.  
These results provide evidence that the modest appearance of a UXL in the widest energy window analysis of the M31 data is driven by mismodeling.

\section{Perseus Cluster Data from Chandra}
\label{sec:chandra-perseus}

Ref.~\cite{Bulbul:2014sua} analyzed Chandra observations of the Perseus cluster to verify that the 3.5 keV line was not an instrumental effect in XMM-Newton MOS. In this section we reanalyze the observations taken with the ACIS-I instrument onboard Chandra.  We do not find evidence for a UXL in any of the analysis variations we perform, finding $t = 0$ in all analyses as described below.

\subsection{Data Reduction}
\label{sec:chandra-perseus-reduction}

We reduce the data with CIAO 4.11 and CALDB 4.8.5, but otherwise identically to Ref.~\cite{Bulbul:2014sua}. We download each ACIS-I observation of the Perseus Cluster with the CIAO task \texttt{download\_chandra\_obsid} and reprocess it with the most recent calibration using \texttt{chandra\_repro}. We mask point sources using \texttt{wavdetect} and use the filtered and masked event files to extract the source spectrum, RMF, and ARF with \texttt{specextract}. We run \texttt{blank\_sky} to create background event files for each observation and normalize that background spectrum such that the 9--12 keV count rate equals that in the observation.
We stack the observation and background spectra and stack the responses weighted by exposure time. We perform the analysis on background subtracted data sets.

\subsection{Likelihood and Model Components}
\label{sec:chandra-perseus-models}

\begin{table*}[htb]{
    \ra{1.3}
    \begin{center}
    \begin{tabular}{c*{9}{C{0.085\textwidth}}}
    \hlinewd{1pt} 
    \textbf{Element} & \textbf{Si} & \textbf{S} & \textbf{S} & \textbf{Ar} & \textbf{Ar} & \textbf{K} & \textbf{K} & \textbf{Ar}  \\ 
    Energy [keV] & 2.506 & 2.62 & 2.88 & 3.124 & 3.32 & 3.472 & 3.511 & 3.617 \\ \hlinewd{1pt} 
    \textbf{Bound} & -- & -- & -- & -- & -- & $10.2$ & $9.3$ & $1.29$ \\ 
    2.5-6 keV Fit & -- & $639_{-15}^{+15}$ & -- & $125_{-9}^{+9}$ & $166_{-8}^{+8}$ & -- & -- & --  \\
    3 keV Fit & -- & -- & -- & $134_{-13}^{+13}$ & $169_{-9}^{+9}$ & -- & -- & -- \\
    1 keV Fit & -- & -- & -- & $228_{-31}^{+31}$ & $181_{-14}^{+14}$ & -- & -- & -- \\
    0.5 keV Fit & -- & -- & -- & -- & $211_{-19}^{+19}$ & -- & -- & -- \\ \hlinewd{1pt} 
    \textbf{Element} & \textbf{Ar} & \textbf{K} & \textbf{Ca} & \textbf{Ca} & \textbf{Ar} & \textbf{Ca} & \textbf{Ca} & \textbf{Cr} \\
    Energy [keV] & 3.685 & 3.705 & 3.861 & 3.902 & 3.936 & 4.107 & 4.584 & 5.682 \\ \hlinewd{1pt}
    \textbf{Bound} & $24$ & $7.8$ & -- & -- & -- & -- & -- & -- \\
    2.5-6 keV Fit & $24$ & -- & -- & $141_{-7}^{+7}$ & -- & $85_{-6}^{+6}$ & -- & $19_{-6}^{+6}$ \\
    3 keV Fit & $24$ & -- & -- & $142_{-7}^{+7}$ & -- & $83_{-6}^{+6}$ & -- & $18_{-6}^{+6}$ \\
    1 keV Fit & $24$ & -- & -- & $190_{-15}^{+15}$ & -- & $164_{-33}^{+33}$ & -- & -- \\
    0.5 keV Fit & $24$ & -- & -- & $290_{-110}^{+110}$ & -- & -- & -- & --  \\ \hlinewd{1pt}
    \end{tabular}\end{center}}
\caption{\label{tab:Chandra_Perseus_Lines} The same as Tab.~\ref{tab:MOS_Lines}, but for the Perseus Chandra data set.}
\end{table*}

Our background model is that in Ref.~\cite{Bulbul:2014sua}, which is nearly identical to the model used for the analysis of the XMM-Newton Perseus observations, discussed in Sec.~\ref{sec:xmm-cluster-models}.  In particular, we use the same list of astrophysical lines as in the Perseus MOS analysis, though  the line dropping procedure is performed self-consistently on the Chandra data. The list of included lines is given in Tab.~\ref{tab:Chandra_Perseus_Lines}. We allow the inferred rest energies of these lines to vary by up to $\delta E = 14.6 \, \mathrm{eV}$ from their expected rest energies. 
We also use the same global optimization procedure with differential evolution for maximizing the likelihood as in the MOS analyses.

\subsection{Data Analysis}
\label{sec:chandra-perseus-analysis}

In this section, we detail our re-analysis of the Chandra Perseus data. The original analysis of these data~\cite{Bulbul:2014sua} was performed over the 2.5-6 keV energy range and found evidence for a UXL at 3.56 keV at the level of $t = 6.2$ (as in the rest of this work, all energies are source-frame) with a flux $(18.6_{-8.0}^{+7.8}) \times 10^{-6}$ cts/cm$^2$/s. The $\chi_\nu^2$ for the null model in~\cite{Bulbul:2014sua} was $158.7/152$ (corresponding to a $p$-value $p=0.34$).\footnote{We note that a typographical error appears in Tab.~5 of \cite{Bulbul:2014sua}, reporting the $\chi_\nu^2$ under the null as $152.6/151$, whereas this is explicitly stated to be the $\chi_\nu^2$ including the signal line in the main text of that reference.}

\begin{figure*}[htb]  
\hspace{0pt}
\vspace{-0.2in}
\begin{center}
\includegraphics[width=0.99\textwidth]{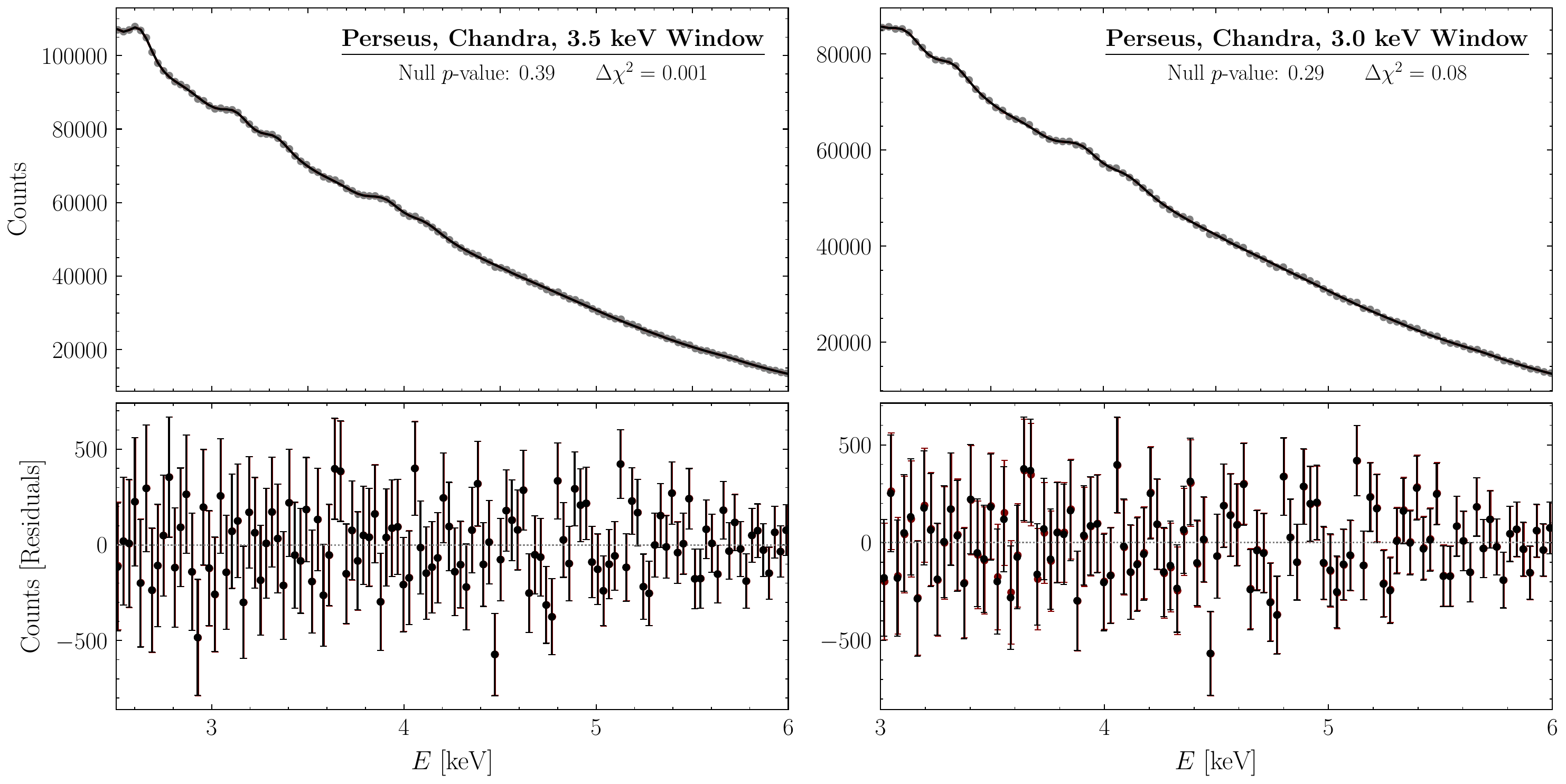}
\includegraphics[width=0.99\textwidth]{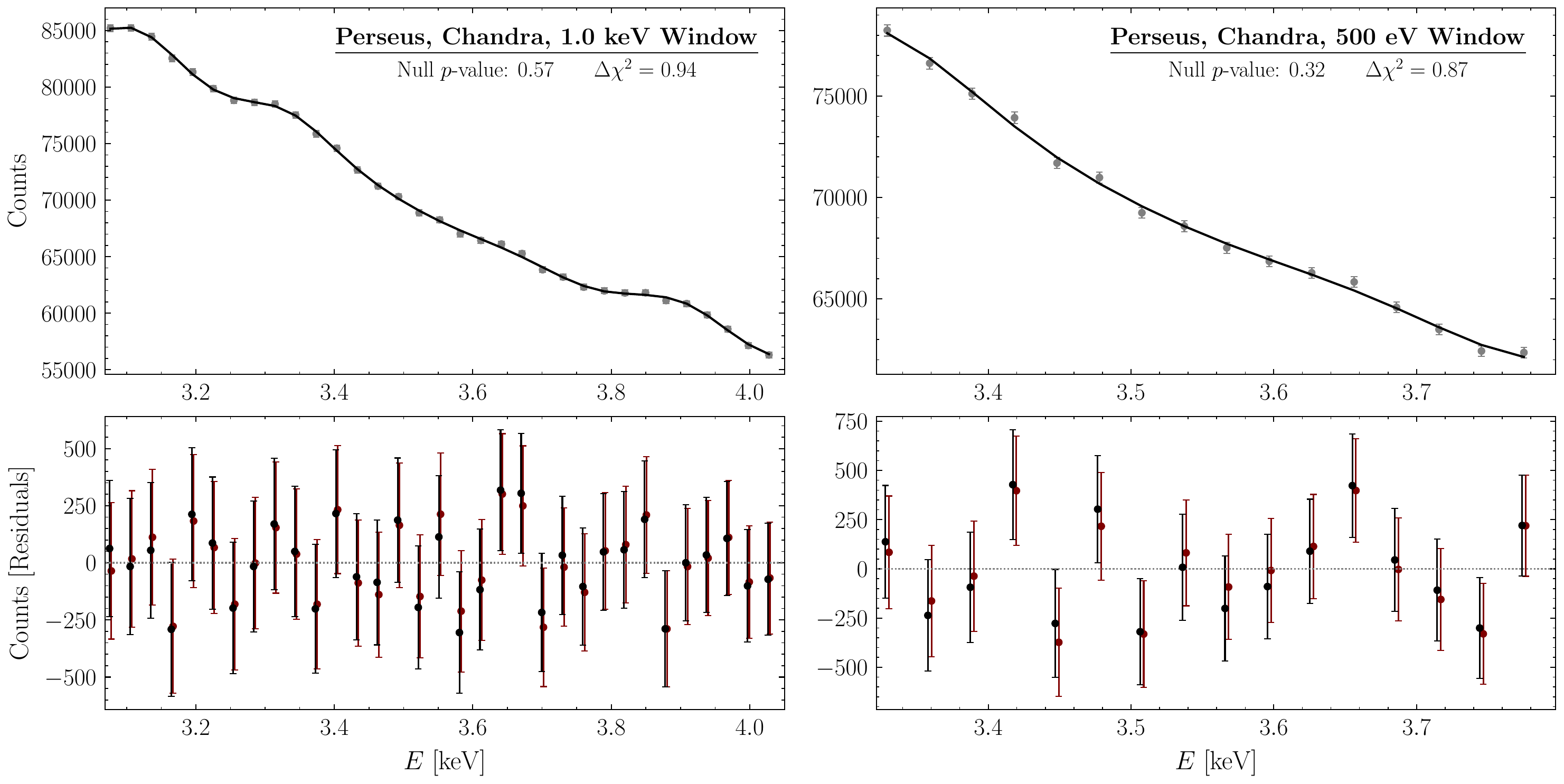}
\caption{The stacked Chandra data of the Perseus cluster (gray points with error bars) along with the best fit null model in each of our analysis windows. In the upper left is the 2.5\textemdash 6 keV window of Ref.~\cite{Bulbul:2014sua}, upper right 3.0 keV, lower left 1.0 keV, and lower right 0.5 keV. The bottom panels illustrate the residuals after subtracting the best-fit null and signal models.  Note that the data in all panels have been down-binned by a factor of two in energy for presentation purposes only.}
\label{fig:Chandra_Perseus_Data}
\end{center}
\end{figure*}

\begin{figure}[htb]  
\hspace{0pt}
\vspace{-0.2in}
\begin{center}
\includegraphics[width=0.49\textwidth]{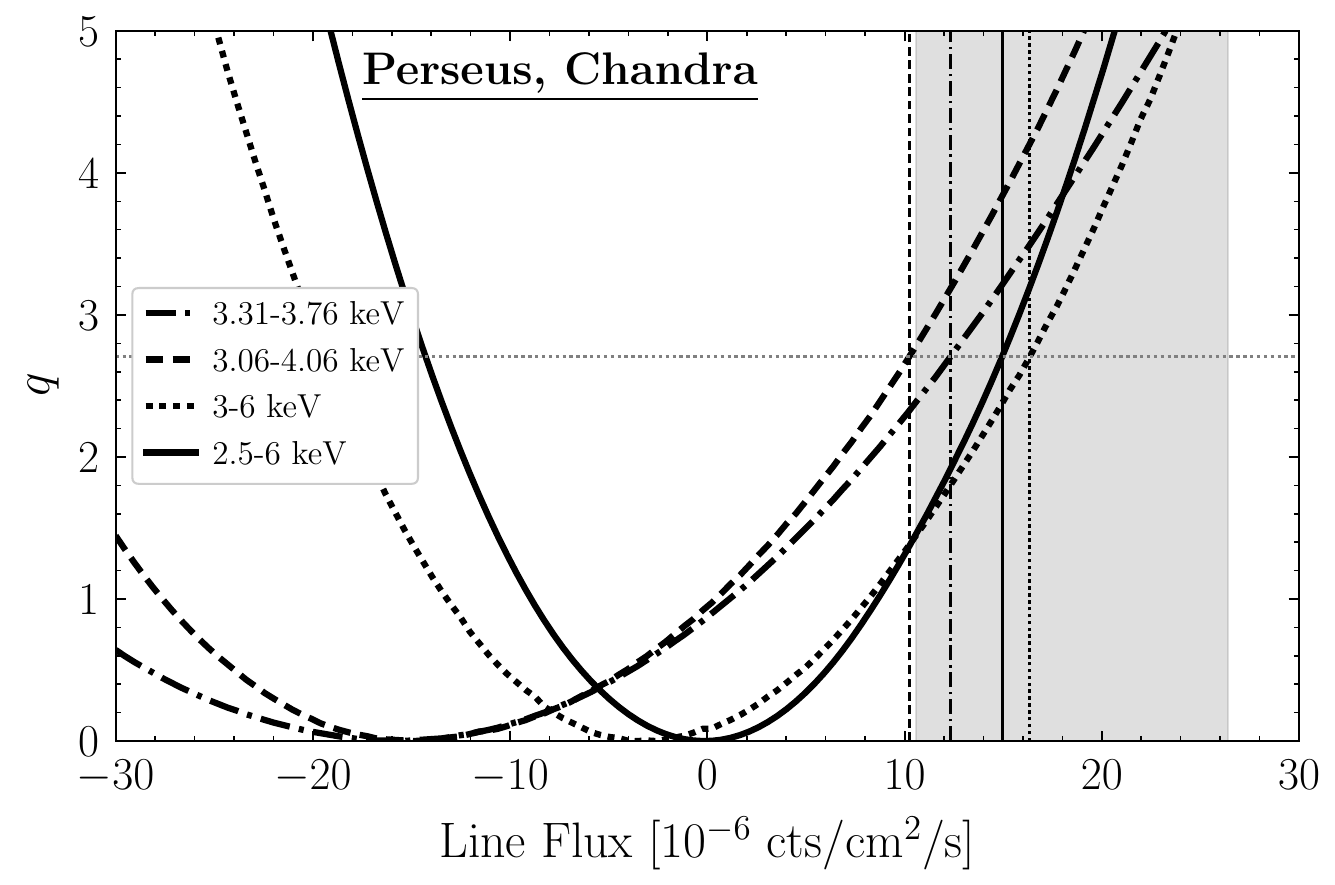}
\caption{The profile likelihood in each of the four analysis windows for the Chandra Perseus analysis: $2.5 - 6$ keV (solid), 3 keV (dashed), 1 keV (dash-dotted), and 0.5 keV (dotted). The 95\% upper limits from each fit are shown as vertical lines with corresponding styles. The 1$\sigma$ best fit region for the 3.5 keV line flux in Ref.~\cite{Bulbul:2014sua} is in shaded gray.}
\label{fig:Chandra_Perseus_Likelihood}
\end{center}
\end{figure}

As in the previous analyses, we profile over positive and negative signal parameters $A$ to construct a profile likelihood, using the global optimization scheme introduced in earlier sections.
Our best-fit null model in this energy window is shown in Fig.~\ref{fig:Chandra_Perseus_Data} along with the Chandra data, in addition to the residuals under the null and signal models.  The profile likelihood is illustrated in Fig.~\ref{fig:Chandra_Perseus_Likelihood}. We find no evidence for a UXL ($t =0$), given that we recover a slightly negative signal flux.
Under the null model, the global optimization finds a $\chi_\nu^2$ of $216.1/211$ ($p=0.39$).\footnote{
We cannot resolve the source of the discrepancy in the number of DOF between our work and~\cite{Bulbul:2014sua}.
}

Now we examine the impact of changing the analysis energy window size. In Fig.~\ref{fig:Chandra_Perseus_Data} we show the best-fit models over the reduced windows of width 3 keV, 1 keV, and 0.5 keV, while in Fig.~\ref{fig:Chandra_Perseus_Likelihood} we show the corresponding profile likelihoods. Going to reduced energy ranges does not qualitatively affect the best-fit line flux, which stays negative, or change the tension between our results and those in~\cite{Bulbul:2014sua}.
A summary of these results is provided in Tab.~\ref{tab:Results}.

\section{Milky Way Survey Fields from Chandra}
\label{sec:survey}

The most recent significant claimed detection of the 3.5 keV line (in 2017) was in blank sky observations of the Milky Way halo taken by Chandra~\cite{Cappelluti:2017ywp}. Ref.~\cite{Cappelluti:2017ywp} found $\sim$3$\sigma$ evidence for a 3.51 keV line in two survey fields --- Chandra Deep Field South (CDFS) and the Chandra-COSMOS Legacy Survey (CCLS) --- when a joint fit is performed to the two data sets. In this section we reanalyze these data sets, implementing identical data reduction and modeling procedures, up to our use of a global optimization procedure and spline interpolation of the \texttt{wabs} absorption model. 

The fiducial analysis of Ref.~\cite{Cappelluti:2017ywp} subtracts from the observed data an instrumental background data set constructed from observations taken when the detector is in a stowed position, {\it i.e.} away from the focal plane and not observing any astrophysical $X$-rays. We find, as in Ref.~\cite{Cappelluti:2017ywp}, that the subtraction of this data set results in poor fits to the data; the fits to the data are dramatically improved without background subtraction.   
By modeling the background, we find no evidence for a 3.5 keV UXL in CDFS and CCLS. The alternate analysis, where we subtract the background data, is discussed in App.~\ref{app:DeepFieldBkgSub}. In that case we do reproduce evidence for a 3.5 keV UXL, but we link this evidence to a deficit near 3.5 keV in the background data itself.

Note that Ref.~\cite{Cappelluti:2017ywp} claims that the background-subtracted analysis has a best-fit UXL line energy of 3.51 keV, while the background-modeled (but not subtracted) analysis has a best-fit line energy of 3.49 keV Since in our analyses below we model instead of subtract the background, we fix the line energy to 3.49 keV.

\subsection{Data Reduction}
\label{sec:survey-reduction}

We perform the data reduction using CIAO 4.14 and CALDB 4.9.8, but otherwise identically to Ref.~\cite{Cappelluti:2017ywp}. This process follows the data reduction of the Chandra Perseus observations in Sec.~\ref{sec:chandra-perseus-reduction}, except that we include the step of deflaring as described in Ref.~\cite{Cappelluti:2017ywp} and keep only observations taken with a focal plane temperature $\leq 153.5$ K and in VFAINT telemetry mode. We do not mask point sources, following the fiducial analysis in Ref.~\cite{Cappelluti:2017ywp}.
Note that we are not able to reproduce the precise set of observations used in Ref.~\cite{Cappelluti:2017ywp}, as no observation list was made available. Although our pre-filtering exposure times agree, when we restrict to those observations with the fits header keyword \texttt{FP\_TEMP} $\leq 153.5$ K and VFAINT telemetry, we are left with only a total exposure time 3.67 Ms and 1.01 Ms in CDFS and CCLS, respectively, as compared with that in Ref.~\cite{Cappelluti:2017ywp} of 5.57 Ms and 3.59 Ms, respectively. With this caveat in mind, we proceed to re-analyze these data following the method of Ref.~\cite{Cappelluti:2017ywp}.

\subsection{Likelihood and Model Components}

\begin{table*}[htb]{
    \ra{1.3}
    \begin{center}
    \begin{tabular}{C{0.15\textwidth}*{3}{C{0.15\textwidth}}}
    \hlinewd{1pt} 
    Element & Au$^*$  & Ti$^*$ & Fe$^*$ \\ 
    Energy [keV] & 2.51  & 4.37 & 6.404 \\ \hlinewd{1pt} 
    2.4-7 keV Fit & $2.5^{+0.4}_{-0.4}$  & $0.8^{+0.3}_{-0.3}$ & $5.8^{+0.6}_{-0.6}$ \\
    3 keV Fit & --  & $0.8^{+0.3}_{-0.3}$ & -- \\
    1 keV Fit & -- & -- & --  \\
    0.5 keV Fit & --  & -- & -- \\
    \hlinewd{1pt} 
    \end{tabular}
    \end{center}}
\caption{\label{tab:Survey_Lines} The same as Tab.~\ref{tab:MOS_Lines}, but for the Chandra CCLS and CDFS data sets. No bounds are put on line intensities.}
\end{table*}

We adopt the model components in the background-modeled analysis of the CDFS and CCLS data sets (see Sec. 4.3 of Ref.~\cite{Cappelluti:2017ywp}). We simultaneously fit to the two data sets a model consisting of an unfolded broken power law (\texttt{bknpower} in \texttt{XSPEC}) to model the instrumental background and an absorbed power law to model the astrophysical background. The broken power law parameters are tied so that the instrumental model is identical in each data set. Absorption is applied using the \texttt{wabs} model, with independent hydrogen column densities $\eta_H$. The hydrogen depths are fixed to $\eta_\mathrm{H,CDFS} = 8.8 \times 10^{19}$ cm$^{-2}$ and $\eta_\mathrm{H,CCLS} = 2.5 \times 10^{20}$ cm$^{-2}$ for CDFS and CCLS, respectively. 
Therefore the continuum is modeled with eight nuisance parameters: the broken power law normalization $I_{\rm bkn}$, its break energy $E_{\rm break}$, its spectral indices below and above the break energy $k_1$ and $k_2$, two power law normalizations $I_{\rm pl}$, and their associated spectral indices $k_{\rm pl}$.
The emission lines, listed in Tab.~\ref{tab:Survey_Lines}, are added to the continuum model, and each have nuisance parameters associated with their rest energy $E$ and intensity $I$ (note that the line widths are fixed in Ref.~\cite{Cappelluti:2017ywp}) and are folded through the detector response. 
All the observed lines were treated as instrumental in Ref.~\cite{Cappelluti:2017ywp}, meaning that their nuisance parameters are tied between the two data sets. Then, the nuisance parameter vector is 
$\bm{\theta} = \{\bm{\theta}_\mathrm{inst.},\{\bm{\theta}_\mathrm{cont.,t}\}_{t=1}^{2},\bm{\theta}_\mathrm{lines}\}$, which are correspondingly defined by:
\es{}{
\bm{\theta}_\mathrm{inst.} &=\{I_\mathrm{bkn}, E_\mathrm{break}, k_1, k_2\} \,, \\
\bm{\theta}_\mathrm{cont.,t} &=\{I_\mathrm{pl,t}, k_\mathrm{pl,t}\} \,, \\
\bm{\theta}_\mathrm{lines.} &= \{\{E_i, I_i\}_{i=1}^{N_\mathrm{inst.}}\} \,.
}
Note that here $t$ is an index over the survey fields.
As usual, the signal model has one model parameter $A$ that controls the flux of the putative 3.5 keV line, modeled with an absorbed zero-width \texttt{gauss} model, and we fix the location of the line to be at its best-fit energy from Ref.~\cite{Cappelluti:2017ywp} of 3.49 keV. Given a set of model parameters, each of the components and the total model prediction per energy bin for a given survey target is constructed in the following way:
\begin{widetext}
\begin{align}
\begin{split}
    \mu_\mathrm{cont.,t}(\bm{\theta}_\mathrm{cont.,t}) &=\mathrm{RSP_t} \star \mathbf{wabs}(\eta_\mathrm{H,t}) \mathbf{powerlaw}(I_\mathrm{pl,t}, k_\mathrm{pl,t}) + \mathbf{bknpower}(k_1, E_\mathrm{break}, k_2, I_\mathrm{bkn,t}) \\
    \mu_\mathrm{lines,t}(A, \bm{\theta}_\mathrm{lines}) &= \mathrm{RSP_t} \star \bigg[ \sum_i^{N_\mathrm{inst.}} \mathbf{gauss}(E_i, 0, I_i) + \mathbf{gauss}(3.49, 0, A)\bigg]\\ 
    \mu_t(A,\bm{\theta}_t) &= \mu_\mathrm{cont.,t}(\bm{\theta}_\mathrm{cont.,t}) + \mu_\mathrm{lines,t}(A,\bm{\theta}_\mathrm{lines}) \,. \\
\end{split}
\end{align}
\end{widetext}
Here we have made explicit that the detector responses differ between the two survey fields, labeled by $t$.

The model is fit simultaneously to the data from the two survey fields, with the line parameters tied between the surveys as indicated. The joint likelihood for the observed number of counts over both data sets $\mathbf{d} = \{\mathbf{d}_\mathrm{CDFS},\mathbf{d}_\mathrm{CCLS}\}$ in the Gaussian limit is given by
\begin{align}
\begin{split}
    \mathcal{L}(\mathbf{d}|A,\bm{\theta}) &= \prod_{t,i}\mathcal{N}\Big(\mathbf{d}_{t,i} | 
    \mu_i = \mathcal{M}_{t,i}(A,\bm{\theta}_\mathrm{t})\Big)
\end{split}
\end{align}
where $i$ runs over the energy bins in the data and $\bm{\theta} = \cup_{t=1}^2 \bm{\theta}_t$ is the full set of nuisance parameters.

\subsection{Data Analysis}
\label{sec:survey-analysis}

\begin{figure*}[htb]  
\hspace{0pt}
\vspace{-0.2in}
\begin{center}
\includegraphics[width=0.99\textwidth]{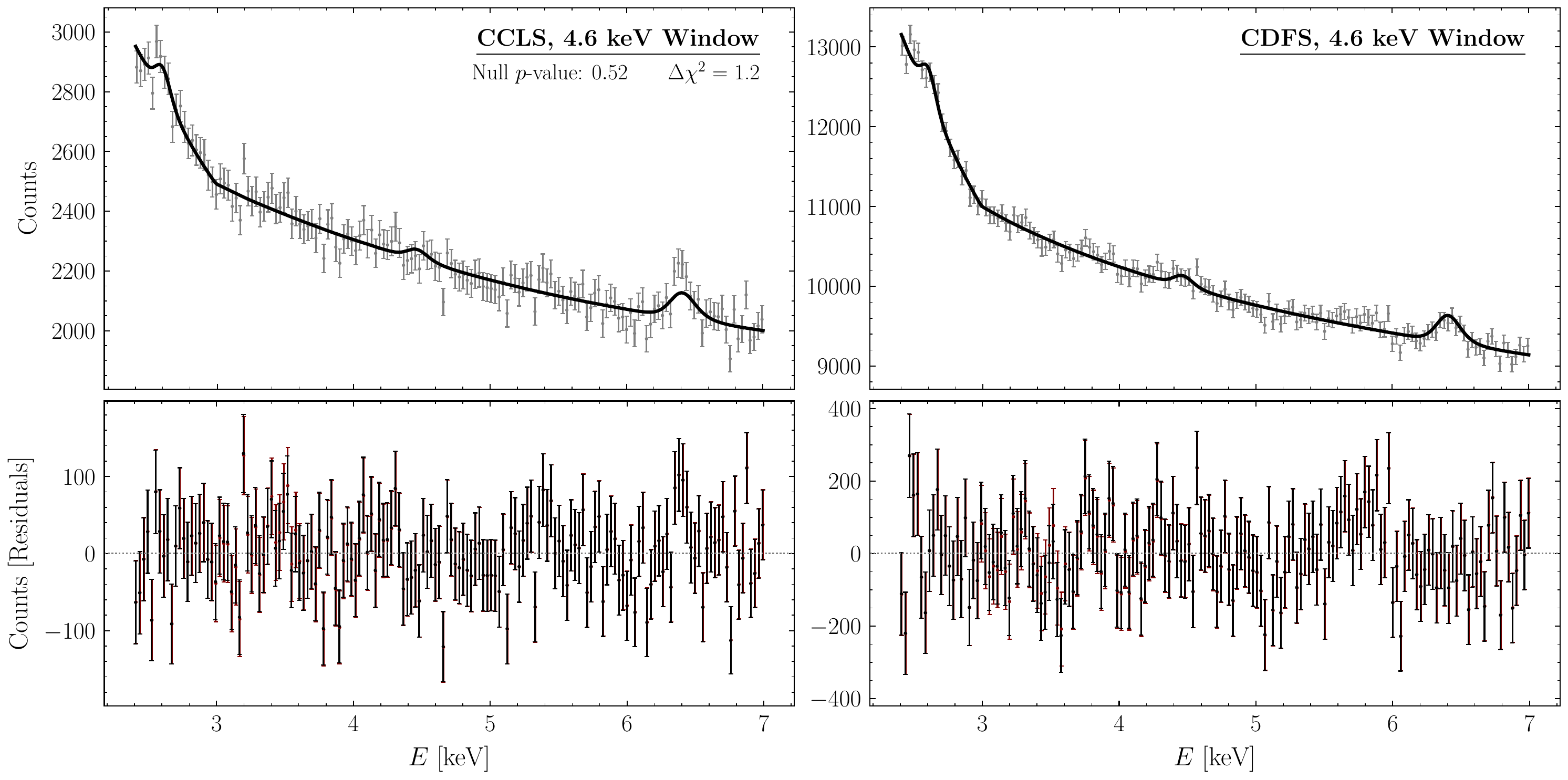}
\caption{\label{fig:survey_fits_wide}
The stacked Chandra CCLS (left) and CDFS (right) data along with the best fit null model in the 2.4--7 keV analysis window of Ref.~\cite{Bulbul:2014sua}. The bottom panels illustrate the residuals after subtracting the best-fit null and signal models.
}
\end{center}
\end{figure*}

\begin{figure*}[htb]  
\hspace{0pt}
\vspace{-0.2in}
\begin{center}
\includegraphics[width=0.99\textwidth]{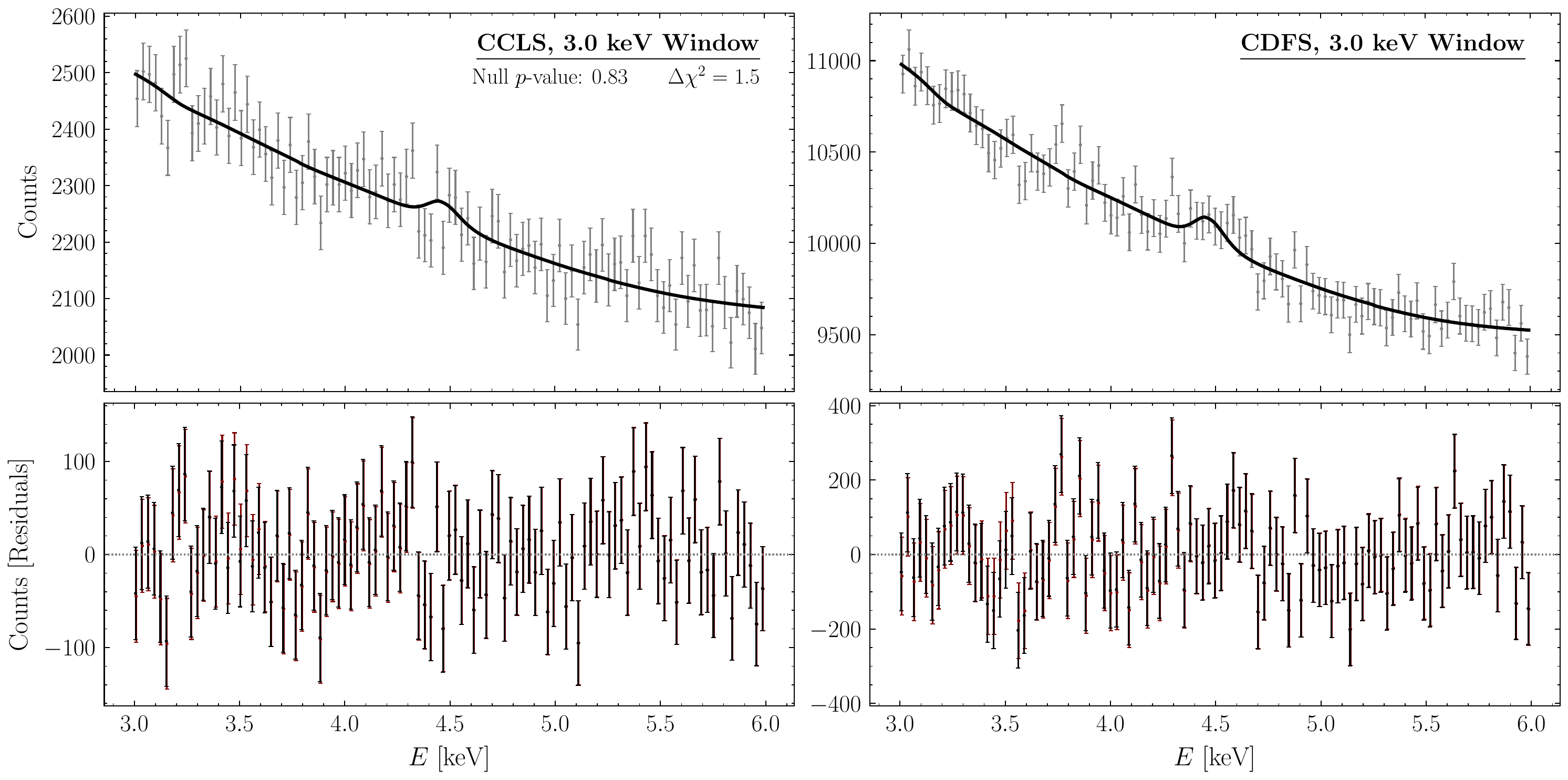}
\caption{\label{fig:survey_fits_E3}
As in Fig.~\ref{fig:survey_fits_wide}, but using an analysis window of 3-6 keV.}
\end{center}
\end{figure*}

\begin{figure*}[htb]  
\hspace{0pt}
\vspace{-0.2in}
\begin{center}
\includegraphics[width=0.99\textwidth]{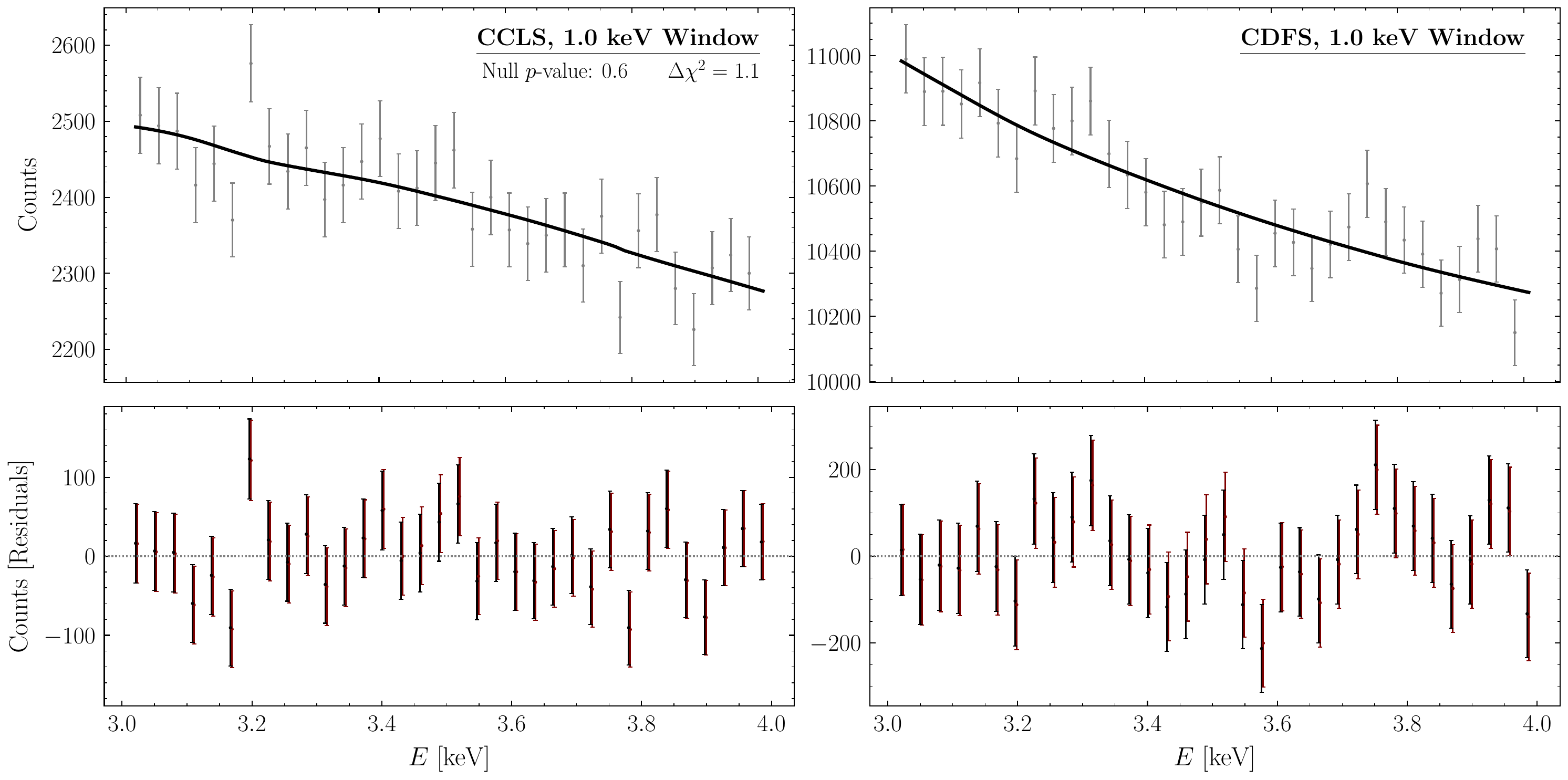}
\caption{\label{fig:survey_fits_E1}
As in Fig.~\ref{fig:survey_fits_wide}, but using an analysis window of 2.99-3.99 keV.}
\end{center}
\end{figure*}

\begin{figure*}[htb]  
\hspace{0pt}
\vspace{-0.2in}
\begin{center}
\includegraphics[width=0.99\textwidth]{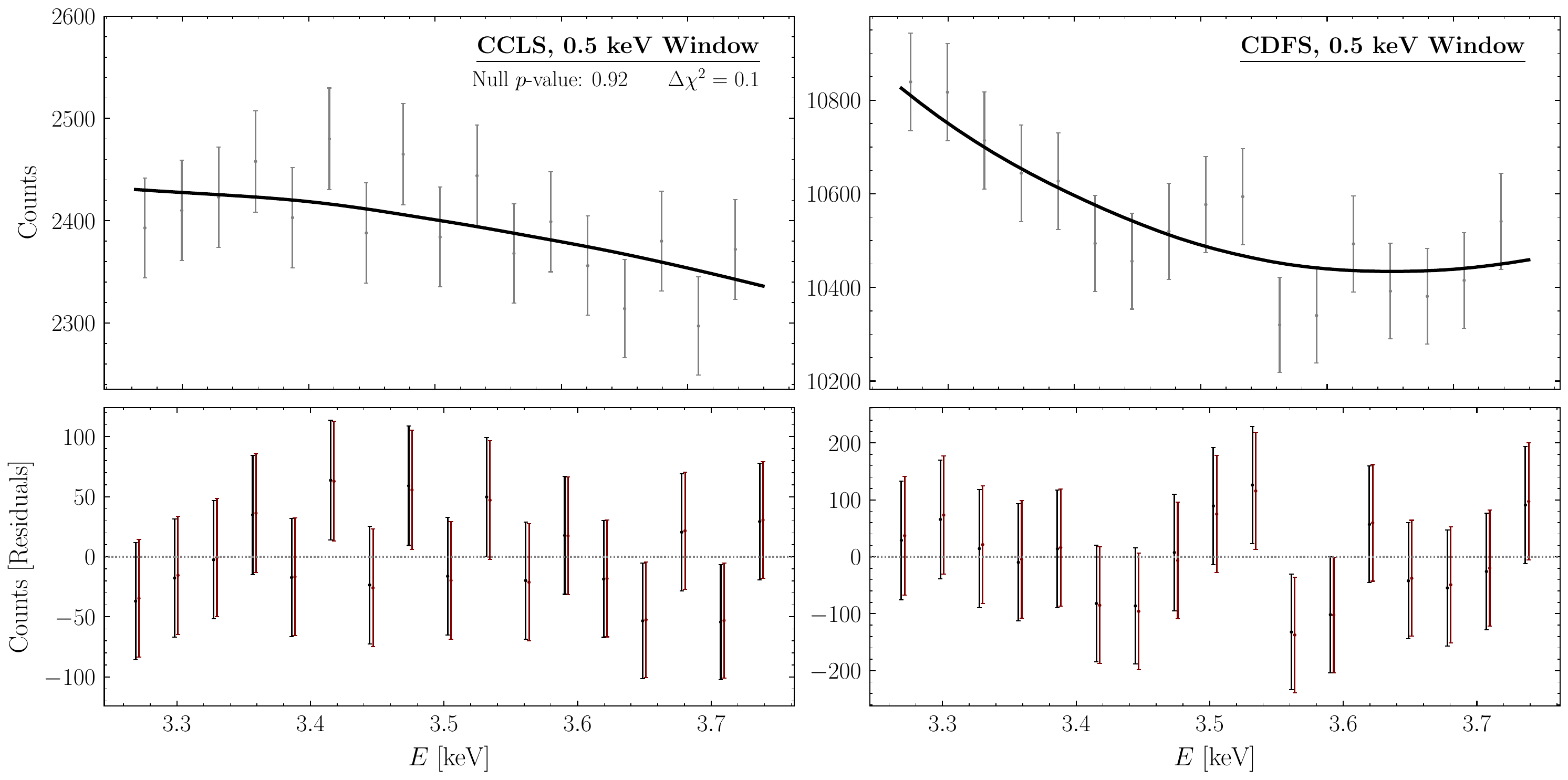}
\caption{\label{fig:survey_fits_E500}
As in Fig.~\ref{fig:survey_fits_wide}, but using an analysis window of 3.24-3.74 keV.}
\end{center}
\end{figure*}

\begin{figure}[htb]  
\hspace{0pt}
\vspace{-0.2in}
\begin{center}
\includegraphics[width=0.49\textwidth]{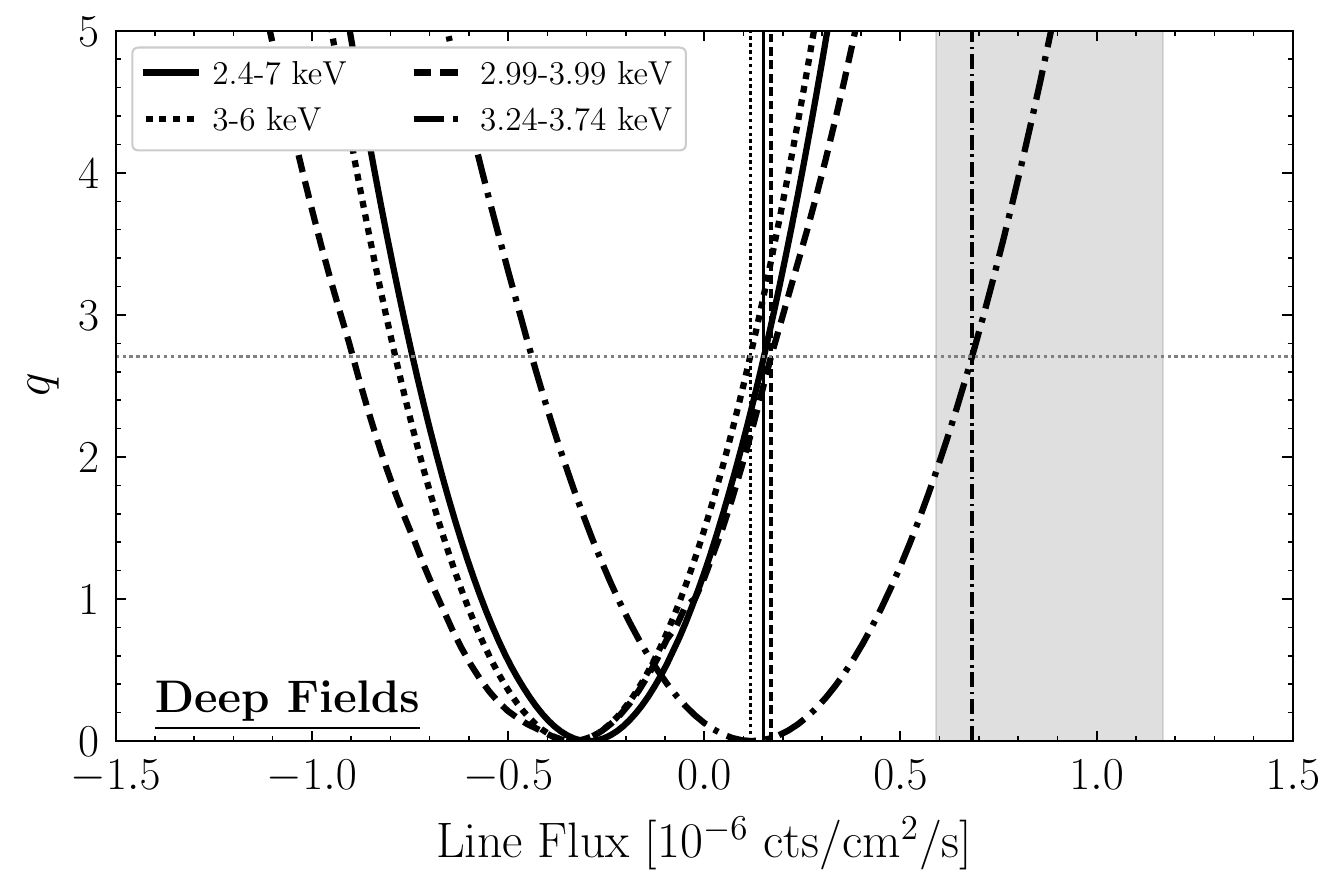}
\caption{\label{fig:survey_ll}
The profiled likelihood for the Chandra Deep Field analysis in each of the four analysis windows: 2.5\textemdash 6 keV (solid), 3 keV (dashed), 1 keV (dash-dotted), and 0.5 keV (dotted). The 95\% upper limits from each fit are shown as horizontal red lines with corresponding styles. The 1$\sigma$ best fit region for the 3.5 keV line flux in Ref.~\cite{Bulbul:2014sua} is in shaded gray.
}
\end{center}
\end{figure}

The original analysis of these data~\cite{Cappelluti:2017ywp} was performed over the 2.4-7 keV energy range. 
That analysis found a best-fit line flux of $0.89_{-0.3}^{+0.3} \times 10^{-6}$ cts/cm$^2$/s, with the signal model preferred over the null hypothesis by $t = 10.2$, corresponding to slightly over 3$\sigma$ evidence in favor of the signal model. As summarized in Tab.~\ref{tab:Results}, we recover a best-fit flux of $-0.32_{-0.27}^{+0.27} \times 10^{-6}$ cts/cm$^2$/s, with $t = 0$.  In Fig.~\ref{fig:survey_fits_wide} we show the best-fit models for both CDFS and CCLS.  The $p$-value associated with the null-hypothesis fit is $p \approx 0.52$, showing no evidence for mismodeling.  Fig.~\ref{fig:survey_ll} illustrates the profile likelihood $q$ as a function of 3.5 keV line flux for the joint fit; our 95\% one-sided upper limit rules out the entire 1$\sigma$ parameter space recovered in~\cite{Cappelluti:2017ywp} for their 3.5 keV line flux. 

Figs.~\ref{fig:survey_fits_E3},~\ref{fig:survey_fits_E1}, and~\ref{fig:survey_fits_E500} show the fits to the data in the 3 keV, 1 keV, and 0.5 keV energy windows. As indicated in the figures and in Tab.~\ref{tab:Results}, all of these fits return acceptable null-hypothesis $p$-values. The corresponding profile likelihoods are illustrated in Fig.~\ref{fig:survey_ll} and all give best-fit fluxes consistent with zero at less than 1$\sigma$ significance.  The fact that there is no evidence for mismodeling in these analyses, which all give results in strong tension with the positive evidence claimed in~\cite{Cappelluti:2017ywp}, suggest that there is no robust evidence for a 3.5 keV line in the Chandra Deep Field survey data sets. On the other hand, as we discuss further in App.~\ref{app:DeepFieldBkgSub}, we are able to reproduce evidence for a 3.5 keV line in the background-subtracted analyses, but: (i) the null-hypothesis $p$-values show clear signs of mismodeling in these cases, and (ii) there is evidence for a 3.5 keV deficit in the background data. 

\section{Discussion}
\label{sec:discussion}

We re-analyze the data used to claim evidence for a 3.5 keV line in the foundational papers on the excess~\cite{Bulbul:2014sua,Boyarsky:2014jta}, which collectively considered bright galaxy clusters and M31, in addition to the later work Ref.~\cite{Cappelluti:2017ywp} that claimed evidence for a 3.5 keV line in Chandra deep-field surveys.  Ref.~\cite{Bulbul:2014sua}, most notably, claims $\sim$4$\sigma$ evidence for a UXL near 3.5 keV from XMM-Newton MOS data taken towards Perseus; in our re-analysis of these same data, following as closely as possible the analysis procedure in~\cite{Bulbul:2014sua}, we find less than a 1$\sigma$ preference for the signal model.  The lack of evidence is robust to going to a narrower analysis energy window, which helps mitigate the possible effects of mismodeling the background emission. Similarly, in a stacked analysis of XMM-Newton MOS data from the clusters Centaurus, Coma, and Ophiuchus, Ref.~\cite{Bulbul:2014sua} finds approximately 4$\sigma$ evidence for a 3.5 keV line. In our joint re-analysis of these data we find that this evidence is driven by the Centaurus cluster; the evidence disappears when going to a narrower analysis energy window. In the narrower windows we are also able to completely exclude the best-fit line flux found in~\cite{Bulbul:2014sua} at more than 95\% confidence.  We are also unable to reproduce the $\sim$3$\sigma$ evidence for a line found by~\cite{Bulbul:2014sua} in an analysis of Chandra data towards Perseus; we find no evidence for a line with their same data and analysis procedure, regardless of the energy window size.

Ref.~\cite{Boyarsky:2014jta} claimed $\sim$3.5$\sigma$ evidence for a line in XMM-Newton MOS M31 data. In our re-analysis of these data, with their same models, we recover an inconsistent and lower best-fit line flux. In all of our analyses we are able to rule out at 95\% confidence the entire 1$\sigma$ best-fit intensities for the UXL found in~\cite{Boyarsky:2014jta}; in our narrowest window and most conservative analyses, the evidence for the 3.5 keV UXL is around 1$\sigma$.
Note that our result here is consistent with that found in~\cite{Jeltema:2014qfa,Jeltema:2014mla}, who also noted a lack of evidence for the 3.5 keV line in M31 when analyzing in a narrower energy window than in~\cite{Boyarsky:2014jta}.  The authors of~\cite{Boyarsky:2014jta} refuted the narrow-window analyses in~\cite{Boyarsky:2014paa} by noting that their power-law background model appears to describe the data to the level of statistical noise over the wide energy range.  On the other hand, as we summarize in Tab.~\ref{tab:Results}, this is not fully true, with the low $p$-value of the null hypothesis fit indicating some level of mismodeling. Moreover, as we show in Sec.~\ref{sec:methods}, the reduced chi-square is not an optimal diagnostic for mismodeling when looking for narrow spectral signatures. In contrast, narrowing the analysis energy window leads to more robust results at the expense of only slightly reduced sensitivity, depending on the size of the reduced energy window.  
As we illustrate in our toy examples, analyzing $X$-ray data over wide energy ranges for narrow spectral features is dangerous because mismodeling of the continuum components can drive artificial evidence for a signal; given how the evidence in favor of the line evolves with shrinking energy-window size for M31, we conclude that the evidence found in the wide-energy-range analysis is likely an artifact of mismodeling.  We are not able to resolve the discrepancy between our lower recovered line flux and detection significance relative to the results claimed in~\cite{Boyarsky:2014jta} for the widest-window analysis.

In our re-analysis of the Chandra deep-field data from~\cite{Cappelluti:2017ywp} we find no evidence for a 3.5 keV UXL. We are only able to find evidence for a 3.5 keV line in these data by performing background subtraction instead of background modeling, but as we show in App.~\ref{app:DeepFieldBkgSub} the background subtraction procedure leads to poor fits to the data and also the background data itself has a significant deficit at 3.5 keV.

One outstanding question left by this work is why we are unable to reproduce the central claims of~\cite{Bulbul:2014sua,Boyarsky:2014jta} when following their claimed analysis procedures. It is possible that these works did not manage to reach the global likelihood maximum, given their use of local optimizers, though we are not able to verify if this indeed the case because absolute $\chi^2_\nu$ comparisons between our results is not meaningful given the stochastic nature of the data reduction procedure, as described further in App.~\ref{app:Randomization}.

Our work strongly suggests that there is no robust evidence for a 3.5 keV line. Note that this is a different and stronger conclusion to that reached in, {\it e.g.},~\cite{Dessert:2018qih}, who claimed that a DM explanation of the line is inconsistent with null results for the line in XMM-Newton blank sky data; here, in contrast, we claim that there never was robust evidence for a line near 3.5 keV in the first place.

There are a number of important implications for our work going into the future, as the next-generation of $X$-ray telescopes, such as eROSITA~\cite{2012arXiv1209.3114M,Dekker:2021bos}, XRISM~\cite{XRISMScienceTeam:2020rvx,Dessert:2023vyl}, {\it Athena}~\cite{2015JPhCS.610a2008B,Neronov:2015kca,Piro:2021oaa}, and LEM~\cite{2022arXiv221109827K,Krnjaic:2023odw}, aim to further improve the sensitivity to decaying DM in the $X$-ray band.  Foremost, wide-energy range parametric frequentist analyses, using physics-based and phenomenological continuum components in addition to lists of possible astrophysical and instrumental lines, towards bright clusters and nearby galaxies, are sub-optimal methods for searching for evidence of DM lines, for a number of reasons. For one, as we have shown, these parametric modeling procedures over large energy ranges are strongly susceptible to mismodeling, which can bias the evidence in favor or against a UXL even if $\chi^2_\nu$ otherwise looks acceptable.  Narrowing the energy range of the parametric analysis to enclose the instrument-broadened line and nearby side-bands is one approach, discussed here, for helping to mitigate mismodeling. The sliding-window analysis approach is commonly applied in other contexts for narrow beyond-the-Standard-Model searches (see, {\it e.g.},~\cite{CMS:2015kne,ATLAS:2017eqx,Fermi-LAT:2015kyq,Foster:2022nva}).  Non-parametric modeling, such as with Gaussian Process (GP) modeling as in~\cite{Frate:2017mai,Foster:2021ngm,ATLAS:2022tnm}, is perhaps an even more robust analysis strategy, as GP models have more freedom to describe features in the data than parametric models, but the correlation length of the GP models can still be restricted to avoid over-degeneracy between the background and signal model.

For DM decay searches it has also been shown that clusters and nearby galaxies are sub-optimal targets for most of the current and planned $X$-ray telescopes (see, {\it e.g.},~\cite{Dessert:2018qih}); instead, Milky Way blank sky observations near the Galactic Center provide both enhanced signal strengths and reduced background rates. Going into the future as analyses collect even larger exposures and statistical errors shrink, it will be even more important to concentrate DM analyses on otherwise empty, pristine regions of the Milky Way, with large expected signal-to-noise ratios, rather than the complicated environments found in bright clusters and nearby galaxies.

\section*{Acknowledgements}
We thank Nicholas Rodd for collaboration at the early stages of this work. We thank Sunayana Bhargava, Nico Cappelluti, Dan Hooper, Tesla Jeltema, Stefano Profumo, Nicholas Rodd, and Tracy Slatyer for helpful discussions and comments on the manuscript. The authors were supported in part by the DOE Early Career Grant DESC0019225 and by computational resources at the Lawrencium computational cluster provided by the IT Division at the Lawrence Berkeley National Laboratory, supported by the Director, Office of Science, and Office of Basic Energy Sciences, of the U.S. Department of Energy under Contract No. DE-AC02-05CH11231. J.F. was supported by a Pappalardo fellowship. C.D. was supported by the National Science Foundation under Grants No. 2210498 and 2210551. C.D. performed this work in part at the Aspen Center for Physics, which is supported by National Science Foundation grant PHY-2210452.

\appendix

\section{Randomization in the XMM-Newton Data Reduction}
\label{app:Randomization}

We explore the effect of the randomization intrinsic to the XMM-Newton data processing by repeating identical data reductions to produce 10 otherwise identical Perseus MOS data sets. We apply our complete global optimization and component-dropping procedure to each of the 10 data sets, studying the distribution of $\chi_\nu^2$ and accepted model components, with results presented in Tab.~\ref{tab:MC_FitDrop_E3}. Despite the large scatter observed in the reduced $\chi^2$, ranging between $585.4/564$ at the smallest to $624.3/564$ at the largest, a high degree of consistency is observed in the accepted model components. In particular, all of the 10 analyzed data sets result in a continuum described by one \texttt{nlapec} model and one unfolded power law. Amongst the line components, the Ar line at 3.685 is most sensitive to randomization in the data reduction as its inclusion typically results in $\Delta\chi^2 \approx 3$, precisely at the threshold for inclusion. In eight of the 10 data sets, this Ar line is excluded, but it is included in the remaining two. With the exception of this near threshold line, the accepted line lists for all 10 data sets are identical. 

\begin{table*}[htb]{
    \ra{1.3}
    \begin{center}
    \begin{tabular}{c*{11}{C{0.07\textwidth}}}
    \hlinewd{1pt} 
    $\chi^2_\nu$ & APEC & Folded & Unfolded & \textbf{Ar} (3.124) & \textbf{Ar} (3.32) & \textbf{K} (3.511) & \textbf{Ar} (3.685) & \textbf{K} (3.705) & \textbf{Ca} (3.902) & \textbf{Ca} (4.107)  & \textbf{Cr} (5.682)\\ \hlinewd{1.5pt}
    614.3/564 & 1 & 0 & 1 & \cmark & \cmark & \cmark & \xmark & \cmark & \cmark & \cmark & \cmark \\ \hline
    605.3/561 & 1 & 0 & 1 & \cmark & \cmark & \cmark & \cmark & \cmark & \cmark & \cmark & \cmark \\ \hline
    585.4/564 & 1 & 0 & 1 & \cmark & \cmark & \cmark & \xmark & \cmark & \cmark & \cmark & \cmark \\ \hline
    605.4/564 & 1 & 0 & 1 & \cmark & \cmark & \cmark & \xmark & \cmark & \cmark & \cmark & \cmark \\ \hline
    624.3/564 & 1 & 0 & 1 & \cmark & \cmark & \cmark & \xmark & \cmark & \cmark & \cmark & \cmark \\ \hline
    600.7/564 & 1 & 0 & 1 & \cmark & \cmark & \cmark & \xmark & \cmark & \cmark & \cmark & \cmark \\ \hline
    619.4/564 & 1 & 0 & 1 & \cmark & \cmark & \cmark & \xmark & \cmark & \cmark & \cmark & \cmark \\ \hline
    589.7/564 & 1 & 0 & 1 & \cmark & \cmark & \cmark & \xmark & \cmark & \cmark & \cmark & \cmark \\ \hline
    598.5/561 & 1 & 0 & 1 & \cmark & \cmark & \cmark & \cmark & \cmark & \cmark & \cmark & \cmark \\ \hline
    594.3/564& 1 & 0 & 1 & \cmark & \cmark & \cmark & \xmark & \cmark & \cmark & \cmark & \cmark \\ \hlinewd{1.5pt}
     -- & 1 & 0 & 1 & \cmark & \cmark & \cmark & \xmark & \cmark & \cmark & \cmark & \cmark \\ \hlinewd{1.5pt}
    \end{tabular}\end{center}}
\caption{A summary of the accepted components of the null model for 10 different, identical up to randomization effects, reductions of the XMM-Newton MOS Perseus data. For each of the 10 data sets, we summarize: the $\chi^2_\nu$, the number of \texttt{nlapec} continua, the number of folded power laws, the number of unfolded power laws, and the lines which are included. With the exception of the 3.685 keV Ar line, which is near the threshold for inclusion/exclusion, the accepted model components are identical across the 10 data sets although the $\chi^2_\nu$ may differ greatly. In the last row, we summarize the null model used in subsequent XMM-Newton Perseus analyses. The 3.685 keV Ar line has been excluded as it was not accepted in most of the data sets. \label{tab:MC_FitDrop_E3}}
\end{table*}

\begin{figure*}[htb]  
\hspace{0pt}
\vspace{-0.2in}
\begin{center}
\includegraphics[width=0.95\textwidth]{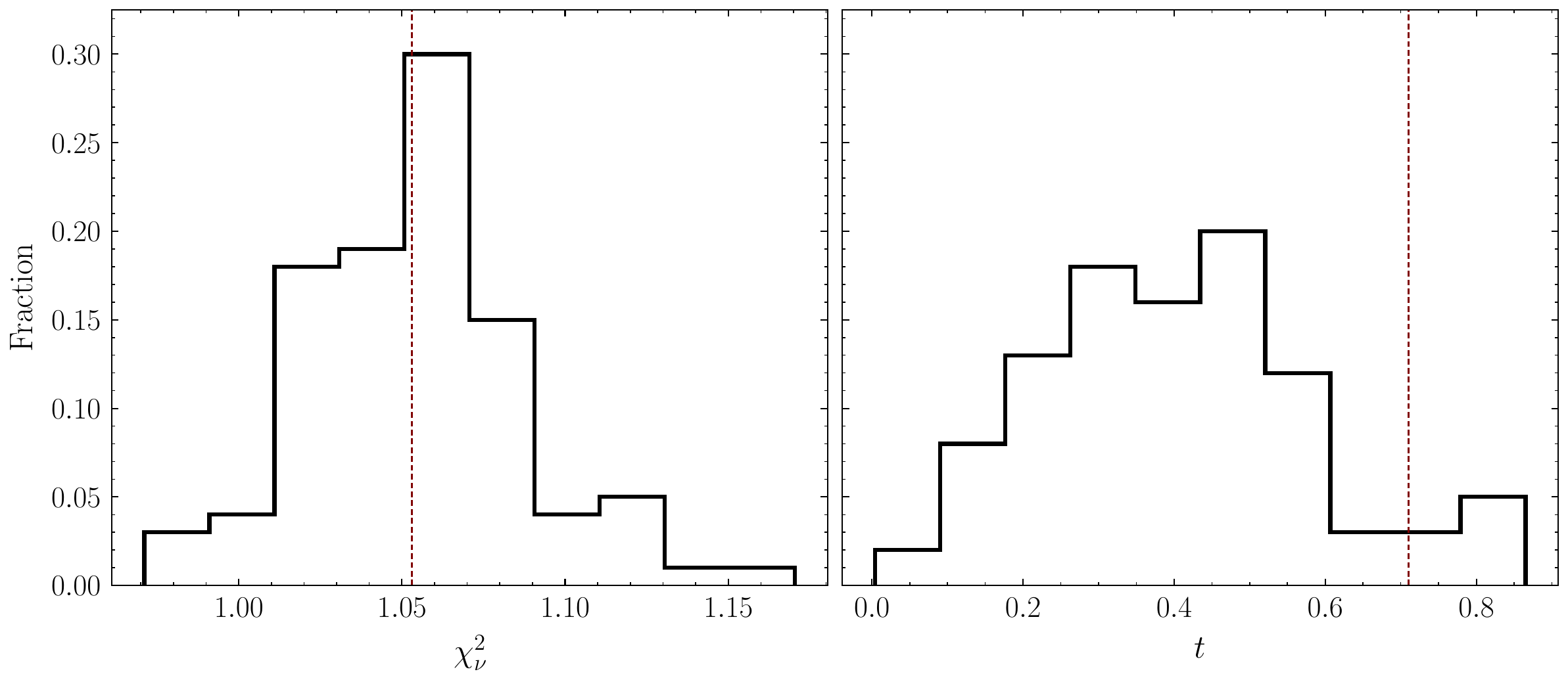}
\caption{\label{fig:RandomizationError}
(\textit{Left}) The distribution of the $\chi^2_\nu$ for our null model over 100 XMM-Newton MOS Perseus data sets which differ only by randomization effects in otherwise identical data reductions. The median $\chi^2_\nu$ associated with our selected reduction realization is indicated by a vertical red dashed line. (\textit{Right}) The distribution of $t$, the profiled likelihood ratio evaluated at the best-fit signal flux over the 100 realizations. The $t$ associated with our selected reduction is again indicated by a red dashed line.}
\end{center}
\end{figure*}

From these 10 fully analyzed data sets, we construct our null model from components which are robustly included across the 10 samples; \textit{i.e.}, we do not include the Ar 3.685 line. The model summary is given in the last row of Tab.~\ref{tab:MC_FitDrop_E3}. With this as our null model, we generate an additional 100 data sets from an identical reduction procedure and analyze them under the signal and null hypotheses. The distribution of the reduced $\chi^2$ obtained under the null model described in Tab.~\ref{tab:MC_FitDrop_E3} is depicted in the left panel of Fig.~\ref{fig:RandomizationError}. In the right panel of Fig.~\ref{fig:RandomizationError}, we present the distribution of the discovery test statistic $t$; \textit{i.e.}, the maximum improvement in the $\chi^2$ associated with the inclusion of a 3.57 keV line with a freely estimated flux. 

We find that the randomization in the data reduction can have relatively dramatic consequences for attempts to assess a goodness-of-fit. Across the 100 data reductions, we find a minimum $\chi_\nu^2$ of $535/564$, corresponding to a $p = 0.80$, and a maximum $\chi^2_\nu$ of $660/564$ corresponding to $p = 0.003$. This significant scatter suggests that it is challenging to directly compare our goodness-of-fits with those obtained in other works since we are unable to access an identical data reduction realization. On the other hand, we find that the distribution of $t$ is more compact, ranging between $0$ and $0.9$, with the significance of possible line detection or non-detection relatively robust across reductions.

In light of these findings, we present in the main text the analysis of the Perseus data reduction which results in the median $\chi^2_\nu$ over the distribution shown in Fig.~\ref{fig:RandomizationError}. However, given that line evidence appears to be stable across reductions, we do not repeat this procedure for other observational targets, instead performing the full model specification and signal analysis with a single data reduction rather than an ensemble. Finally, there is randomization in the Chandra data reductions, but the energy randomization seed is fixed by default and a deterministic algorithm is used for sky pixel randomization, so that in practice the data reduction procedure is deterministic.

\section{Milky Way Survey Fields with Background Subtraction}
\label{app:DeepFieldBkgSub}

In this Appendix we reanalyze the Chandra data of the MW survey fields CDFS and CCLS after subtracting the instrumental background as measured by observations when the detector was not exposed to the sky. As we are performing direct background subtraction, we do not include the unfolded broken power law; otherwise, the data preparation, model components, and energy ranges used in the analysis are identical to those in Sec.~\ref{sec:survey}.

\begin{figure*}[htb]  
\hspace{0pt}
\vspace{-0.2in}
\begin{center}
\includegraphics[width=0.99\textwidth]{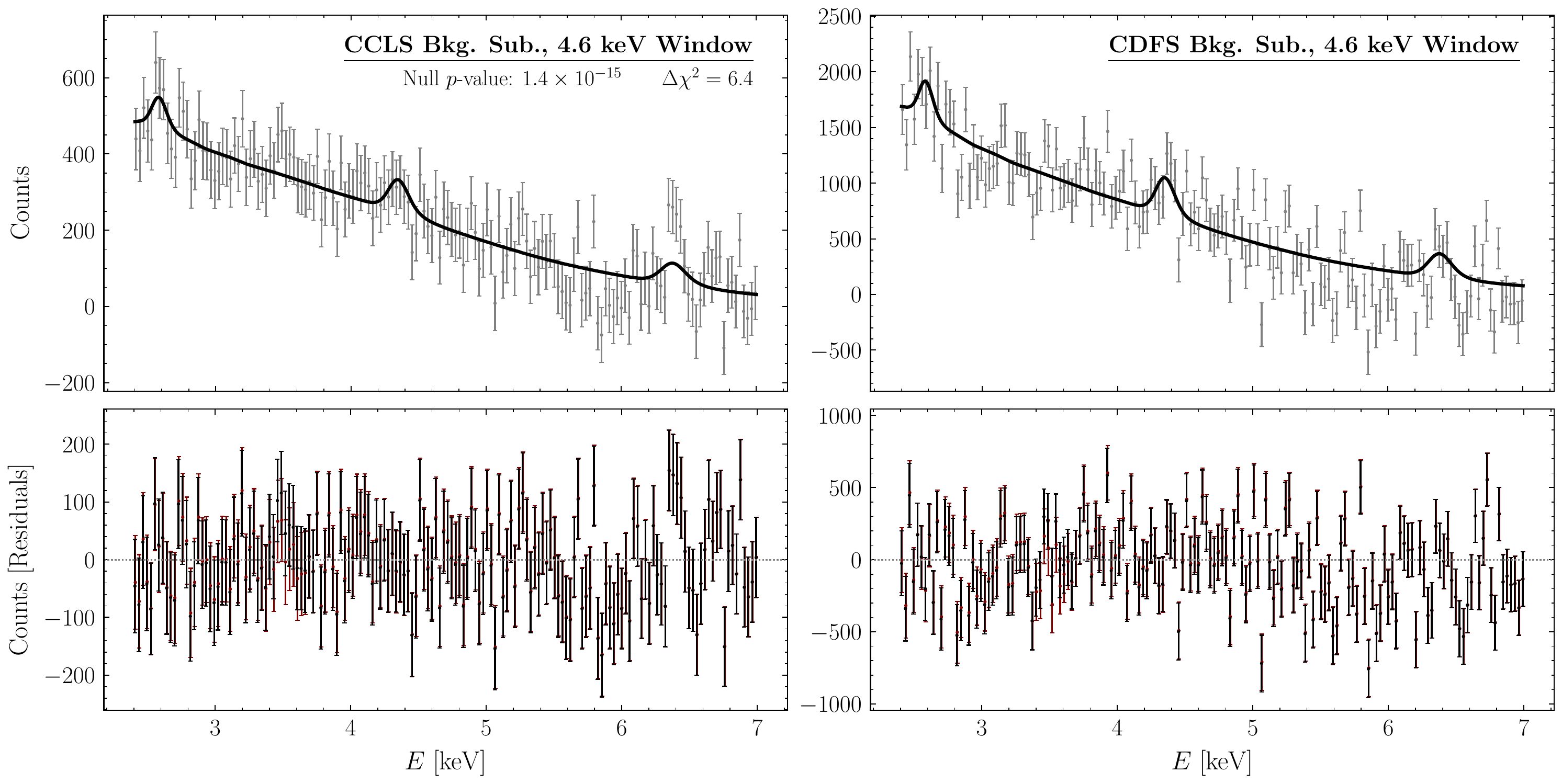}
\includegraphics[width=0.99\textwidth]{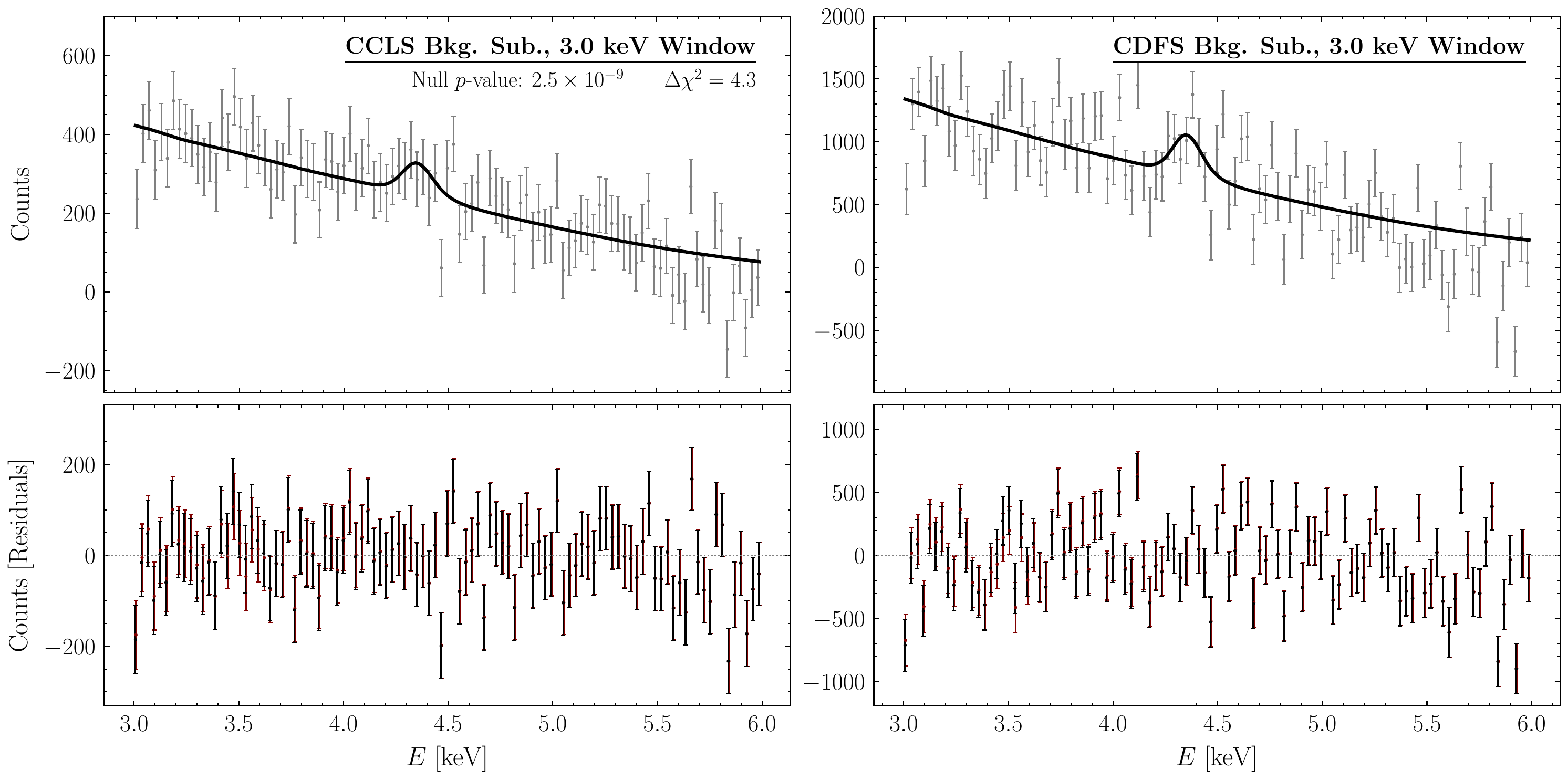}
\caption{\label{fig:survey_fits_bkgsub_wide}
As in Fig.~\ref{fig:survey_fits_wide} (\textit{above}) and Fig.~\ref{fig:survey_fits_E3} (\textit{below}), but using background subtraction.}
\end{center}
\end{figure*}

\begin{figure*}[htb]  
\hspace{0pt}
\vspace{-0.2in}
\begin{center}
\includegraphics[width=0.99\textwidth]{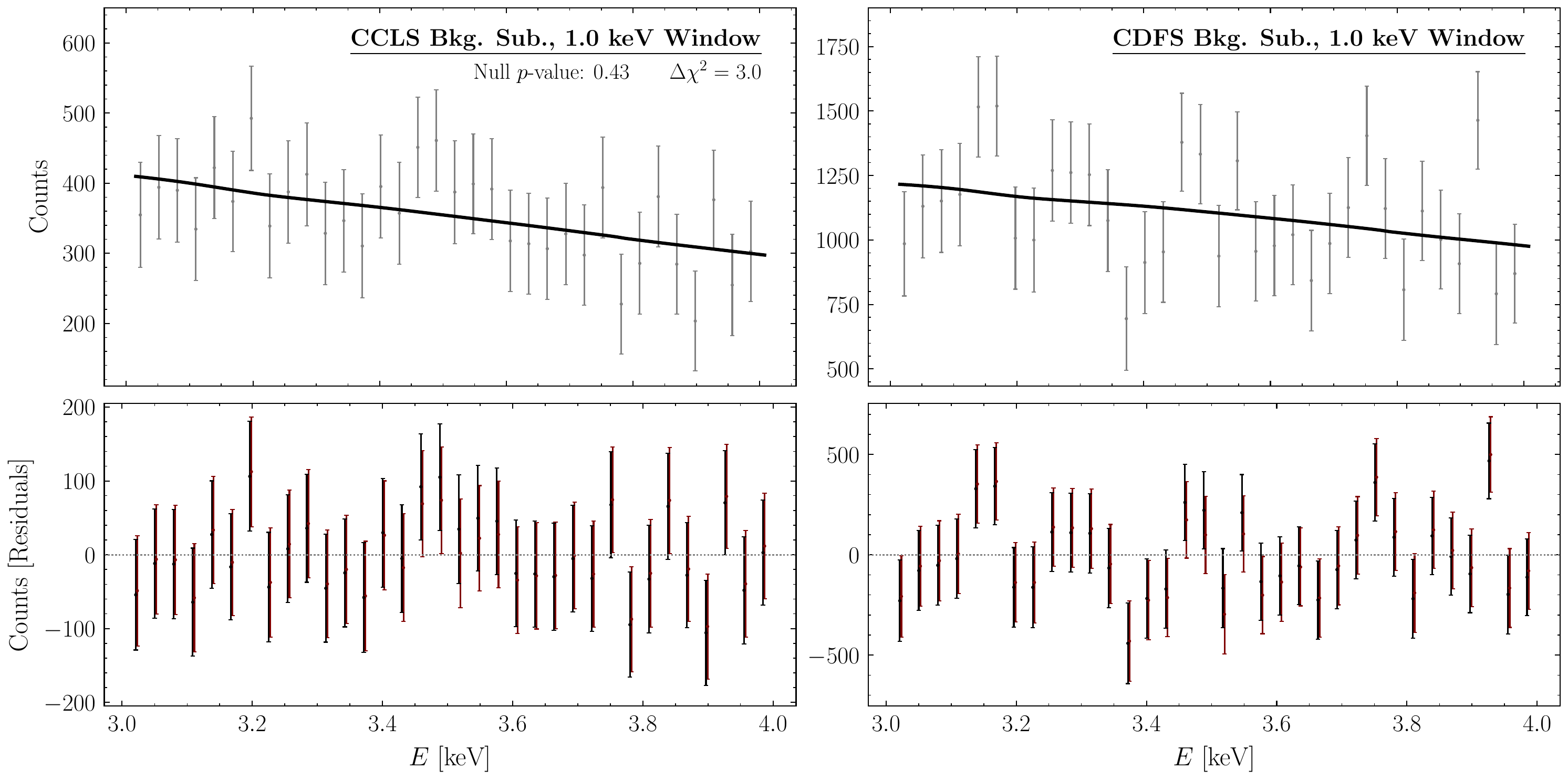}
\includegraphics[width=0.99\textwidth]{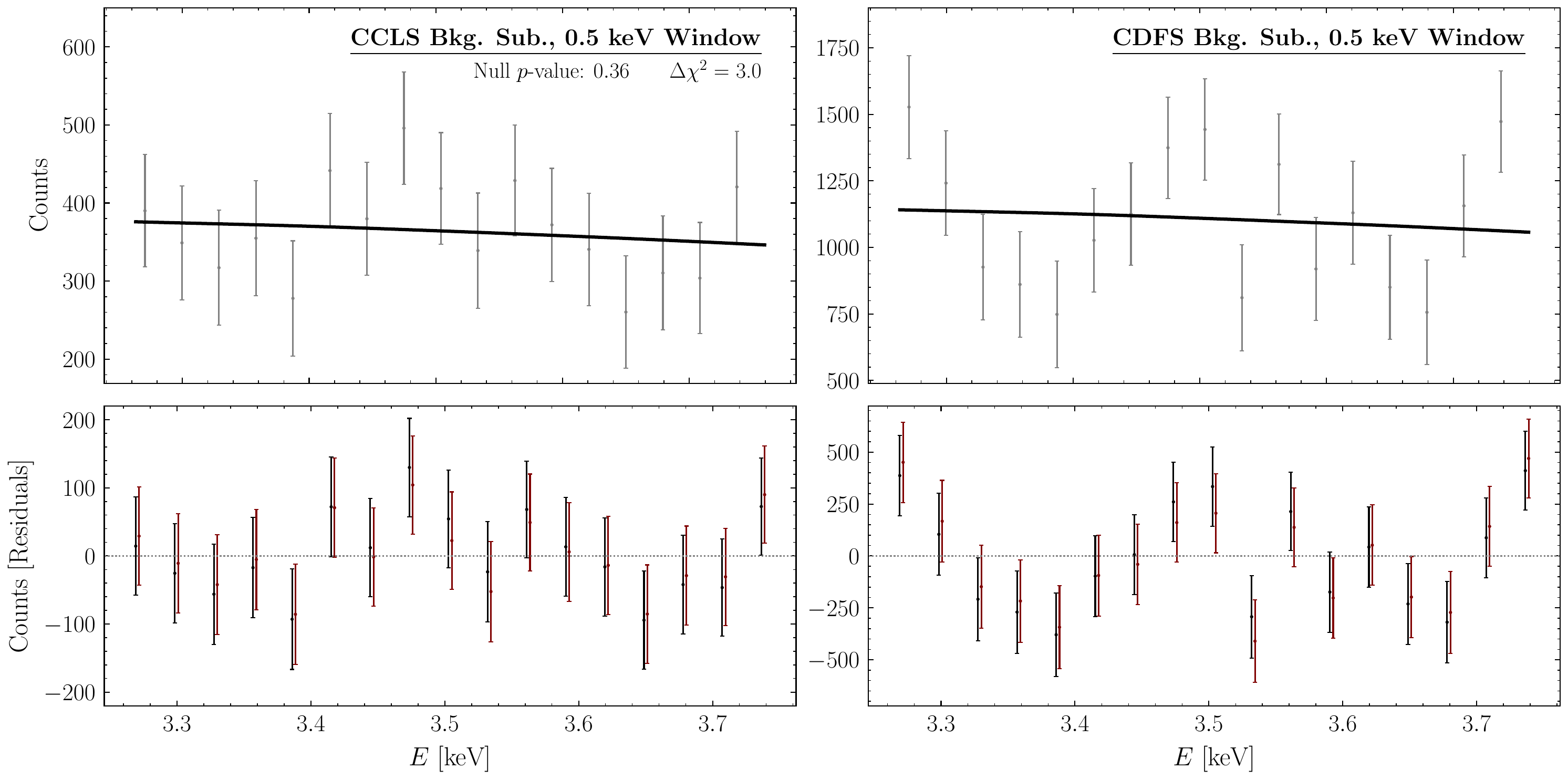}
\caption{\label{fig:survey_fits_bkgsub_E1}
As in Fig.~\ref{fig:survey_fits_E1} (\textit{above}) and Fig.~\ref{fig:survey_fits_E500} (\textit{below}), but using background subtraction.}
\end{center}
\end{figure*}

\begin{figure*}[htb]  
\hspace{0pt}
\vspace{-0.2in}
\begin{center}
\includegraphics[width=0.49\textwidth]{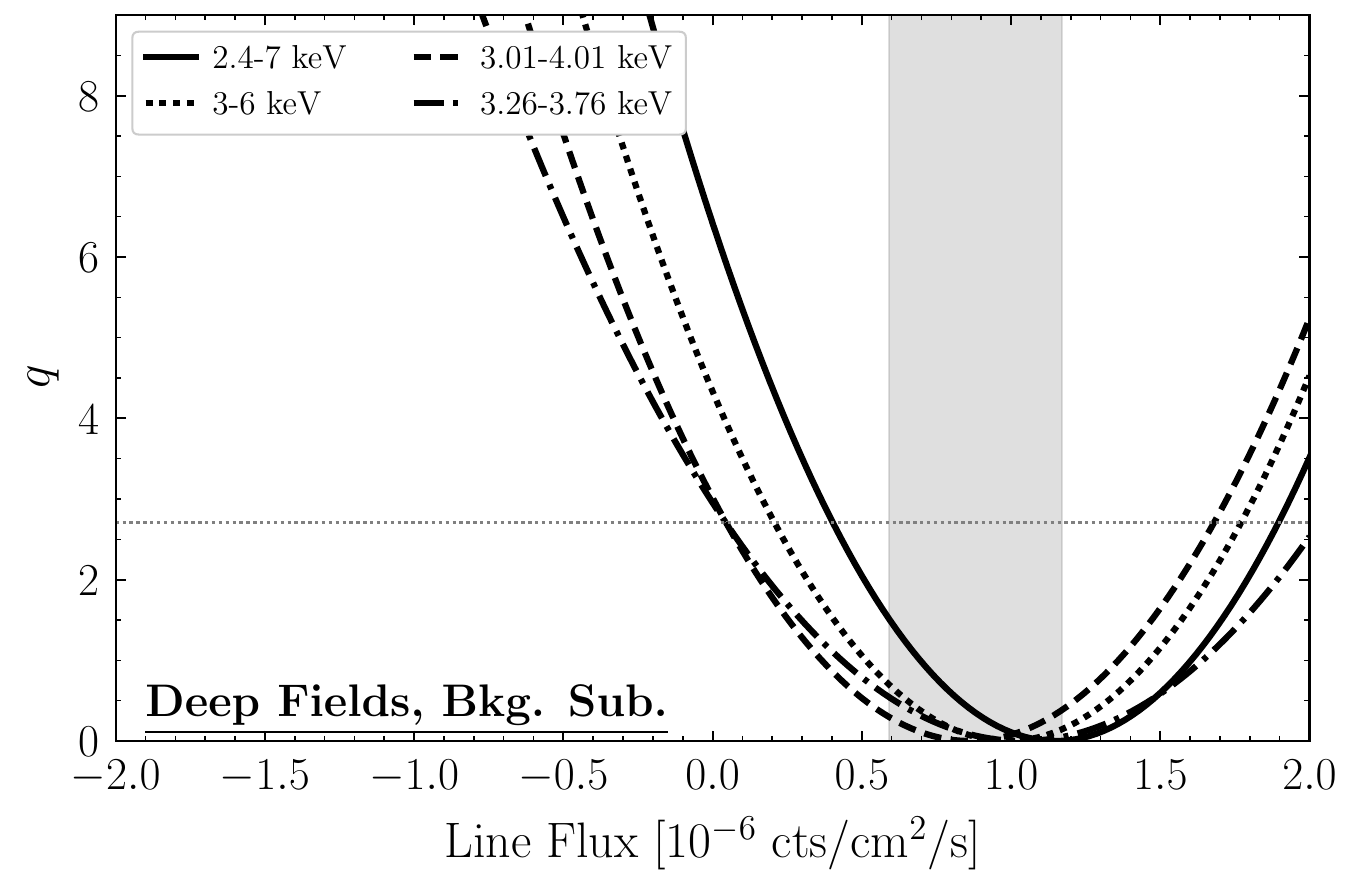}\includegraphics[width=0.49\textwidth]{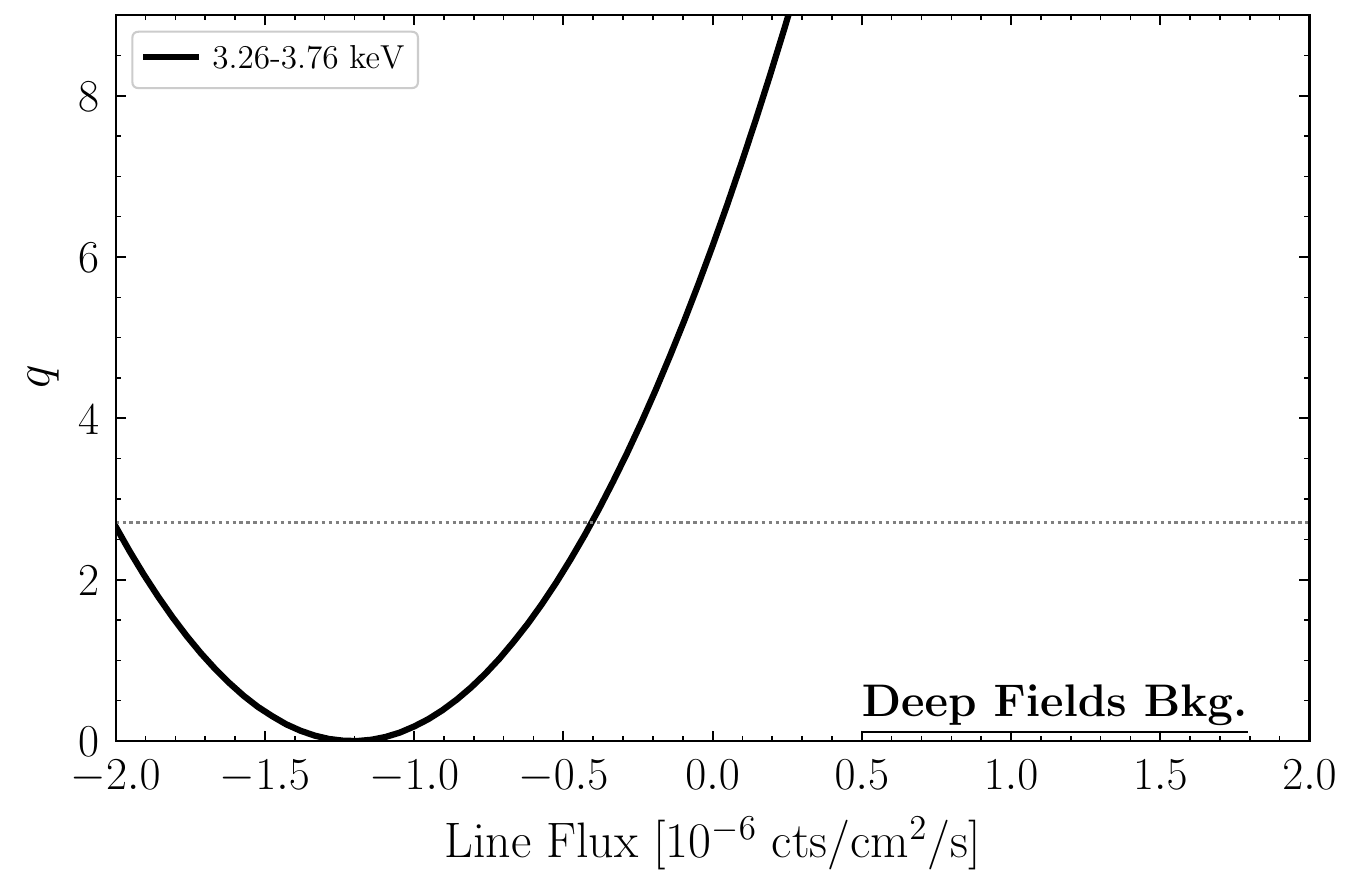}

\caption{\label{fig:survey_ll_bkgsub}
(\textit{Left}) As in Fig.~\ref{fig:survey_ll}, but for the analysis using background subtraction. (\textit{Right}) 
As in Fig.~\ref{fig:survey_ll}, but for the 500 eV window analysis of the background data.}
\end{center}
\end{figure*}

The original analyses of these data~\cite{Cappelluti:2017ywp} found a best-fit line flux  $0.89_{-0.3}^{+0.3} \times 10^{-6}$ cts/cm$^2$/s using a 2.4-7 keV energy range at a rest energy of 3.51 keV. We recover a consistent best-fit flux of $1.1_{-0.4}^{+0.4} \times 10^{-6}$ cts/cm$^2$/s though at a reduced significance $t \sim 6.4$.  In Figs.~\ref{fig:survey_fits_bkgsub_wide} and \ref{fig:survey_fits_bkgsub_E1}, we present the best-fit models for both CDFS and CCLS over our different analysis energy windows. The profiled likelihoods $q$ as functions of the $3.51$ keV line flux joined over the data sets are illustrated in the left panel of  Fig.~\ref{fig:survey_ll_bkgsub}. 

On the other hand, we note that both our fit and the fit in~\cite{Cappelluti:2017ywp} show serious signs of mismodeling. As quoted in Tab.~\ref{tab:BkgSubResults} and Figs.~\ref{fig:survey_fits_bkgsub_wide} and \ref{fig:survey_fits_bkgsub_E1}, the $p$-values of both our null model fit and the null model fit in~\cite{Cappelluti:2017ywp} are at or below $10^{-8}$, meaning that it is virtually impossible that the null model (or the signal model) describes the data to the level of statistical noise. We note that a poor fit to the data is expected due to the construction of the background dataset, in which multiple near-duplicates of the observed stowed data are summed, leading to underestimated uncertainties (see Sec.~\ref{sec:survey-reduction}). Furthermore, the uncertainties on the stowed dataset itself are likely underestimated due to the injection of fake events copied from one CCD chip to another~\cite{Hickox:2005dz}. These considerations strongly motivate shrinking the analysis window to mitigate mismodeling. As quantified by the $p$-values presented in Tab.~\ref{tab:BkgSubResults}, the fits over smaller energy windows are able to describe the data increasingly well, at least as quantified through the $p$-value of the null hypothesis fit. For example, the 0.5 keV window analysis has a $p$-value around 0.25.  On the other hand, the analysis in the narrowest energy window, still produces a $\sim$2$\sigma$ preference for the signal hypothesis.

\begin{table*}[!]{
    \ra{1.3}
    \begin{center}
    \begin{tabular}{@{\extracolsep{4pt}}L{0.11\textwidth}rP{0.15\textwidth}*{4}{P{0.15\textwidth}}@{}}
    \hlinewd{1pt}
    & & \textbf{Original} & \multicolumn{4}{c}{\textbf{This work}} \\
    \cline{3-3}\cline{4-7} \multicolumn{2}{l}{\textbf{Analysis Range}} & Full & Full & 3-6 keV & 1 keV & 0.5 keV \\ \hlinewd{1pt}
    \multirow{5}{*}{\parbox{0.11\textwidth} {\raggedright\textbf{Chandra\\Deep Field}}} & $\chi_\nu^2$ & $527.0/298$ & 942.7/622 & 595.3/406 & 136.2/134 & 69.7/66 \\ 
    & $p$ & $3\times10^{-15}$ & $1.4 \times 10^{-15}$& $2.5\times10^{-9}$ &  $0.43$& $0.36$ \\ 
    & $\hat{A}$ & $0.89_{-0.3}^{+0.3}$ & $1.1^{+0.4}_{-0.4}$ & $1.0^{+0.5}_{-0.5}$ & $0.9^{+0.5}_{-0.5}$ & $1.0^{+0.6}_{-0.6}$ \\ 
    & $t$ & $10.2$ & 6.4 & 4.3 & 3.0 & 3.0 \\ 
    & $A^{95}$ & \textemdash & $1.9$ & $1.8$ & $1.7$ & $2.1$ \\ \hlinewd{1pt}  
    \end{tabular}
    \end{center}}\caption{The same as Tab.~\ref{tab:Results}, but for the background-subtracted Chandra Deep Field analyses. \label{tab:BkgSubResults}}
\end{table*}

\subsection{Analysis of Background Data for Milky Way Survey Fields}
\label{app:DeepFieldBkgAnalysis}

Though we do not find a line when analyzing the data from the Milky Way Survey Fields without background subtraction in Sec.~\ref{sec:survey}, we do find one at moderate significance when performing background subtraction, as previously described, even in our narrowest analysis window where the $p$-value obtained under the null is not obviously disqualifying. This suggests that the presence of the line in the background subtracted data may be an artifact of the background subtraction itself.

\begin{figure*}[htb]  
\hspace{0pt}
\vspace{-0.2in}
\begin{center}
\includegraphics[width=0.99\textwidth]{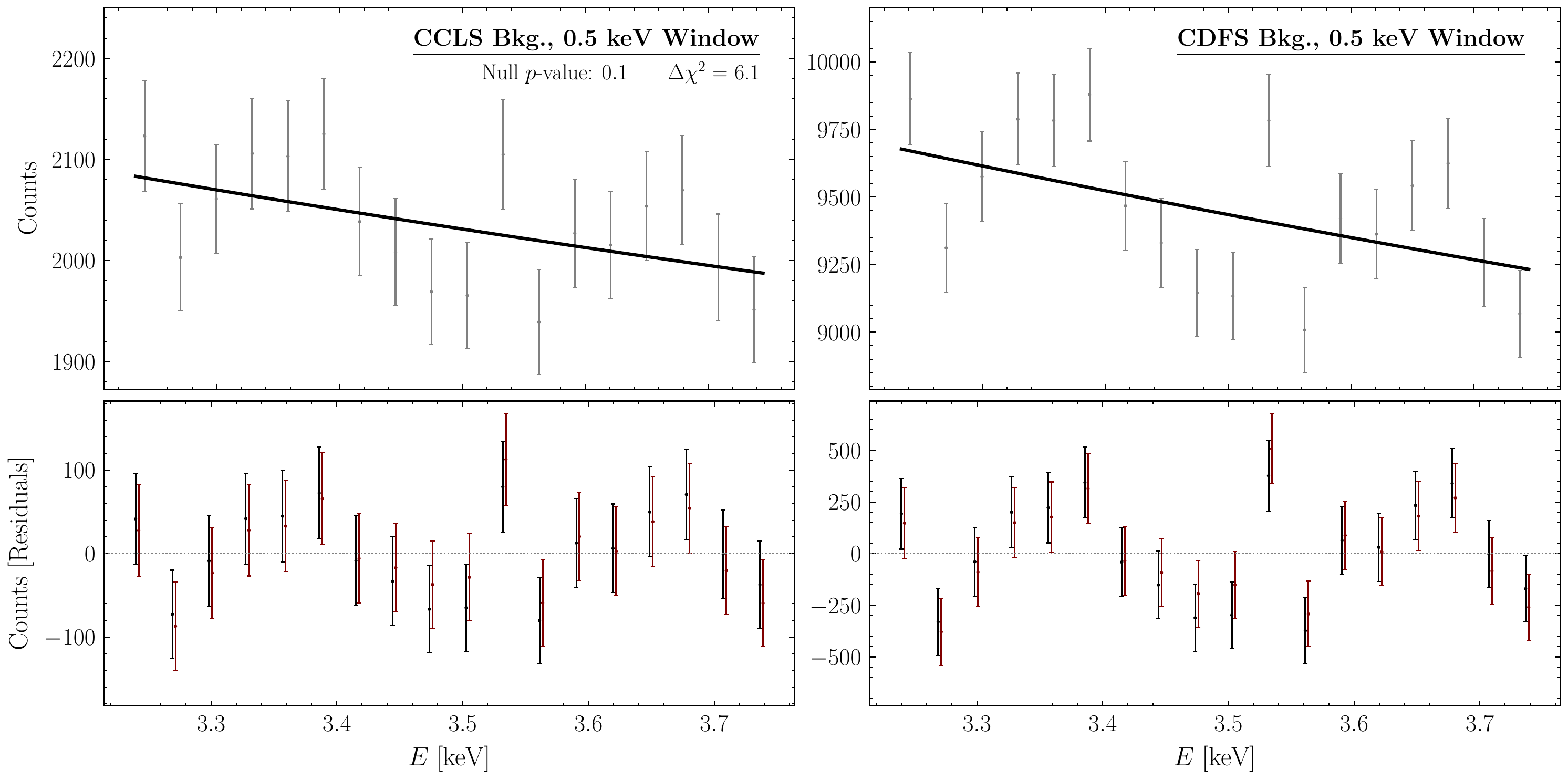}
\caption{The background data for Chandra CCLS (left) and CDFS (right) along with the best-fit null model for a 500 eV window centered at 3.51 keV. The bottom panels illustrate the residuals after subtracting the best-fit null and signal models.\label{fig:survey_fits_bkg_E500}}
\end{center}
\end{figure*}

We pursue this hypothesis by analyzing the background data which is subtracted from the observational data using a 500 eV window centered on the putative 3.51 keV line location. As the background-subtracted and background-unsubtracted analyses of the Survey Field data differ only in the exclusion/inclusion of an unfolded broken power law, we model the CDFS and CCLS continua with an unfolded power law. A broken power law is unnecessary as the fitted break energy found in Sec.~\ref{sec:survey} is outside our analysis window. No background lines are within our window, so we perform our analysis with only a candidate 3.51 keV line which is allowed to take positive or negative fluxes. 

The fits to these data are presented in Fig.~\ref{fig:survey_fits_bkg_E500}, with associated profiled likelihood in the right panel of Fig.~\ref{fig:survey_ll_bkgsub}. We note that the CCLS and CDFS background data are highly similar as they are produced from nearly identical sets of observations. Subtracting these data and adding errors in quadrature treats them as statistically independent, and thereby likely overestimates statistical precision in a background subtracted analysis. We find $q = 6.1$ for a line flux of approximately $-1.5\times 10^{-6}$ cts/cm$^2$/s, corresponding to $2.5\sigma$ evidence of a deficit. We conclude that the moderate significance excess in the background-subtracted analysis likely has its origin in the subtraction of a nearly equal significance deficit in the background data.

\clearpage

\bibliography{xmm}

\end{document}